\documentclass[pre,aps,eqsecnum,preprint]{revtex4}
\usepackage{graphicx}
\usepackage[dvips]{epsfig}
\usepackage{color}
\definecolor{red}{rgb}{1,0,0}
\definecolor{green}{rgb}{0.13,0.55,0.13}
\definecolor{blue}{rgb}{0,0,1}

\begin{document}

\title{Casimir interaction of rod-like particles in a two-dimensional critical system}

\author{ E. Eisenriegler$^{1}$ and T. W. Burkhardt$^{2}$ }

\affiliation{
$^1$Theoretical Soft Matter and Biophysics, Institute of Complex
Systems,\\
Forschungszentrum J\"ulich, D-52425 J\"ulich, Germany\\
$^2$Department of Physics, Temple University,
Philadelphia, PA 19122, USA}

\date{\today}

\begin{abstract}
We consider the fluctuation-induced interaction  of two thin, rod-like particles or ``needles" immersed in a two-dimensional critical fluid of Ising symmetry right at the critical point. Conformally mapping the plane containing the needles onto a simpler geometry in which the stress tensor is known, we analyze the force and torque between needles of arbitrary length, separation, and orientation. For infinite and semi-infinite needles we utilize the mapping of the plane bounded by the needles onto the half plane, and for two needles of finite length the mapping onto an annulus.  For semi-infinite and infinite needles the force is expressed in terms of elementary functions, and we also obtain analytical results for the force and torque between needles of finite length with separation much greater than their length. Evaluating formulas in our approach numerically for several needle geometries and surface universality classes, we study the full crossover from small to large values of the separation to length ratio. In these two limits the numerical results  agree with results for infinitely long needles and with predictions of the small-particle operator expansion, respectively. 

\end{abstract}

\maketitle

\section{Introduction}
\label{intro}
Two objects immersed in a near-critical fluid, for example
colloidal particles in a binary liquid mixture near the critical
point of miscibility, experience a long-range, fluctuation-induced
force
\cite{directmeasurement,directmeasurement2,FdG,Krech,BDT,Gambassi}.
Changes in the positions of the objects alter the space available
to the critically fluctuating fluid, and hence its free energy,
giving rise to an effective interaction of the objects.  In
analogy with the Casimir effect in quantum electrodynamics
\cite{KG,BMM,emig2008,gies}, this is known as the critical or
thermodynamic Casimir interaction.

The critical Casimir interaction displays a high degree of
universality, i.e. is largely independent of microscopic details
\cite{FdG,Krech,BDT,Gambassi}. It only depends on universal
properties of the fluid, the universality class of the boundary
between the fluid and the immersed particles, and geometrical
properties of the particles, such as their size, shape, and
relative position.

Particles immersed in a solution \cite{polcollint,ee2006,SHKD,KTL}
of long, flexible polymer chains or particles to which a polymer
chain is attached \cite{Magh,KTL} experience a similar Casimir
interaction due to fluctuations of the polymers \cite{dGbook}.
Fluctuations of the superfluid order parameter also lead to
critical Casimir forces, and this has been detected in wetting
films of
${^4}$He \cite{GCh1,Helium,O1}.% and $^3$He/$^4$He mixtures
%\cite{GCh2,MGD}

Binary liquid mixtures belong to the Ising universality class. The
surfaces of the immersed particles generally attract one of the
two components of the mixture preferentially,  corresponding to
($+$ or $-$) boundary conditions  in the Ising model. A surface
prepared to suppress the preference corresponds to free-spin
boundary conditions \cite{NHB}. In the terminology of surface
critical phenomena these two surface universality classes
\cite{hwd} are known as ``normal'' ($+$ or $-$) and ``ordinary''
($O$).

In studies of critical Casimir interactions, systems with planar
walls and systems with spherical particles have received the most
attention \cite{FdG,O1,Hasenbusch,evste,ber}. For non-spherical
particles the Casimir interaction depends on their orientation as
well as their separation, i.e., there is a torque as well as a
force. Recently the universal scaling form of the Casimir interaction
of a prolate uniaxial ellipsoid and a planar wall,
with pairs $++$ or $+-$ of boundary conditions on the two
surfaces, was calculated within mean-field theory by Kondrat et
al. \cite{kondrat}.

In this paper we derive exact results for the Casimir interaction
of two rod-like particles in a two-dimensional critical system in
the Ising universality class. The following considerations provide
some motivation:

(i) Recent experiments suggest that  biological membranes are
tuned  close to a critical point of miscibility in two dimensions
\cite{baumgartveatch}. The possibility of critical Casimir
interactions between inclusions in the membrane has been studied
by Machta et al. \cite{machta2012}.

(ii) Systems at the critical point are generally invariant not
only under scale transformations, but also under conformal or
angle-preserving coordinate transformations \cite{Cardy}. This has
far-reaching consequences for the Casimir interaction of two
particles, especially in two spatial dimensions
\cite{ber,VED,Bim}.

In general dimension $d$ the region outside two spherical
particles with arbitrary radii and separation can be conformally
mapped onto the region bounded by two concentric spheres using
homogeneous translations, rotations, and dilatations and the
inversion. Burkhardt, Eisenriegler, and Ritschel have shown \cite{ber} that
this mapping determines the asymptotic form of the Casimir
interaction both for large and small separation of the spheres in
an arbitrary critical medium, not necessarily Ising-like, in
arbitrary spatial dimension $d$.

In $d=2$ the conformal group is much richer than in general $d$.
Conformal mappings are generated by analytic functions, and
the doubly connected region outside two particles of {\it
arbitrary } shape can be conformally mapped onto the annulus
bounded by two concentric circles or, equivalently, onto the
surface of a cylinder of finite circumference and length. The
Casimir interaction of the particles in an infinite,
two-dimensional, Ising-like critical medium follows from the free
energy of the critical Ising model on the cylindrical surface
\cite{Cardy,ber,VED,Bim}, which Cardy \cite{Cardystrip} has derived in analytic form
for all aspect ratios and for all pairs of
boundary conditions $+$, $-$, and $O$ at the ends of the
cylindrical surface. Burkhardt and Eisenriegler \cite{ber} and
Machta et al. \cite{machta2012} followed this route in evaluating
the Casimir interaction of two particles with circular shape.
Bimonte et al. \cite{Bim} have given a general analysis of
asymptotic properties of the Casimir interaction  in critical
two-dimensional conformal field theories for two particles of
arbitrary shape, based on the mapping of the portion of the plane
outside the particles onto an annulus.

Being interested in the Casimir torque as well as the force, we
consider anisotropic particles. The rod-like particles in our
study have negligible width compared with their length, and we
model each particle as a segment of a straight line. This shape is
extremely simple and highly-anisotropic. Following Ref. \cite{VED}
we refer to the particles as ``needles" \cite{Stephan}. The approach we use,
which makes use of conformal mappings, is not limited to these
needles but is applicable, in principle, to particles in two
dimensions of arbitrary shape.

In Section \ref{forcetorque} and Appendix \ref{shiro} we show how
the force and torque between the particles are related to the stress
tensor \cite{Cardy}, which for our purpose is the quantity most
convenient to work with \cite{ber,VED,Bim}.

In Section \ref{infseminf} we calculate the critical Casimir force
between a semi-infinite needle and an infinite needle and between
two semi-infinite needles, with arbitrary relative position. These
are instructive cases to begin with since the calculations can be
carried out analytically, without special functions. The key step
is to generate by a conformal mapping $z(w)$ the complex $z$ plane with the
two embedded needles from the upper half $w$ plane
with the corresponding needles on the real axis.

In Section \ref{finfin} we discuss the more complicated
Schwarz-Christoffel transformation required for two needles of
finite length. It generates the complex $z$ plane, with an embedded
needle between points $z_1$ and $z_2$ and a second needle between
points $z_3$ and $z_4$, from an annulus, with circular needles on
the outer and inner boundaries.

In Section \ref{five} detailed results for the Casimir force and
torque between needles of finite length are presented for several
configurations of the needles. The results are consistent with
predictions of Vasilyev et al. \cite{VED}, who have studied the
Casimir interaction of the needles with Monte Carlo simulations
and calculated them  in certain symmetric cases with conformal
invariance methods, but without the generality of the  approach
considered below. The results of Section \ref{five} are also
asymptotically consistent with the predictions for 
infinite or semi-infinite needles derived in Section
\ref{infseminf}. Since the torque diverges for needles of infinite length,
checking its asymptotic behavior is more subtle and is addressed in 
Appendix C. 

For large separation of the needles in comparison with their
lengths \cite{distsmall}, the  numerical results of Section
\ref{five} reproduce the predictions of the ``small-particle
operator expansion" (SPOE). This expansion, which is reviewed in
Appendix \ref{SPE}, has proved to be extremely useful in studies
of the critical Casimir interaction and is similar in spirit to
the operator product expansion \cite{WK} in field theory. Large
needle separation corresponds to a small ratio of inner to outer
radius of the annulus, and the corresponding expansions in Section
\ref{finfin} and Appendix \ref{SPE} allow us to check the
agreement with the SPOE analytically.

The final section of the paper, Section \ref{concl},  contains a summary and
concluding remarks.
\newpage
\section{FORCE, TORQUE, AND THE STRESS TENSOR}
\label{forcetorque}
In this section the force and torque acting on needles in the $z$ plane are expressed in terms of the average complex stress tensor $\langle T(z) \rangle$. In subsequent sections these three quantities are calculated for several two-needle geometries of interest. In each case $\langle T(z) \rangle$  is obtained through a conformal mapping $z(w)$ of a simpler geometry in the $w$ plane, for which $\langle T(w) \rangle$ is known, onto the desired geometry in the $z$ plane, using the fundamental transformation property \cite{Cardy}
\begin{eqnarray} \label{stressstress}
\langle T(z) \rangle \, = \, {1 \over (z'(w))^{2}} \, \left[
\langle T(w) \rangle \, - \, {1 \over 24} \, S(w) \right].
\end{eqnarray}
Here $S(w)$ is the Schwarzian derivative
\begin{eqnarray} \label{Schwarz}
S(w) = {z'''(w)z'(w) - (3/2) (z''(w))^{2} \over (z'(w))^2} \equiv
{d^2 \over dw^2} \ln {dz \over dw} - {1 \over 2} \Bigl( {d \over
dw} \ln {dz \over dw} \Bigr)^2 \, ,
\end{eqnarray}
and the primes denote derivatives.

The transformation of the free energy under an arbitrary infinitesimal coordinate transformation is reviewed in Appendix A. According to Eq. (\ref{shiftz}) the components $f_x$ and $f_y$ of the force on needle
$I$ due to a second needle $II$ are given by
\begin{eqnarray} \label{force}
(f_x, \, f_y)/(k_B T) = -({\rm Im}, \, {\rm Re})\tau,
\end{eqnarray}
where
\begin{eqnarray} \label{force'}
\tau ={1\over\pi} \int_{{\cal C}_{I}} dz \, \langle T(z) \rangle =  {1\over\pi} \int_{{\cal C}} dw \, z'(w)\langle T(z) \rangle.
\end{eqnarray}
The  integration path ${\cal C}_{I}$ in the $z$ plane
encloses needle $I$ in a clockwise fashion, leaving needle $II$
outside, and ${\cal C}$ is the corresponding path in the $w$  plane, which  maps onto ${\cal C}_I$ under the conformal transformation
$z(w)$. With the help of  Eq. (\ref{stressstress}),
$\tau$ can be expressed as
\begin{eqnarray} \label{force''}
\tau = {1\over\pi}\int_{{\cal C}} dw \, {1 \over z'(w)} \, \left[ \langle
T(w) \rangle \, - \, {1 \over 24} \, S(w) \right]\,
\equiv \, \tau^{(T)} + \tau^{(S)} \, .
\end{eqnarray}

We define the torque $\Theta$ on a needle $I$ with fixed length, extending from $z_{1}$
to $z_{2}$, and forming an angle  $\Phi_{I} \equiv \Phi_{12}
= {\rm arg}(z_{1}-z_{2})$ with the $x$ axis, by 
\begin{equation}
\Theta=-(\partial/ \partial \Phi_{I}) \delta F \,.\label{definitionoftorque}
\end{equation}
Here $\delta F$ is the free energy of interaction \cite{delF}, and the derivative is taken for an infinitesimal rotation of needle $I$ about its midpoint with the midpoint  $z_{I} ={1\over 2}\thinspace(z_{1}+z_{2})$ fixed. According to Eqs. (\ref{definitionoftorque}) and
(\ref{rotatez}), the torque may be written as
\begin{eqnarray} \label{turn}
\Theta \, = \, - k_B T \, {\rm Re} \, \theta\,,
\end{eqnarray}
where
\begin{eqnarray} \label{turn'}
\theta \, = \, {1\over\pi}\int_{{\cal C}_{I}} dz \, \langle T(z)
\rangle \left(z-z_{I}\right)
\end{eqnarray}
and the integration path ${\cal C}_{I}$ is the same as in Eq. (\ref{force'}).
Since
\begin{eqnarray} \label{thetaw''}
z-z_{n} \, = \, \int_{w_n}^{w} d\tilde{w} \, (dz/d\tilde{w}) \,
\equiv \, \zeta_{n}(w) \, ; \; n=1,2 \, ,
\end{eqnarray}
where $w_{n}$ is the point in the $w$ plane which maps onto needle endpoint $z_{n} \equiv
z(w_{n})$, the torque may be  expressed as
\begin{eqnarray} \label{thetaw}
\Theta \, = \, - k_B T \, {\rm Re}\left(\theta^{(T)} \, + \, \theta^{(S)}\right)\,,
\end{eqnarray}
where
\begin{eqnarray} \label{thetaw'}
\Bigl\{\theta^{(T)} \, , \, \theta^{(S)} \Bigr\} = {1\over 2\pi}\int_{{\cal C}}
dw {1 \over dz/dw} \,\Biggl\{ \langle T(w) \rangle \, ,
\,  -  {1 \over 24} \, S(w) \Biggr\}
\Bigl[ \zeta_{1}(w)+\zeta_{2}(w) \Bigl]\,,
\end{eqnarray}
in terms of $\langle T(w) \rangle$.
Note that the contributions $\tau^{(S)}$ in Eq. (\ref{force''}) and $\theta^{(S)}$  in Eq. (\ref{thetaw'}) which involve
the Schwarzian derivative are purely geometrical and do not depend on the surface universality classes of the needles.

\newpage

\section{INTERACTIONS OF INFINITE AND SEMI-INFINITE NEEDLES}
\label{infseminf}

\subsection{Force between a semi-infinite and an infinite needle} \label{semiinhalf}
In this and subsequent sections we use the notation $z=r_{x}+ir_{y}$ and $w=\rho_{u}+i\rho_{v}$ for the complex variables $z$ and $w$ and their real and imaginary parts. 

The conformal transformation $z(w)$, where
\begin{eqnarray}
&&z'(w)={\cal A} e^{i\alpha}w
^{-\alpha/\pi-1}(w-1),\label{is1}\\
&&z(w)=\pi
{\cal A} e^{i\alpha}w^{-\alpha/\pi}\left({w\over \pi-\alpha}+{1\over\alpha}\right),
\label{infandsemiinfmapping}
\end{eqnarray}
${\cal A}$ is a positive real constant, and $0<\alpha<\pi$, considered in  \S 12.1 of Kober \cite{Kob}, maps the upper half
$w$ plane, with semi-infinite needles along the positive and
negative $u$ axes, onto the upper half $z$ plane with two transformed
needles. Needle $II$, the image of the negative $u$ axis, is infinitely long
and corresponds to the entire $x$ axis. Needle $I$, the image of the
positive  $u$ axis, is semi-infinite and extends from the point
\begin{equation} \label{end}
z(1)={\pi^2\over \alpha(\pi-\alpha)}{\cal A}e^{i\alpha}\label{z(-1)}\label{z(1)infsemi}
\end{equation}
to $\infty$, forming an angle $\alpha$ with the $x$ axis.

The integrand  in expressions (\ref{force}), (\ref{force'}), and (\ref{force''}) for the Casimir force
follows from Eqs. (\ref{Schwarz}), (\ref{is1}), and the corresponding stress tensor \cite{BX}
\begin{equation} \label{t12}
 \langle T(w)\rangle={{\tilde t}\over w^2}={(0,0,{1\over 2},{1\over 16})\over w^2}\label{t12etc}
\end{equation}
for $(OO,++,+-,O+)$  boundary conditions on the two needles and is given by
\begin{eqnarray}
&& z'(w)\thinspace\langle T(z)\rangle= -{e^{-i\alpha}\over 48 \pi^2 {\cal A}}\thinspace{w^{\alpha/\pi-1}\over (w-1)^3}\nonumber\\
&&\qquad\times\left[\alpha(2\pi-\alpha)w^2 -2(2\pi-\alpha)(\pi+\alpha)w+\pi^2-\alpha^2-48\pi^2
{\tilde t}(w-1)^2\right].\label{integrand1}
\end{eqnarray}
In Eq. (\ref{force'}) integrating clockwise along the edges of needle $I$ in the $z$ plane (path ${\cal C}_I$) corresponds to integrating along 
the $u$ axis from $\rho_{u}=0$ to $+\infty$ in the $w$ plane (path ${\cal C}$). Combining Eqs. (\ref{force'}) and (\ref{integrand1}), evaluating the integral, and making use of Eqs. (\ref{force}) and (\ref{z(1)infsemi}), we obtain $f_x=0$, as expected since needle $II$ is infinite, and \cite{intform}
\begin{equation}
\tau=-{f_y\over k_BT}={1\over 96\thinspace r_{y}(1)}\thinspace
{(2\pi-\alpha)(\pi+\alpha)-96\pi^2 {\tilde t}\over
\alpha(\pi-\alpha)}\,. \label{fy3}
\end{equation}
Here  $r_{y}(1)={\rm Im}[z(1)]$  is the distance of the tip of needle $I$ from needle $II$, and we have used the relation $r_{y}(1)=\pi^2 {\cal A}(\sin\alpha)/[\alpha(\pi-\alpha)]$, which follows from Eq. (\ref{z(1)infsemi}).

As expected, $f_y$  in Eq. (\ref{fy3}) is an even function of the
deviation $\gamma=\alpha-{1\over 2}\pi$ from perpendicular
orientation of the needles and diverges in the limit
$\gamma\to\pm{1\over 2}\pi$, corresponding to parallel needles.  From the values of ${\tilde t}$ in Eq.
(\ref{t12etc}), it follows that  the force between the needles is attractive for  $OO$ and $++$ boundaries and
repulsive for  $+-$ and $O+$ boundaries, with the strongest force in the $+-$ case.

\subsection{Force between two semi-infinite needles} \label{semisemi}

The conformal transformation $w(z)$, where
\begin{eqnarray} 
&&z'(w)={\cal B}e^{i\alpha}w
^{-\alpha/\pi-1}(w-1)(w+b),\label{tn1}\\
&&z(w)=\pi
{\cal B}\left[ e^{i\alpha}w^{-\alpha/\pi+1}\left({w\over
2\pi-\alpha} +{b-1\over \pi-\alpha}+{b\over\alpha w}\right)\right.\nonumber\\&&\qquad\qquad\qquad
-\left. b^{-\alpha/\pi+1} \left({b\over 2\pi-\alpha}-{b-1\over
\pi-\alpha}+{1\over\alpha}\right)\right],\label{twoneedlemapping}
\end{eqnarray} where ${\cal B}$ and $b$ are positive real constants and $0<\alpha<\pi$,  considered in \S 12.3 of Kober \cite{Kob}, maps the
upper half $w$ plane, with semi-infinite needles along the
positive and negative $u$ axes, onto the full $z$ plane with two
embedded semi-infinite needles. Needle $II$, the image of the negative $u$ axis, coincides with the positive $x$ axis. Needle $I$, the image of the positive $u$ axis,  extends from the point
\begin{equation} \label{end'}
z(1)=-{\cal B}\pi^2\,{\left[\alpha-b(2\pi-\alpha)\right]e^{i\alpha}+b^{-\alpha/\pi+1}(2\pi-\alpha-b\alpha)\over \alpha(\pi-\alpha)(2\pi-\alpha)}\label{z(1)semisemi}
\end{equation}
to $\infty$, forming an angle $\alpha$ with needle $II$.

The integrand  in expressions (\ref{force}), (\ref{force'}), and (\ref{force''}) for the Casimir force
follows from Eqs. (\ref{Schwarz}), (\ref{tn1}), and (\ref{t12etc}) and
is given by
\begin{eqnarray} 
z'(w)\langle T(z)\rangle&=&{e^{-i\alpha}\over
48\pi^2 {\cal B}}\,{w^{\alpha/\pi-1}\over(w-1)^3(w+b)^3}\,\left\{(3\pi-\alpha)(\pi-\alpha)w^4+2(1-b)\alpha(3\pi-\alpha)w^3\right.\nonumber\\
  &&+\left[-\alpha(2\pi-\alpha)+2b(5\pi^2+4\pi\alpha-2\alpha^2)-b^2\alpha(2\pi-\alpha)\right]w^2\nonumber\\&&
 - 2b(1-b)(2\pi-\alpha)(\pi+\alpha)w- b^2(\pi^2-\alpha^2)\nonumber\\ &&\left.
+48\pi^2{\tilde t}(w-1)^2(w+b)^2\right\}\,.\label{integrand2}
\end{eqnarray}
Integrating along path ${\cal C}$ in Eq. (\ref{force'}) again amounts to integrating along the $u$ axis from $\rho_{u}=0$ to $+\infty$ in the $w$ plane, passing above the pole at $w=1$. Combining Eqs. (\ref{force'}) and (\ref{integrand2}), evaluating the integral, and making use of Eq. (\ref{force}), we obtain \cite{intform}
\begin{eqnarray}
&&\tau=-{f_y+if_x\over k_BT}={1\over 96 \pi^2 {\cal B}(1+b)^3\sin\alpha}\times \nonumber\\
&&\times\left\{\left[\alpha(3\pi-\alpha)+2(3\pi-\alpha)(\pi+\alpha)b+(2\pi-\alpha)(\pi+\alpha)b^2-96\pi^2(1+b)^2{\tilde t}\right]+e^{-i\alpha} b^{\alpha/\pi-1}\right.\nonumber \\
&&\times\left.\left[(2\pi-\alpha)(\pi+\alpha)+2(3\pi-\alpha)(\pi+\alpha)b+\alpha(3\pi-\alpha)b^2 - 96\pi^2 (1 + b)^2 {\tilde t}\right]\right\},\label{fxy}
\end{eqnarray}
where the parameters ${\cal B}$ and $b$ are related to the endpoint $r_{x}(1),r_{y}(1)$  of needle $I$ by Eq. (\ref{z(1)semisemi}).

Since needle $II$ corresponds to the positive $x$ axis, one expects to recover the results of the preceding subsection, in which needle $II$ is infinite, in the limit $r_{x}(1)\to+\infty$ with $r_{y}(1)$ and $\alpha$ fixed. According to  Eq. (\ref{z(1)semisemi}) this limit is achieved on substituting ${\cal B}={\cal A}/b$ in the equation and then taking the limit $b\to\infty$ with ${\cal A}$ fixed. In this limit the derivative (\ref{tn1}) reduces to Eq. (\ref{is1}), the integrand (\ref{integrand2}) reduces to (\ref{integrand1}), and the Casimir force (\ref{fxy}) reduces to (\ref{fy3}).

\newpage

\section{INTERACTIONS OF NEEDLES OF FINITE LENGTH}
\label{finfin}

In this section the approach for infinite and semi-infinite needles is extended to needles of finite length. The region
outside two finite needles, which is doubly connected, is generated from an annulus bounded by two concentric circles,
for which the thermal average of the stress tensor is known.

\subsection{Two needles of finite length} \label{twofinite}

\subsubsection{Conformal mapping} \label{map}

An arbitrary configuration of two non-overlapping needles $I$ and
$II$ with finite lengths $D_{I}$ and $D_{II}$ in the $z$ plane can
be generated by a conformal transformation $z(w)$ of the
Schwarz-Christoffel type which maps the interior of the annulus
$h<|w|<1$ in the $w \equiv |w|e^{i\varphi}$ plane onto the region
outside the two needles in the $z$ plane. The desired mapping is a special
case of the mapping onto the region outside two non-overlapping polygons derived by Akhiezer in
1928 and given at the end of  \S 48 of Ref. \cite{Akh}. In the special case in which the polygons reduce to needles, the mapping $z(w)$ has the  derivative
\begin{equation}
z'(w)={\mu\over w^2}\;{\prod \limits_{\ell=1}^{4} \vartheta_1 \left( (2\pi i)^{-1} \ln(w/a_\ell) \right)\over \vartheta_{1}^{2} \left( (2\pi i)^{-1} \ln (w/c)\right)\thinspace\vartheta_{1}^{2} 
\left((2\pi i)^{-1} \ln (wc)\right)}\,,\label{zprimetheta1}
\end{equation} 
in terms of the elliptic theta functions $\vartheta_{1}$ and constants $\mu$, $a_\ell$, and $c$ defined in \cite{Akh}. 

Substituting  $a_\ell=w_\ell$, $c=Ch^{1/2}$, $\mu=\left(w_1 w_2 w_3 w_4\right)^{-1/2}A$ in Eq. (\ref{zprimetheta1}), and using the expression for $\vartheta_1$ in Table IX of \cite{Akh}, we obtain the useful product representation
\begin{eqnarray} \label{dzdw}
&&z'(w) =  {A \over w^{2}} \times \\
&& \times \prod \limits_{k=1}^{\infty} {\prod \limits_{\ell=1}^{4}
(1-h^{2k-2}w/w_{\ell})(1-h^{2k}w_{\ell}/w) \over
(1-h^{2k-5/2}w/C)^{2}(1-h^{2k+1/2}C/w)^{2} (1-h^{2k-3/2}w
C)^{2}(1-h^{2k-1/2}/(w C))^{2}} \nonumber \, .
\end{eqnarray}
We will see that $h\ll 1$ for needles short in comparison with their separation, and in this regime the representation (\ref{dzdw}) is especially convenient \cite{but}.

The complex constant $A$ in Eq. (\ref{dzdw}) corresponds to a homogeneous
rotation and dilatation, the positive real constant $C$, with $h^{1/2}<C<h^{-1/2}$, characterizes the value $Ch^{1/2}$
of $w$ which is mapped to $z=\infty$, and the points $w_{1}=e^{i
\varphi_{1}}, \, w_{2}=e^{i \varphi_{2}}$ and $w_{3}=he^{i
\varphi_{3}}, \, w_{4}=he^{i \varphi_{4}}$ on the outer and inner
boundary circles of the annulus are the pre-images of the endpoints
$z_{1}, \, z_{2}$ and $z_{3}, \, z_{4}$ of needles $I$ and
$II$, respectively. This is evident from the changes
$dz$ corresponding to displacements  $dw=d(e^{i\varphi})$ and $dw=h \, d(e^{i\varphi})$ along the outer and inner boundaries of the annulus,
for which Eq. (\ref{dzdw}) implies
\begin{eqnarray} \label{out}
dz &=& \, {i \,
A \over h} C^{2} e^{-i(\varphi_{3}+\varphi_{4})} \, {\cal
G}(\varphi;\varphi_{1},\varphi_{2}) \, {\cal P}(\varphi; \,
\varphi_{1},\varphi_{2}; \, \varphi_{3},\varphi_{4}; \, C; \, h)
\, d\varphi
\end{eqnarray}
and
\begin{eqnarray} \label{in}
dz&=& \, {i
\, A \over h} \, {\cal G}(\varphi;\varphi_{3},\varphi_{4}) \,
{\cal P}(\varphi; \, \varphi_{3},\varphi_{4}; \,
\varphi_{1},\varphi_{2}; \, C^{-1}; \, h) \, d\varphi \, ,
\end{eqnarray}
respectively. Here
\begin{eqnarray} \label{calG}
{\cal G}(\varphi;\varphi_{J},\varphi_{K}) \, &=& \, e^{-i\varphi}
\, \Bigl( 1-e^{i(\varphi-\varphi_{J})} \Bigr)\, \Bigl(
1-e^{i(\varphi-\varphi_{K})} \Bigr) \nonumber \\
&=& \, -4 \, e^{-i(\varphi_{J}+\varphi_{K})/2} \, \sin
{\varphi-\varphi_{J} \over 2} \, \sin {\varphi-\varphi_{K} \over
2}
\end{eqnarray}
and
\begin{eqnarray} \label{out'}
{\cal P}(\varphi; \, \varphi_{1},\varphi_{2}; \,
\varphi_{3},\varphi_{4}; \, C; \, h) \, \equiv \, \prod
\limits_{k=1}^{\infty} {\prod \limits_{n=1,2}
|1-h^{2k}e^{i(\varphi-\varphi_{n})}|^{2} \, \prod \limits_{m=3,4}
|1-h^{2k-1}e^{i(\varphi-\varphi_{m})}|^{2} \over
|1-h^{2k-1/2}C^{-1}e^{i\varphi}|^{4}|1-h^{2k-3/2}Ce^{i\varphi}|^{4}}
\, ,
\end{eqnarray}
which is always positive.

As $\varphi$ varies, the argument of $dz$ in Eq.
(\ref{out}) stays constant except at
$\varphi=\varphi_{1}$ and $\varphi=\varphi_{2}$ where, due to the
factor ${\cal G}(\varphi;\varphi_{1},\varphi_{2})$, it changes by
$\pi$. This corresponds to constant slopes along the two sides of needle $I$ with endpoints
$z_{1}$ and $z_{2}$. Analogous results for needle $II$ follow from Eq. (\ref{in}).

Moreover, moving counterclockwise inside the annulus close to the outer and inner boundary circle corresponds to encircling needle $I$ clockwise and needle $II$ counterclockwise, respectively, in the $z$ plane. This is most easily verified near the needle tips $z_\ell$, where, due to Eq. (\ref{dzdw}), $dz \equiv z'(w) dw = {\rm const} \times (w-w_\ell) dw$, since $w-w_\ell$ turns 180 degrees clockwise and counterclockwise, respectively, on passing point $w_\ell$ on the outer and inner boundary.

Without loss of generality and for later convenience we assume
\begin{eqnarray} \label{philhat}
-\pi < \varphi_{\ell} \leq \pi \; ; \quad\ell=1,2,3,4 \,
\end{eqnarray}
for the arguments of the four pre-images $w_\ell$ of the needle
ends.

The mapping $z=z(w)$ is required to be single-valued, so that the displacement $z(w_{a})-z(w_{b})=\int_{w_{b}}^{w_{a}} (dz/dw) dw$ for any two points
$w_{a}$ and $w_{b}$ in the annulus is independent of the integration path.
For $w_{a}=w_{b}$ the integral must vanish, even if the path
encloses the inner boundary circle or the singularity at
$w=Ch^{1/2}$. To ensure this, we require that the integrals 
\begin{eqnarray} \label{condi}
\int_{\varphi=-\pi}^{\varphi=\pi} \Biggl({dz \over dw} dw
\Biggr)_{w=e^{i \varphi}} \, = \, 0 \, , \quad
\int_{\varphi=-\pi}^{\varphi=\pi} \Biggl({dz \over dw} dw
\Biggr)_{w=h e^{i \varphi}} \, = \, 0 \,
\end{eqnarray}
around the outer and inner boundary
circles vanish. On inserting (\ref{out})-(\ref{out'}) in
(\ref{condi}) and discarding $\varphi$-independent complex factors,
the conditions (\ref{condi}) imply the vanishing of two real functions of
the six real parameters $\varphi_{1}, .. , \varphi_{4}, C, h$.
This leaves four independent parameters (apart from the complex
constant $A$), consistent with the four degrees of freedom
needed to specify (apart from homogeneous translations,
rotations, and dilatations) a configuration of two needles, for example,
the two lengths and the two angles the needles form with the
vector between their midpoints. According to Eqs. (\ref{out}) and (\ref{in}), the second of the conditions (\ref{condi})
follows from the first on exchanging the pairs
$\varphi_{1}, \varphi_{2}$ and $\varphi_{3}, \varphi_{4}$ and
replacing $C$ by $C^{-1}$. This leads from one
allowed parameter set to another. Explicit expressions for
small $h$ are given in Eqs. (\ref{varphiN'}) below.

In conformity with above remarks, we use the notation
\begin{eqnarray} \label{shortnot}
z_{1}-z_{2}=z_{12}=|z_{12}| \, e^{i\Phi_{12}} \equiv D_{I} \,
e^{i\Phi_{I}} \, &,& \, z_{3}-z_{4}=z_{34}=|z_{34}| \, e^{i\Phi_{34}}
\equiv D_{II} \, e^{i\Phi_{II}} \, , \nonumber \\
(z_{1}+z_{2})/2=z_{I} \, &,& \, (z_{3}+z_{4})/2 =z_{II} \, ,
\end{eqnarray}
and
\begin{eqnarray} \label{shortnot'}
z_{I}- z_{II}=z_{I,II}
\end{eqnarray}
for the needle vectors, the positions of their midpoints,
and their separation vector.

The needle vectors $z_{12}$ and $z_{34}$ follow from the first and
second integrals in Eq. (\ref{condi}) on replacing the lower and upper
limits $-\pi$ and $\pi$ by $\varphi_{2}$ and $\varphi_{1}$ and
by $\varphi_{4}$ and $\varphi_{3}$, respectively. This yields
\begin{eqnarray} \label{needvec}
z_{12}&=&e^{-i(\varphi_{3}+\varphi_{4})} (iA/h) C {\cal
N}(\varphi_{1},\varphi_{2}; \, \varphi_{3},\varphi_{4}; \, C; \,
h) \, , \nonumber \\
z_{34}&=&(iA/h) C {\cal N}(\varphi_{3},\varphi_{4}; \,
\varphi_{1},\varphi_{2}; \, C^{-1}; \, h)\, ,\label{ea}
\end{eqnarray}
where, on using Eqs. (\ref{calG}), (\ref{out'}),
\begin{eqnarray} \label{needvec'}
{\cal N}(\varphi_{1},\varphi_{2}; \, \varphi_{3},\varphi_{4}; \,
C; \, h) \, = \, e^{-i(\varphi_{1}+\varphi_{2})/2} \, {\rm sgn\thinspace}(\varphi_{1}-\varphi_{2}) \, P(\varphi_{1},\varphi_{2}; \,
\varphi_{3},\varphi_{4}; \, C; \, h) \, .
\end{eqnarray}
Here
\begin{eqnarray} \label{needvec''}
&&P(\varphi_{1},\varphi_{2}; \, \varphi_{3},\varphi_{4}; \, C; \,
h)= \nonumber \\
&& \qquad=4C \, \Big| \int_{\varphi_{2}}^{\varphi_{1}}d\varphi \;
\sin {\varphi-\varphi_{1} \over 2} \, \sin {\varphi-\varphi_{2}
\over 2} \; {\cal P}(\varphi; \, \varphi_{1},\varphi_{2}; \,
\varphi_{3},\varphi_{4}; \, C; \, h)\Big| \, ,\label{eb}
\end{eqnarray}
where ${\cal P}$ is given in Eq. (\ref{out'}). 

For the angle enclosed by the two needles, Eqs. (\ref{ea})-(\ref{eb})  imply the simple
relation
\begin{eqnarray} \label{enclangle''}
e^{i(\Phi_{12}-\Phi_{34})} \, = \, e^{-i(\varphi_{1}+ \varphi_{2}+
\varphi_{3}+\varphi_{4})/2} \, {\rm sgn\thinspace}(\varphi_{1}- \varphi_{2})
\, {\rm sgn\thinspace}(\varphi_{3}-\varphi_{4}) \, .
\end{eqnarray}
Note that in the sector (\ref{philhat}) the complex numbers
$e^{-i(\varphi_{1}+\varphi_{2})/2}$ and
$e^{-i(\varphi_{3}+\varphi_{4})/2}$ and the signs of $
\varphi_{1}- \varphi_{2}$ and $\varphi_{3}-
\varphi_{4}$ are uniquely determined by $w_{1}, w_{2}$ and by
$w_{3}, w_{4}$, respectively.
%As long as we consider needles with
%identical surface properties on both rims the free energy of
%interaction, force, and torque will not change by rotating one
%needle by 180 degrees about its center and the signs in
%(\ref{enclangle''}) will not affect these quantities.

For the ratio of needle lengths Eqs.
(\ref{needvec})-(\ref{needvec''}) yield
\begin{eqnarray} \label{lengthrat}
D_{I}/D_{II} \equiv |z_{12}|/|z_{34}|=P(\varphi_{1},\varphi_{2};
\, \varphi_{3},\varphi_{4}; \, C; \, h)/P(\varphi_{3},\varphi_{4};
\, \varphi_{1},\varphi_{2}; \, C^{-1}; \, h) \, ,
\end{eqnarray}
so that exchanging the pairs $\varphi_{1}, \varphi_{2}$ and
$\varphi_{3}, \varphi_{4}$ and replacing $C$ by $C^{-1}$ changes
$D_{I}/D_{II}$  to its inverse. For the special parameter sets $C=1$
with either $\varphi_{3}=\varphi_{1}$, $\varphi_{4}=\varphi_{2}$
or $\varphi_{3}=-\varphi_{1}$, $\varphi_{4}=-\varphi_{2}$, the two
needles have equal lengths $|z_{12}|=|z_{34}|$. For the second set
this follows from Eq. (\ref{out'}), which implies ${\cal P}(\varphi; \,
\varphi_{1},\varphi_{2}; \, -\varphi_{1},-\varphi_{2}; \, C; \,
h)={\cal P}(-\varphi; \, -\varphi_{1},-\varphi_{2}; \,
\varphi_{1},\varphi_{2}; \, C; \, h)$, yielding
$P(\varphi_{1},\varphi_{2}; \, -\varphi_{1},-\varphi_{2}; \, 1; \,
h)=P(-\varphi_{1},-\varphi_{2}; \, \varphi_{1},\varphi_{2}; \, 1;
\, h)$.

We mention another, rather obvious, property of the transformation
(\ref{dzdw}): Changing the parameters from
$(\varphi_{1},\varphi_{2};\varphi_{3},\varphi_{4};C;h)$ to
$(-\varphi_{1},-\varphi_{2};-\varphi_{3},-\varphi_{4};C;h)$, i.e.,
changing all four $w_\ell$ to $w_\ell^{\star}$, leads from one
single-valued mapping to another, in which the needle
configuration is changed from $(z_{12},z_{34};z_{I,II})$ to
$(z_{12}^{\star},z_{34}^{\star};z_{I,II}^{\star})$, assuming $A$ is real. Here 
and below an asterisk denotes complex
conjugation. 

Except for the enclosed angle it is, in general, not obvious how to
choose the parameters in the transformation (\ref{dzdw}) 
to generate a given configuration of the two needles. Here we list some simple classes (A)-(E) of needle configurations which only require a parameter search in a reduced subspace. We choose the vector between the needle centers to be parallel to the real axis, so that $z_{I,II}=|z_{I,II}|$, with needle $I$ to the right of needle $II$, and refer to the ratios $|z_{12}|/|z_{I,II}|$ and $|z_{34}|/|z_{I,II}|$ as the ``reduced needle lengths''.
%For
%simplicity we choose the locations $z_{I}=1/2$ and $z_{II}=-1/2$
%for the two needle midpoints so that $z_{I,II}=1$.

(A) Symmetric-perpendicular configurations of two needles of arbitrary reduced lengths with the symmetry
of the letter T, corresponding to Fig. \ref{FigABCD}-(A): To be specific, we consider the needle vectors
$z_{12}=i|z_{12}|$ and $z_{34}=-|z_{34}|$. These configurations can be generated from 
parameters in the subspace
$(\varphi_{1},\varphi_{2},\varphi_{3},\varphi_{4})=(-|\varphi_{1}|,
|\varphi_{1}|, 0, \pi)$ where $w_{2}=w_{1}^{\star}, \, w_{3}=h,
\, w_{4}=-h$. The reason is that on choosing $A$ real, the
integrals over (\ref{dzdw}) from $w=w_{2}$ to $w=w_{3}$ and from
$w=w_{1}$ to $w=w_{3}$ (and likewise those from $w=w_{2}$ to
$w=w_{4}$ and from $w=w_{1}$ to $w=w_{4}$) are complex conjugates, implying the properties
$z_{2}-z_{3}=(z_{1}-z_{3})^{\star}$ and
$z_{2}-z_{4}=(z_{1}-z_{4})^{\star}$. Since the expression multiplying $d\varphi$ in  Eq. (\ref{in}) is
an odd function of $\varphi$, the integral in Eq. (\ref{condi})
around the inner circle vanishes for all values of $\varphi_{1}$,
$C$, and $h$. The requirement that the integral over the outer circle 
vanish implies a relation $\varphi_{1}=\psi(C,h)$, leaving $C$ and $h$ free to generate given
values for the two reduced needle lengths. Note, finally, that the general
enclosed angle relation (\ref{enclangle''}) is satisfied, since both of its sides equal $-i$ in the above subspace of parameters
in the annulus and for the needle configuration in which $e^{i
\Phi_{12}}=i$, $e^{i \Phi_{34}}=-1$.
 
(B) Symmetric-parallel configurations of two needles with arbitrary reduced lengths perpendicular to the vector between
their centers, corresponding to Fig. \ref{FigABCD}-(B): To be specific we choose $z_{12}=i|z_{12}|$ and
$z_{34}=i|z_{34}|$, so that the needles are parallel to the imaginary axis.
These configurations are generated by
$(\varphi_{1},\varphi_{2},\varphi_{3},\varphi_{4})=(-|\varphi_{1}|,
|\varphi_{1}|, -|\varphi_{3}|,|\varphi_{3}|)$, i.e., by
$w_{2}=w_{1}^{\star}$, $w_{4}=w_{3}^{\star}$ since, as in case (A), for real values of the parameter $A$ there is reflection symmetry about the real
axis, and the integrals over (\ref{dzdw}) from $w=w_{2}$ to
$w=w_{3}$ and from $w=w_{1}$ to $w=w_{4}$ (and likewise those from
$w=w_{2}$ to $w=w_{4}$ and from $w=w_{1}$ to $w=w_{3}$) are
complex conjugates, implying the properties
$z_{2}-z_{3}=(z_{1}-z_{4})^{\star}$ and
$z_{2}-z_{4}=(z_{1}-z_{3})^{\star}$. The vanishing of the two
integrals (\ref{condi}) implies relations
$\varphi_{1}=\chi(C,h)$ and $\varphi_{3}=\omega(C,h)$, and the
two parameters $C$ and $h$ can be adjusted to generate the two reduced
needle lengths. The enclosed angle relation (\ref{enclangle''}) is
satisfied, with both sides equal to 1.

(C) Configurations of two needles which are mirror symmetric about the imaginary
axis, corresponding to Fig. \ref{FigABCD}-(C): Here  $z_{12}=|z_{12}|e^{i \Phi_{12}}$, $z_{34}=|z_{12}|e^{i
(\pi-\Phi_{12})}=-{z}_{12}^{\star}$. The needles have the same arbitrary reduced length, and with no loss of generality the
angle $\pi-2\Phi_{12}$ between them can be restricted to values between 0 and $\pi$. 

(D) Antiparallel configurations of two needles with the same arbitrary reduced length, corresponding to Fig. \ref{FigABCD}-(D): Here
$z_{12}=|z_{12}|e^{i \Phi_{12}}=-z_{34}$, where the angle
$\Phi_{12}$ is arbitrary \cite{rotateD}.

The needle configurations (C) and (D) are generated by $C=1$ in
both cases  and by $\varphi_{3}=\varphi_{1}, \,
\varphi_{4}=\varphi_{2}$ in case (C) and
$\varphi_{3}=-\varphi_{1}, \, \varphi_{4}=-\varphi_{2}$ in case
(D). In both subspaces the lengths $|z_{12}|$, $|z_{34}|$ of the
two needles are equal for arbitrary values of the three parameters
$\varphi_{1}$, $\varphi_{2}$, and $h$, and the two conditions in
(\ref{condi}) reduce to a single
condition (see the remarks below Eqs. (\ref{lengthrat}) and
(\ref{condi}), respectively). This leaves two free parameters, which
can be adjusted to generate the given common reduced length of the needles
and the angle $\Phi_{12}$ needle $I$ forms with the distance
vector $z_{I,II}$ between the needle midpoints. In case (C) the enclosed
angle relation (\ref{enclangle''}) predicts
$e^{2i\Phi_{12}}=-e^{-i(\varphi_{1}+\varphi_{2})}$, and in case (D)
it is satisfied since both sides equal $-1$.

Typical configurations from classes (A)-(D) are shown in Fig. \ref{FigABCD}. Classes (B), (C), and (D) encompass two
particularly simple needle configurations for which the conformal mapping 
can be found in the literature:

(i) Collinear needles of equal length with
$\Phi_{12}=0$ and $\Phi_{34}=\pi$ are generated by $(w_{1}, \,
w_{2}, \, w_{3}, \, w_{4})=(1, \, -1, \, h, \, -h)$ and $C=1$,
which is a special case of both (C) and (D). In this case the two
conditions (\ref{condi}) are satisfied, since both integrands
(\ref{out}) and (\ref{in}) are odd functions of $\varphi$. The reduced
needle length is determined by the parameter
$h$. The corresponding conformal transformation $z(w)$ is
discussed in Refs. \cite{Akh,Kob}.
%, and the stress tensor averages
%$\langle T(z) \rangle$ for this needle configuration are given in
%Ref. \cite{VED}.

(ii) Two parallel needles of equal length with
$\Phi_{12}=\Phi_{34}=\pi /2$, i.e., a configuration with the
symmetry of the letter H: This needle geometry is generated by $(\varphi_{1}, \,
\varphi_{2}, \, \varphi_{3}, \, \varphi_{4})=(-|\varphi_{1}|, \,
|\varphi_{1}|, \, -|\varphi_{1}|, \, |\varphi_{1}|)$ and $C=1$,
which is a special case of both (B) and (C). The two parameters
$|\varphi_{1}|$ and $h$ are chosen to satisfy the two {\it
identical} conditions (\ref{condi}) and to generate a given reduced needle
length. The conformal transformation leading to this needle
configuration is considered in some detail in Refs.
\cite{Akh,Kob}.

(E) Widely separated needles: Needles with lengths $|z_{12}|, \,
|z_{34}|$ much smaller than their separation $|z_{I,II}|$ are
generated by Eq. (\ref{dzdw}) on choosing $C$ of order 1 and $h \ll
1$. While detailed results for the mapping in cases (A)-(D) for
needles of arbitrary length can only be obtained numerically (see
Section \ref{five}), for widely separated needles analytic results may be derived by expanding in terms of 
the small parameter $h^{1/2}$.
Using the two conditions (\ref{condi}) to express $\varphi_{2}$
and $\varphi_{4}$ in terms of the four free parameters
$\varphi_{1}, \, \varphi_{3}, \, C$, and $h$ leads to
\begin{eqnarray} \label{varphiN'}
&&\varphi_2 = \varphi_{1}-\pi\thinspace {\rm sgn}\thinspace(\varphi_1-\varphi_2) +
G(\varphi_1 ; \, \varphi_3 ; \, C; \, h)\,, \nonumber \\
&&\varphi_4 = \varphi_{3}-\pi\thinspace {\rm
sgn\thinspace}(\varphi_{3}-\varphi_4)+G(\varphi_3 ; \, \varphi_1 ; \,C^{-1} ; \, h)\,,
\end{eqnarray}
where
\begin{eqnarray} \label{G}
&&G(\varphi_1 ; \, \varphi_3 ; \, C ; \, h)  = 4 h^{1/2} C \sin
\varphi_1 + 4 h C^{2} \sin(2\varphi_1) -\nonumber\\
&&\;\; - 4 h^{3/2} \Bigl\{  \textstyle{1\over 3}C^{3}
\left(7-16\cos^{2} \varphi_1\right) \sin \varphi_1+C^{-1} [2 \sin \varphi_3 \cos(\varphi_1 -
\varphi_3) - \sin \varphi_1] \Bigr\} \, ,
\end{eqnarray}
apart from terms of order $h^{2}$. Equations (\ref{varphiN'}) reflect
the symmetry mentioned below Eq. (\ref{condi}) and are consistent with our assumption (\ref{philhat}). The
dependence of the needle configuration on the four free parameters
is given by
\begin{eqnarray} \label{rationl''}
{z_{12} \over z_{I,II}} \, = \, R(\varphi_{1}, \, \varphi_{3}, \,
C,\, h) \, , \quad {z_{34} \over z_{I,II}} \, = \, -
R(\varphi_{3}, \, \varphi_{1}, \, C^{-1}, \, h)
^{\star}\,,
\end{eqnarray}
where
\begin{eqnarray} \label{Ratio}
&&R(\varphi_{1}, \, \varphi_{3}, \, C,\, h) = 4 C h^{1/2}
e^{-i\varphi_1} \times \nonumber \\
&&\qquad\qquad\qquad \times \Bigl\{ 1 - 2 i h^{1/2} C \sin \varphi_1 - h [C^{-2}
(1+2e^{2i \varphi_{3}}) + C^{2}] + {\cal O}(h^{3/2}) \Bigr\} \, .
\end{eqnarray}
The symmetry embodied in Eqs. (\ref{rationl''}), which we have
checked within the $h$-expansion, is expected to hold for arbitrary
$h$. Starting from a set
$(\varphi_{1},\varphi_{2};\varphi_{3},\varphi_{4};C,h)$ of six
parameters obeying the two conditions (\ref{condi}) and replacing 
it by $(\varphi_{3},\varphi_{4};\varphi_{1},\varphi_{2};C^{-1},h)$
corresponds to reflecting the two needle configuration about the symmetry
axis of the needle midpoints, i.e., about the imaginary axis of
the $z$-plane if we choose $z_{I}=|z_{I}|$ and $z_{II}=-|z_{I}|$ by
adjusting $A$ appropriately. This is consistent with the above
discussion of the reflection-invariant needle configuration (C), for
which $C=1$.

For the ratio of needle vectors, Eqs. (\ref{rationl''}) and
(\ref{Ratio}) imply
\begin{eqnarray} \label{endendratio}
&&{z_{12} \over z_{34}} = C^{2} \Bigl[ 1+2h
\Bigl(C^{2}\cos^{2}\varphi_{1} - C^{-2}\cos^{2}\varphi_{3}\Bigr)
\Bigr] \times \nonumber \\
&& \qquad\quad\times \exp\left(- i\left \{\textstyle{\varphi_{1}+\varphi_{3}+\pi+{1\over 2}[G(\varphi_{1};\varphi_{3};C;h)
+G(\varphi_{3};\varphi_{1};C^{-1};h)]}
\right\}\right)+\nonumber\\
&&\qquad\quad +{\cal O}(h^{3/2}),
\end{eqnarray}
where the phase factor and modulus are consistent with the
enclosed angle relation (\ref{enclangle''}) and the ratio
of needle lengths (\ref{lengthrat}), respectively. We
also note the relation
\begin{eqnarray} \label{midmidmu}
z_{I,II}&=& A \; h^{-3/2} C e^{-2i \varphi_3} \Bigl\{ 1 + 2
h^{1/2} C^{-1} (-e^{i
\varphi_3}+e^{-i \varphi_3}) + \nonumber \\
&+& h \bigl[ C^{-2} (2 e^{2i \varphi_3}-3+4e^{-2i \varphi_3})
+C^{2} (1+2e^{-2i \varphi_1}) \bigr] + {\cal O}(h^{3/2}) \Bigr\}\,,
\end{eqnarray}
which determines the value of $A$ needed to generate a given $z_{I,II}=|z_{I,II}|$.

\subsubsection{Force and torque} \label{foandto} 

The force and torque on needle $I$ due to needle $II$ can be evaluated using Eqs.
(\ref{Schwarz})-(\ref{force''}) and (\ref{turn})-(\ref{thetaw'}),
respectively. The stress tensor average in the annulus
\cite{tmeaning} was determined by Cardy \cite{Cardystrip} and
can be written as
\begin{eqnarray} \label{Tann}
\langle T(w) \rangle \equiv \langle T(w) \rangle_{\rm annulus} =
{1 \over 2w^{2}} \, t(h)\,,
\end{eqnarray}
where
\begin{eqnarray} \label{Tann'}
t(h) &\equiv& h {d \over d h} \ln \Biggl\{ \Bigl(1+ [({\cal
S}_{11}+{\cal S}_{21})/2, \, {\cal S}_{11}, \, {\cal S}_{21}, \,
{\cal S}_{22}] \Bigr) \prod_{n=1}^{\infty}(1-h^{2n})^{-1} \Biggr\}
\end{eqnarray}
for the combinations $OO, \, ++, \, +-, \, O+$ of
universality classes of the two needles \cite{localT}. Here
\begin{eqnarray} \label{Spq}
{\cal S}_{pq} \equiv {\cal S}_{pq}(h)=\sum \limits_{r=2}^{\infty}
h^{(r^{2}-1)/24}
\, \sin{\pi p r \over 3}\sin{\pi q r \over 4} \Big/ \Bigl(\sin{\pi
p \over 3}\sin{\pi q \over 4}\Bigr) \,,
\end{eqnarray}
where the series converges for $h<1$. For the integration path
${\cal C}$ in (\ref{force''}) and (\ref{thetaw'}), which goes around
the inner boundary circle counterclockwise, it is most convenient
to use the circle ${\cal C}_{c}$ given by $w=Ch^{1/2} e^{i
\varphi}$, which passes through the pre-image $w=Ch^{1/2}$ of
$z=\infty$.

 Unlike the force and torque contributions $\tau^{(T)}$ and
$\theta^{(T)}$ in (\ref{force''}) and (\ref{thetaw'}), which depend, via Eqs.
(\ref{Tann})-(\ref{Spq}), on the surface universality classes of
the two needles, the contributions $\tau^{(S)}$ and
$\theta^{(S)}$, which involve $z'(w)$ and the
Schwarzian derivative (\ref{Schwarz}), are solely determined by the
geometric configuration of the needles and in this sense
``hyper-universal''. This was already mentioned at the end of Sec.
\ref{forcetorque}, and it applies to the semi-infinite and infinite needles of Sec.
\ref{infseminf}. The occurrence of a hyper-universal term in the
free energy of interaction of a non-circular particle with 
other particles in a near-critical two-dimensional system is well known from the ``small-particle
operator expansion'' (SPOE). As discussed in Refs. \cite{ee2006,e} and
Appendix \ref{operator} below, the hyper-universal interaction arises from the stress tensor in the operator expansion corresponding to the
particle, in our case a needle. The hyper-universal term in the expansion depends
on the orientation of the needle, is proportional to the square of
its length, which is the smallest power involving its
orientation dependence, and it reproduces the results
corresponding to the $h \to 0$ contributions of $\tau^{(S)}$ and
$\theta^{(S)}$, as we show in Eqs. (\ref{hypf})-(\ref{hypfreeeng})

In general, the force vector is neither parallel nor
antiparallel to the vector $z_{I,II}$ between the needle
midpoints, as seen, for example, in Eq. (\ref{hypf}) below. However,
for the symmetric perpendicular and parallel configurations in (A)
and (B) and for the mirror symmetric configurations (C), the
force clearly points along $z_{I,II}$ or $-z_{I,II}$. Detailed numerical results for
force and torque in cases (A)-(D) are reported in
Sec. \ref{five}. Here we give a few analytic results for the case
(E) of two short needles or, equivalently, two widely separated needles.

According to Eqs. (\ref{rationl''}), (\ref{Ratio}), this regime corresponds to
small $h$, and  in leading order
\begin{eqnarray} \label{tausmall}
\tau^{(T)}={2i \over z_{I,II}}t(h \ll 1) \, , \quad \tau^{(S)}={2i
\over z_{I,II}}h^{2}e^{2i(-\varphi_{1}+\varphi_{3})} \,,
\end{eqnarray}
for the two contributions to the force on needle $I$ in
(\ref{force}). For more details see Eqs. (\ref{SCc})-(\ref{taufromS'}) in Appendix
\ref{SPE}. On using (\ref{Tann'}) and
(\ref{Spq}), we obtain
\begin{eqnarray} \label{tOOO+}
t(h \ll 1)\, = \, [h, \, -h]
\end{eqnarray}
for needle pairs of type $[OO,O+]$, while for pairs of type
$[++,+-]$
\begin{eqnarray} \label{t+++-}
t(h \ll 1)\, = \, {1 \over 8} \, \Biggl[\sqrt{2}h^{1/8} \Bigg/
\Biggl(1+\sqrt{2}h^{1/8} \Biggr), \quad -\sqrt{2}h^{1/8} \Bigg/
\Biggl(1-\sqrt{2}h^{1/8} \Biggr) \Biggr] \,,
\end{eqnarray}
for $h$ values which are small compared to 1 without
requiring $h^{1/8}$ to be small. The force $(f_{x},f_{y})$ on
needle $I$ in (\ref{force}) is thus dominated by the contribution
from $\tau^{(T)}$. In the remainder of this subsection we again assume that 
$z_{I,II} \equiv |z_{I,II}|$, so that the distance vector $z_{I,II}$
between needle centers is parallel to the $x$-axis. Using $h \to
|z_{12}||z_{34}|/(16 |z_{I,II}^{2}|)$ due to (\ref{rationl''}),
(\ref{Ratio}), one finds 
\begin{eqnarray} \label{flead}
f_{x}/k_{B}T \, = \, - \, {1 \over |z_{I,II}|} \times \Biggl[\pm
{1 \over 8} {D_{I}D_{II} \over |z_{I,II}|^{2}} \, , \; \pm {1
\over 4} {(D_{I}D_{II})^{1/8} \over|z_{I,II}|^{1/4}} \Bigg/
\Biggl( 1 \pm {(D_{I}D_{II})^{1/8} \over|z_{I,II}|^{1/4}} \Biggr)
\Biggr]
\end{eqnarray}
and $f_{y}/(k_{B} T)=0$ for the force components in leading order.
Here the upper and lower signs describe the needle
universality classes $[OO,++]$ and $[O+,+-]$, respectively, and $D_{I} \equiv |z_{12}|$ and $D_{II} \equiv |z_{34}|$
are the needle lengths introduced in Eq. (\ref{shortnot}). For needles with equal (unequal)
universality classes the force is antiparallel (parallel) to the
distance vector $z_{I,II}$, i.e., attractive (repulsive), as
expected. As in a multi-pole expansion, the shape anisotropy does
not appear in the leading ``monopole'' contribution (\ref{flead}), in
which the force is independent of the needle orientations $\Phi_{12}$
and $\Phi_{34}$, but it appears in higher order in the
needle lengths. Unlike the corresponding higher order contributions
from $\tau^{(T)}$, which also depend on the needle universality classes,
the contributions from $\tau^{(S)}$ are hyper-universal, as
mentioned above. Equations (\ref{rationl''})
and (\ref{Ratio}) imply $e^{-i\varphi_{1}} \to e^{i\Phi_{12}}$ and
$e^{i\varphi_{3}} \to -e^{i\Phi_{34}}$, and the lowest-order
hyperuniversal term, given by the second expression in
(\ref{tausmall}), leads to
\begin{eqnarray} \label{hypf}
(f_{x}^{(S)}, \, f_{y}^{(S)})/(k_{B} T) \, = \, - {D_{I}^{2}
D_{II}^{2} \over 2^{7} |z_{I,II}|^{5}} \,  \Biggl( \cos\Bigl(
2(\Phi_{12}+\Phi_{34})\Bigr) \, , \; -\sin\Bigl(
2(\Phi_{12}+\Phi_{34})\Bigr) \Biggr)\,.
\end{eqnarray}
As expected, the force is unchanged
on rotating a needle through 180 degrees.

We now turn from Eq. (\ref {hypf}) to the hyperuniversal
contribution $- {\rm Re} \, \theta^{(S)}$ to the reduced torque $\Theta/(k_B T)$,
introduced in Eqs. (\ref{turn})-(\ref{thetaw'}). Calculating $\theta^{(S)}$
by means of the mapping (\ref{dzdw}), one obtains
\begin{eqnarray} \label{hyptor}
\theta^{(S)} \, &\to& \, -ih^{2} \,
e^{2i(-\varphi_{1}+\varphi_{3})}
\; , \nonumber \\
{\rm Re} \, \theta^{(S)} \, &\to& \, {D_{I}^{2} D_{II}^{2} \over
2^{8} |z_{I,II}|^{4}} \, \sin\Bigl( 2(\Phi_{12}+\Phi_{34})\Bigr)
\, 
\end{eqnarray}
in leading order. For more details see the paragraph containing Eqs. (\ref{zeta}),
(\ref{zeta321}) in Appendix \ref{SPE}.

A detailed discussion of the force and torque for two short
needles, based on the SPOE, is given in Appendix \ref{optwodist}. With
this entirely different approach we confirm the leading
behavior (\ref{flead}) for the force and obtain
\begin{eqnarray} \label{hypfreeeng}
\delta F^{({\rm hu})}/(k_{B}T) \, = \, -{D_{I}^{2} D_{II}^{2}
\over 2^{10}} \Biggl( e^{2i(\Phi_{12}+\Phi_{34})} \, {1 \over
z_{I,II}^{4}} \, + \, {\rm c.c.} \Biggr)
\end{eqnarray}
for the hyperuniversal (hu) contribution to the free energy of
interaction \cite{delF} of the needles, which agrees with the
results for the force components in (\ref{hypf}) and the
torque in (\ref{hyptor}).

%\newpage

\subsection{Interaction of a finite and a semi-infinite
needle} \label{finsem}

Consider the case in which needle 12 has a finite length
$D_{I} \equiv |z_{12}|$ but needle 34 is semi-infinite, with $z_{3} = z_{\rm e}$ and $z_{4}=\infty$. This needle geometry is generated by  Eq. (\ref{dzdw}) in the limit  $C=h^{1/2}$ and $w_{4}=h$ in which the pre-images $w_{4}$ and
$Ch^{1/2}$ of $z=z_{4}$ and $z=\infty$, respectively, coincide, so that
\begin{eqnarray} \label{semi+fin}
z'(w) \, = \, {A \over h^{2}} \prod \limits_{k=1}^{\infty}
\Biggl( {(1-h^{2k-3}e^{-i\varphi_{\rm
e}}w)(1-h^{2k+1}e^{i\varphi_{\rm e}}/w) \over
(1-h^{2k-3}w)^{3}(1-h^{2k+1}/w)^{3}} \times \nonumber \\
\times \prod \limits_{n=1,2} (1-h^{2k-2}e^{-i
\varphi_{n}}w)(1-h^{2k}e^{i \varphi_{n}}/w) \Biggr) \,. 
\end{eqnarray}
This implies 
\begin{eqnarray} \label{semi+fin''}
dz \,  &=& \,
(d\varphi) \, i A \, e^{-i \varphi_{\rm e}} \, {\cal
G}(\varphi;\varphi_{1},\varphi_{2}) \times \nonumber \\
&& \times \prod \limits_{k=1}^{\infty}
{|1-h^{2k-1}e^{i(\varphi-\varphi_{\rm e})}|^{2}\prod
\limits_{n=1,2} |1-h^{2k}e^{i (\varphi-\varphi_{n})}|^{2} \over
|1-h^{2k-1}e^{i\varphi}|^{6}}
 \,
\end{eqnarray}
and
\begin{eqnarray} \label{semi+fin'}
dz\, &=& \,
-(d\varphi) \, {i A \over h} \, e^{-i\varphi_{\rm e}/2} \,
{\sin[(\varphi-\varphi_{\rm e})/2] \over 4 \sin^{3}(\varphi/2)}
\times
\nonumber \\
&& \times \prod \limits_{k=1}^{\infty}
{|1-h^{2k}e^{i(\varphi-\varphi_{\rm e})}|^{2} \prod
	\limits_{n=1,2} |1-h^{2k-1}e^{i(\varphi- \varphi_{n})}|^{2} \over
	|1-h^{2k}e^{i\varphi}|^{6}}   \, ,
\end{eqnarray}
for displacements  $dw=d(e^{i\varphi})$ and $dw=h \, d(e^{i\varphi})$ around the outer and inner boundaries of the annulus, respectively.
Here $\cal G$ is defined in Eq. (\ref{calG}), and $w_{3} \equiv w_{\rm
e}=he^{i \varphi_{\rm e}}$ is the pre-image of $z_{\rm e}$. The
behavior (\ref{semi+fin'}) near $\varphi=0$ implies that the
semi-infinite needle extends from $z=z_{\rm e}$ to $z=s |\infty|$
along the tangential unit vector $s=-i(A/|A|)e^{-i \varphi_{\rm e}
/2} \, {\rm sgn\thinspace}\varphi_{\rm e}$ with $\varphi_{\rm e} \equiv
\varphi_{3}$ obeying (\ref{philhat}). For convenience we choose
the semi-infinite needle to coincide with the positive real axis, i.e.,
$z_{\rm e}=0$ and $s=1$, so that
\begin{eqnarray} \label{fill}
A=i|A|e^{i \varphi_{\rm e}/2} \, {\rm sgn\thinspace}\varphi_{\rm e} \, .
\end{eqnarray}
From Eq. (\ref{enclangle''}) we obtain
\begin{eqnarray} \label{enclangle'''}
e^{i \Phi_{12}}=-e^{-i(\varphi_{1}+\varphi_{2}+\varphi_{\rm e})/2}
\, {\rm sgn\thinspace}(\varphi_{1}-\varphi_{2}) \, {\rm sgn\thinspace}\varphi_{\rm
e} \,,
\end{eqnarray}
since $\Phi_{34}=\pi$, $\varphi_{4}=0$, and
$\varphi_{3}=\varphi_{\rm e}$.

Apart from homogeneous dilatations, the needle configuration is
determined by three parameters - the length ratio
$|z_{12}|/|z_{I}|$ and the two angles ${\rm arg} z_{I}$ and
$\Phi_{12}$ which $z_{I}$ and $z_{12}$ form with the
semi-infinite needle. Here $z_{I} \equiv r_{I,x}+ir_{I,y}$ is the
vector from $z_{\rm e}=0$ to the midpoint of needle 12. Correspondingly, there are, apart from $|A|$, three
independent mapping parameters. Since the derivative
$dz/dw$ is analytic in the interior of the annulus, imposing the requirement
\begin{eqnarray} \label{Acond'}
I_{\rm outer} \equiv \int_{{\cal C}_{\rm outer}} d w (dz/dw)=0
\end{eqnarray} on the four parameters $h, \, \varphi_{1}, \, \varphi_{2}, \,
\varphi_{\rm e}$ ensures that the mapping $z=z(w)$
is single valued.

Now consider the case of a finite needle which is much shorter than its distance from the closest point of the semi-infinite needle, so that 
$|z_{12}|^{2} \ll |z_{I}|(|z_{I}|-r_{I,x})$. Explicit
results for the force and torque in this regime can be obtained by expanding in terms
of $h$ and are expected to agree with the SPOE. In
the remainder of this subsection this is checked in leading order.

For small $h$ the constraint (\ref{Acond'}) reads
\begin{eqnarray} \label{semiconst}
\varphi_{2}=\varphi_{1}-\pi\thinspace {\rm
	sgn\thinspace}(\varphi_{1}-\varphi_{2})+2h[3 \sin
\varphi_{1}-\sin(\varphi_{1}-\varphi_{\rm e})]+{\cal O}(h^2) \,,
\end{eqnarray}
yielding in terms of independent parameters the needle configuration
\begin{eqnarray} \label{semiAAA}
z_{I} \, \to \, {|A| \over 4h|\sin\varphi_{\rm e}/2|} \, e^{-i
	\varphi_{\rm e}}\,,
\end{eqnarray}
\begin{eqnarray} \label{semiAA}
z_{12} \, \to \, -4|A| ie^{-i\varphi_{1}} \, e^{-i \varphi_{\rm
		e}/2} \, {\rm sgn\thinspace}\varphi_{\rm e} \,,
\end{eqnarray}
to leading order in $h$. Since $ie^{-i\varphi_{1}}$ equals
$e^{-i(\varphi_{1}+\varphi_{2})/2} {\rm
sgn\thinspace}(\varphi_{1}-\varphi_{2})$ in leading order, Eq.
(\ref{semiAA}) is consistent with (\ref{enclangle'''}). Equations
(\ref{semiAAA}) and (\ref{semiAA}) allow us to express $|A|, \, h, \, \varphi_{1}, \, \varphi_{\rm e}$ in
terms of needle parameters, and we note that
\begin{eqnarray} \label{exmu}
|A| \to |z_{12}|/4
\end{eqnarray}
and
\begin{eqnarray} \label{exh}
h \to {1 \over 8} {|z_{12}| \over
\sqrt{2|z_{I}|(|z_{I}|-r_{I,x})}}
\end{eqnarray}
for use below.

The force $(f_{x}, f_{y})$ on the 12 needle follows from Eqs.
(\ref{force}) and (\ref{force''}) on substituting the 
derivative (\ref{semi+fin}), integrating along a circle infinitesimally larger than the inner boundary circle of the annulus, avoiding the singularity at
$w=w_{\rm e}=he^{i\varphi_{\rm e}}$. 

The leading contribution comes from $\tau^{(T)}$
and is given by
\begin{eqnarray} \label{12semitau}
\tau^{(T)} /t(h) \, \to \, {1 \over 4 |z_{I}|} {e^{i \varphi_{\rm
e}/2} (e^{i\varphi_{\rm e}} - 3) \over \sin (\varphi_{\rm e} /2)}\,,
\,
\end{eqnarray}
yielding
\begin{eqnarray} \label{12semiforce}
\Bigl[(f_{x}-if_{y})/(k_B T)\Bigr]/t(h) \, \equiv \, i\tau /t(h)
\, \to \,- \, {3z_{I}-|z_{I}| \over 2 z_{I} (z_{I}-|z_{I}|)} \,,
\end{eqnarray}
since $z_{I}/|z_{I}| \to  e^{-i\varphi_{\rm e}}$, see Eq.
(\ref{semiAAA}). The leading contribution to the force follows from Eq. 
(\ref{12semiforce}) on replacing $t(h)$ by $t(h \ll 1)$ in Eqs.
(\ref{tOOO+}), (\ref{t+++-}) and on replacing $h$ by the right hand side of Eq.
(\ref{exh}). As in Eq. (\ref{flead}), the leading contribution is
independent of the orientation $\Phi_{12}$ of the small needle. The leading dependence on orientation comes from $\tau^{(S)}$ in (\ref{force''}) in higher order in $h$. In Appendix \ref{opsmallsemi} we use the SPOE to calculate the leading isotropic and angle-dependent contributions to the force and the leading contribution to the torque on the 12 needle. The SPOE prediction is in complete agreement with the $h$-expansion result for the force given in Eq. (\ref{12semiforce}).

%\newpage

\subsection{Interaction of a finite and an infinite needle}\label{fineedhapl} 

On setting $w_{3}=w_{4}=Ch^{1/2}=h$ in Eq. (\ref{dzdw}), 
both $z_{3}$ and $z_{4}$ become infinite, so that the 34
needle takes the form of an infinite needle or boundary line. For  $A=-i|A|$, the infinite needle coincides with the boundary ${\rm Im}z=0$ of the upper
half plane. The derivative of the transformation is given by
\begin{eqnarray} \label{Ahalfdzdw}
z'(w) \, = \, -i |A| {1 \over h^{2}} \prod
\limits_{k=1}^{\infty} {\prod \limits_{n=1,2} (1-h^{2k-2}e^{-i
\varphi_{n}}w)(1-h^{2k}e^{i \varphi_{n}}/w) \over
(1-h^{2k-3}w)^{2}(1-h^{2k+1}/w)^{2}} \,,
\end{eqnarray}
so that
\begin{eqnarray} \label{Ahalfdzdw''}
dz \,  = \, (d\varphi)
\, |A| \, {\cal G}(\varphi;\varphi_{1},\varphi_{2}) \, \prod
\limits_{k=1}^{\infty} {\prod \limits_{n=1,2} |1-h^{2k}e^{i
(\varphi-\varphi_{n})}|^{2} \over |1-h^{2k-1}e^{i\varphi}|^{4}}
 \,
\end{eqnarray}
and 
\begin{eqnarray} \label{Ahalfdzdw'}
dz \, = \,
-(d\varphi) \, {|A| \over h} \, {1\over 4 \sin^{2}(\varphi/2)}
\prod \limits_{k=1}^{\infty} {\prod \limits_{n=1,2}
|1-h^{2k-1}e^{i(\varphi- \varphi_{n})}|^{2} \over
|1-h^{2k}e^{i\varphi}|^{4}}   \,
\end{eqnarray}
for displacements  $dw=d(e^{i\varphi})$ and $dw=h \, d(e^{i\varphi})$ along the outer and inner boundary circles, respectively. Thus, a counterclockwise path around the inner circle corresponds to a path along the real axis from $+\infty$ to $- \infty$, and a counterclockwise path around the outer circle to a clockwise path around the 12 needle along its edges. The function  ${\cal G}$ is defined by Eq.  (\ref{calG}).

As in the case (\ref{semi+fin}) of a finite needle interacting
with a semi-infinite needle,
the constraint $I_{\rm outer}=0$ ensures that
the mapping (\ref{Ahalfdzdw}) is single-valued, so that only two of
the three parameters $h, \, \varphi_{1}, \, \varphi_{2}$ are
independent. They can be adjusted to fix the angle $\Phi_{12}$
between needle and boundary and the ratio of the length
$|z_{12}|$ of the needle and its distance to the boundary. 
According to the argument leading to Eq. (\ref{enclangle''}), the unit vector characterizing the direction of $z_{12}$ is given by
\begin{eqnarray} \label{Phi12}
e^{i \Phi_{12}} \equiv {z_{12} \over |z_{12}|} = e^{-i(
\varphi_{1}+ \varphi_{2})/2} \, {\rm sgn\thinspace}(\varphi_{1}-
\varphi_{2}) \, .
\end{eqnarray}
%

%\newpage
%
\subsubsection{Distant needle} \label{halfdist}
A needle far from the boundary in comparison with its length corresponds to $h \ll 1$. The products in (\ref{Ahalfdzdw})-(\ref{Ahalfdzdw'}) can be
expanded in powers of $h$, and Eq. (\ref{Acond'}) yields
\begin{eqnarray} \label{Acond''}
&&\varphi_{2}=\varphi_{1}-\pi\thinspace {\rm
sgn\thinspace}(\varphi_{1}-\varphi_{2})+g(\varphi_{1};h)
\, , \nonumber \\
&&g(\varphi_{1};h) \equiv 4h \sin\varphi_{1}+4h^{2}
\sin(2\varphi_{1})+{\cal O}(h^{3})\,,
\end{eqnarray}
which is consistent with the constraint (\ref{semiconst}) in Sec. \ref{finsem} on setting $\varphi_{\rm e}=0$ there. For the vector $z_{12}$ between
the ends of the needle and for the distance $r_{I,y} \equiv {\rm Im}z_{I}\equiv
{\rm Im} (z_{1}+z_{2})/2$ of the needle center from the boundary,
one obtains
\begin{eqnarray} \label{Aneedlehalfplane}
&&z_{12} \to 4|A| (1+2h^{2}\cos^{2}\varphi_{1}) \, e^{-i
\{ \varphi_{1}+[-\pi+g(\varphi_{1};h)]/2 \}} \, + \, {\cal O}(h^{3}) \, , \nonumber \\
&&r_{I,y} \to {|A| \over 2 h} \{1+2h^{2}[1+2
\cos(2\varphi_{1})]+{\cal O}(h^{3})\} \, .
\end{eqnarray}
The direction of the needle, given by the phase factor of $z_{12}$ in Eq. (\ref{Aneedlehalfplane}),
is consistent with the general expression on the right hand side
of Eq. (\ref{Phi12}).
%
%\newpage
%
\subsubsection{Force and torque}
The force and the torque which the boundary exerts on needle 12 are
again given by Eqs. (\ref{force})-(\ref{force''}) and
(\ref{turn})-(\ref{thetaw'}) together with
(\ref{Tann})-(\ref{Spq}), except that now $dz/dw$ and $S(w)$ follow
from (\ref{Ahalfdzdw}). It is most convenient to use the
integration path ${\cal C}$ in Eqs. (\ref{force''}) and (\ref{thetaw'})
along the inner boundary circle of the annulus. Since there is no force 
$f_{x}$ parallel to the boundary, the imaginary part of $\tau$ must vanish.

For a distant needle the force is determined by
$\tau^{(T)}$ for $h\ll1$. In this regime Eqs. (\ref{force''}), (\ref{Tann}), and
(\ref{Ahalfdzdw'}) imply $\tau^{(T)}/t(h) \to 2h/|A|$, and using Eqs.
(\ref{Aneedlehalfplane}), one obtains
\begin{eqnarray} \label{AtauThalf}
{f_{y} \over k_{B}T} \to - {1 \over r_{I,y}} t(h \ll 1) \, , \quad
h \to {|z_{12}| \over 8 r_{I,y}}\,,
\end{eqnarray}
where $t(h \ll 1)$ is taken from Eqs. (\ref{tOOO+}), and (\ref{t+++-}). As expected, this result is in 
agreement with (\ref{12semiforce}) in Sec. \ref{finsem} for $r_{I,x} \to + \infty$. An orientation 
dependence of the needle only appears in higher order and is determined explicitly for the needle 
geometry considered here with the SPOE in Appendix \ref{opsmallhalf} . 
As for $\tau^{(S)}$, we have checked
that its leading $h$ power is higher than $h^{3}$, so that
$\tau^{(S)}$ does not contain a term proportional to $|z_{12}|^{2}
/r_{I,y}^{3}$. This is consistent with the vanishing of the stress-tensor
average in the half plane and the absence in the SPOE of a hyper-universal
contribution $\propto |z_{12}|^{2}$ to the free energy, force, and
torque.

The leading contribution to the torque $\Theta$ is determined by
$\theta^{(T)}$ and is given by
\begin{eqnarray} \label{smallhalfthetaT}
{\Theta\over k_B T}\to -{\rm Re} \, \theta^{(T)} \to 6 h^{2}
\sin(2 \varphi_{1})\thinspace t(h \ll 1); \;\;\sin(2 \varphi_{1})
\to\sin(2\Phi_{12}),\;\; h \to {|z_{12}| \over 8 r_{Iy}}\,,
\end{eqnarray}
where the $h$-expansion of ${\rm Re} \, \theta^{(T)}/t(h)$ is
derived in Appendix \ref{hsmallhalf} and Eqs.
(\ref{Phi12})-(\ref{Aneedlehalfplane}) have been used. Equations
(\ref{AtauThalf}) and (\ref{smallhalfthetaT}) are consistent
with the SPOE results in Appendix \ref{opsmallhalf}.

\newpage

\section{RESULTS FOR ARBITRARY NEEDLE LENGTHS} \label{five}

We now consider some simple needle geometries in which the needle length is neither very large nor very small compared to the distance between the needles. Calculating the force and torque requires the full machinery described in Sec. \ref{finfin} for arbitrary values of the mapping parameter $h$ in the interval $0<h<1$. Unlike the completely analytic approaches for semi-infinite needles in Sec. \ref{infseminf} and for short needles (small $h$ expansion) in Sec. \ref{finfin}, we now resort to numerical evaluation, which, however, yields results over the entire range from small to large needle lengths \cite{fxcoll}. Actually, we restrict our attention to needle configurations for which the six mapping parameters are restricted to subspaces of lower dimension. These include the configurations (A)-(D) introduced between Eqs. (\ref{lengthrat}) and (\ref{varphiN'}) and the configurations of a finite needle in the half plane, discussed in Sec. \ref{fineedhapl}. 

(A) For the symmetric-perpendicular configuration (A) defined in Sec. \ref{map} and shown in Fig. \ref{FigABCD}, the force component $f_y$ and the torque on needle $I$ vanish, due to symmetry. The component $f_x$ is attractive ($f_x<0$) for needle university classes $OO, ++ $ and repulsive $(f_x>0$) for classes $+-,O+$. We consider the case of equal needle lengths $D$, in which $Df_x/(k_B T)$, apart from the universality classes, only depends on  ${\tilde c}=c/D$, where $c=z_{I}-z_4$ is the minimum distance between the needles. In Fig. \ref{FigAfx} our exact numerical results for $Df_x/(k_B T)$ in the region $10^{-2} < {\tilde c} < 10^2$ are indicated by full points. For large and small ${\tilde c}$ there is excellent agreement with the asymptotic behavior  (\ref{perpforce}), (\ref{perpforce'}) for short needles and with the results of Sec. \ref{infseminf} for semi-infinite needles, respectively. In the latter limit, ${\tilde c} \to 0$, and the force $f_x$ becomes independent of $D$ and is given by $f_y$ in Eq. (\ref{fy3}) with $\alpha=\pi /2$ and $r_y (1)=c$.

(B) In the symmetric-parallel configuration (B) defined in Sec. \ref{map} and shown in Fig. \ref{FigABCD}, the force component $f_y$ and the torque also vanish by symmetry. We have evaluated $f_x$ numerically in two special cases, B1 and B2. In case B1, which is denoted by (ii) in Sec. \ref{map} and belongs to classes (B) and (C), both needles have the same length $D$. In case B2 we denote the length of needle $I$ by $D$ and choose the length of needle $II$ equal to the needle separation $c=z_I-z_{II}$. The dependence of $c f_x$ on $D/c$ in both cases, B1 and B2, is shown in Fig.  \ref{FigBfx}. Again our numerical results (points) merge nicely with the asymptotic expressions (curves) for small and large $D/c$. For small $D/c$ these follow from Eqs. (\ref{mirrorforce}) and (\ref{mirrorforce'}) with $\alpha=0$ in case B1 and from Eqs. (\ref{funequsymmpar''})-(\ref{funequsymmpar'}) in case B2. For large $D/c$  case B1 reduces to an infinitely long strip, and $cf_x/(k_{B}T) \to \Delta D/c$, where $\Delta=\pi(\tilde{t}-1/48)$, with $\tilde{t}$ given in Eq. (\ref{t12}), is the corresponding Casimir amplitude \cite{tmeaning}. For large $D/c$ case B2 reduces to a needle $II$ parallel to the boundary of a half plane, a geometry considered in the last two paragraphs of this section, and $cf_x /(k_B T)$ is the same as $Df_y /(k_B T)$ for $\tilde{B}=1$ and $\Phi=0$ in Fig. \ref{FigE1} . 

(C) For the mirror-symmetric configuration (C) defined in Sec. \ref{map} and shown in Fig. \ref{FigABCD}, the force component $f_y$ vanishes for all angles $\pi-2\Phi_{12} \equiv \alpha$ between the needles. The component $f_x$ and the torque $\Theta$ are non-zero, with the exception of the torque at $\alpha=0$ and $\pi$. Figure \ref{FigCfxtheta} shows $f_x$ and $\Theta$ for needles forming an angle $\alpha=\pi /5= 36^\circ$ and with lengths ranging from short to long. The quantities $cf_x /(k_B T)$ and $\Theta/(k_B T)$ are plotted as functions of $D/c \equiv 1/{\tilde c}$ where the minimum separation $c=z_{24}=z_2 -z_4$ of the needles is  the distance between the two lower needle ends.  For small $D/c$ the numerical data (full points) merge nicely with the results of the small needle expansion given in Eqs. (\ref{mirrorforce}), (\ref{mirrorforce'}). For large $D/c$ the data for the ($D$-independent) force are in excellent agreement with the corresponding force for mirror-symmetric semi-infinite needles following from Eqs. (\ref{end'}), (\ref{fxy}) with $b=1$. The torque appears to vary linearly with $D$ for large $D$, in agreement with the analytic argument at the end of Appendix \ref{strongover}.  
 
(D) Next we consider the force and torque for the antiparallel configuration defined in Sec. \ref{map}. We consider needles of equal length antiparallel and parallel to the $x$-axis \cite{rotateD}, illustrated by Fig. \ref{FigABCD}-(D'), with $z_{34}=-z_{12}=D$. The configuration is uniquely specified by the value of $D$, the vertical separation of the needles $W=r_{1,y} -r_{4,y}$ and the relative horizontal displacement $r_{4,x} -r_{1,x}$. For the fixed ratio $(r_{4,x} -r_{1,x})/W =1.4$, Figs. \ref{FigDfxfy} and \ref{FigDtheta}, show our numerical results (points) for $Wf_x /(k_B T)$, $Wf_y /(k_B T)$, and $\Theta/(k_B T)$ as functions of $D/W$. For small $D/W$ we show the small needle prediction following from Eqs. (\ref{antiforcexrot})-(\ref{antitorque'}). For large $D/W$ the perpendicular force component $f_y$ is dominated by the usual Casimir force for a long strip \cite{Cardy}, so that $Wf_y /(k_B T) \to (D/W)\Delta$ with $\Delta$ from \cite{tmeaning}. This and the behavior of the ($D$-independent)  parallel force component $f_x$ and of the torque $\Theta$ for large $D/W$, derived in Eqs. (\ref{parforce}), (\ref{paraxlambda}) and (\ref{toroverlapequal'}), respectively, are also indicated by full lines in the figures. 

Finally we consider the force and torque on a single needle in the upper half plane for various ratios ${\tilde B} \equiv r_{I,y}/D$ of the distance of the needle center from the boundary to the needle length and for various angles $\Phi_{12} \equiv \Phi$ between the needle and the boundary. The torque vanishes by symmetry for $\Phi=0$ and $\Phi= \pm \pi /2$. The results for ${\tilde B}=10$ and $0< \Phi < \pi/2$, shown in Fig. \ref{FigE10},  are in perfect agreement with the predictions (\ref{fhalfplane}) and (\ref{fhalfplane'}) of the operator expansion for a distant needle. In this case $f_y -f_y |_{\Phi = \pi /4}$ and $\Theta$ are odd and even functions, respectively, of $\Phi- \pi/4$. Figure \ref{FigE1} shows corresponding results for the intermediate distance ratio ${\tilde B}=1$,  where the minimum distance between the needle and the boundary, which corresponds to the perpendicular orientation $\Phi=\pi /2$, is half the length of the needle. As expected, there are significant deviations from our operator expansion of low order, in particular for the force near $\Phi= \pi /2$. 

For ${\tilde B}<1/2$ the needle touches the boundary before attaining the perpendicular orientation, and the force and torque diverge. Figure \ref{FigE433} shows the case ${\tilde B}=\sqrt{3}/4=0.433$ for $0<\Phi<\pi /3$, with diverging results as $\Phi$ approaches the angle $\pi /3$ and the distance $r_{2,y}={1\over 2}D(\sin{\pi\over3}-\sin\Phi)$ of the needle tip $z_2$ from the boundary shrinks to zero. Since the divergence is a local effect, for $r_{2,y}<<D$ one expects $f_y$ to be independent of $D$ and the same as the force (\ref{fy3}) on a semi-infinite needle with the same endpoint $r_{2,y} \equiv r_y (1)$ and angle $\Phi \equiv \alpha$ in the notation of Subsec. \ref{semiinhalf}. From Eq. (\ref{fy3}) we obtain
$Df_y/k_BT\approx -{1\over 12}(5-216\thinspace\tilde{t})\left({\pi\over 3}-\Phi\right)^{-1}$ for the leading divergent term, which is plotted in Fig. \ref{FigE433}. The exact numerical data (points) in the figure are in excellent agreement with this prediction, and for all four sets of boundary conditions it gives an astonishingly good fit over the entire range $0<\Phi<\pi/2$. Heuristic arguments (see last two paragraphs of Appendix \ref{strongover}) suggest that the torque also diverges as $\left({\pi\over 3}-\Phi\right)^{-1}$, and the exact numerical data in Fig. \ref{FigE433} appear to support this.

\section{SUMMARY AND CONCLUDING REMARKS} \label{concl}

The Casimir interaction of particles immersed in a binary liquid mixture near a critical point of miscibility has a long range and universal character, and non-spherical particles experience both a force and a torque. We consider the interaction of two needle shaped particles right at the critical point of a two-dimensional fluid in the Ising universality class. While particular needle configurations have been considered before \cite{VED}, the approach of this paper allows us to calculate the interaction for two needles of arbitrary lengths, separations, and orientations for various combinations of surface universality classes \cite{hwd}. 

As in earlier work \cite{ber,e,ee2006,VED,Bim}, we utilize the conformal invariance of two-dimensional critical systems and generate the needle geometry of interest from a simpler standard geometry by means of a conformal mapping. As outlined in Sec. \ref{forcetorque} and Appendix \ref{shiro}, we work with the stress tensor, which has well-understood conformal transformation properties, is known in the simple standard geometry, and determines the force and torque in the needle geometry of interest.

In Sec. \ref{infseminf} we consider arbitrary configurations of an infinite and a semi-infinite needle and of two semi-infinite needles and obtain the results for the force given in Eqs. (\ref{end}), (\ref{fy3}) and Eqs. (\ref{end'}), (\ref{fxy}), respectively. The simple form of the force follows from the simplicity of the stress tensor in the standard geometry and of the mapping generating the needles. The region outside the needles is simply connected, and the standard geometry is the upper half plane with the  two needles on the $x$ axis.     

For needles of finite length the space bounded by the needles is doubly connected, and the standard geometry is an annulus with circular needles on its boundaries. The stress tensor in the annulus is known from Cardy's work \cite{Cardystrip} and summarized in Eqs. (\ref{Tann})-(\ref{Spq}). The mapping onto the two-needle geometry is a special case of Akhiezer's formula \cite{Akh} for mapping the annulus onto the region outside two nonoverlapping polygons, and its derivative is given by Eq. (\ref{dzdw}) in Sec. \ref{twofinite}. Two conditions (\ref{condi}) are imposed to ensure that the mapping is single valued. Searching for values of the six parameters $h,C,\varphi_1,..,\varphi_4$ in Eq. (\ref{dzdw}) that satisfy these two conditions and generate a given needle configuration is a formidable task. The simple relation (\ref{enclangle''}), which expresses the angle enclosed by the two needles in terms of the sum $\varphi_1+...+\varphi_4$ reduces the space of parameters in which one must search, and we have found some simple configurations (A)-(D) of the needles, discussed in Sec. \ref{map} and shown in Fig. \ref{FigABCD}, in which the space can be further reduced. In Secs. \ref{finsem} and \ref{fineedhapl} we analyze the special case in which  one of the needles has a finite length and the other is semi-infinite or infinite.

We have put the finite-needle approach to work in two ways.\\
\noindent (a) First of all we have analyzed the case of needles with separation much greater than their lengths analytically. In this regime the inner radius $h \ll 1$ of the annulus is much smaller than the outer radius of 1, and both the small $h$ expansion and the small-particle operator expansion (SPOE) yield information on the force and torque.  Beginning in paragraph (E) of Sec. \ref{map} and continuing in Secs. \ref{foandto}, \ref{finsem}, \ref{fineedhapl}, and Appendix \ref{SPE}, we show the consistency of these two approaches. For example, the surface-class-independent (hyperuniversal) contribution to force and torque arising via (\ref{force})-(\ref{thetaw'}) from the Schwarzian derivative of the mapping is provided within the SPOE (\ref{sme}) by the stress-tensor operator (\ref{tildeT}).\\
\noindent (b) Secondly, by using the same conformal mapping approach and evaluating formulas numerically, we have studied the force and torque over the full range from small to large values of the ratio of needle length to needle separation. Results for several types of needle configurations (see Fig. \ref{FigABCD}) and several combinations of universality classes are shown in Figs. \ref{FigAfx}-\ref{FigE433} and discussed in Sec. \ref{five}. In all cases the force is attractive for $OO$ and $++$ boundaries and repulsive for $+-$ and $O+$.  For small and large values of the needle length to separation ratio, the numerical results are in excellent agreement with the SPOE results and the results for needles of infinite length in Sec. \ref{infseminf}, respectively. The linear dependence of the torque on long needles on the needle length, discussed in Appendix \ref{strongover}, is also confirmed by the data in Figs. \ref{FigCfxtheta} and \ref{FigDtheta}.  

In this paper we have concentrated on needles in two dimensions with identical surface universality classes on both sides of the needle. The approach for needles of infinite length in Sec. \ref{infseminf} can easily be extended to needles with different boundary conditions on the two sides, for example, a needle along the $x$-axis with its upper edge in the class $-$ and its lower edge in the class $+$. This is discussed in Appendix \ref{diffrims}, and explicit results are given for the force between (i) two semi-infinite collinear needles and (ii) a semi-infinite needle perpendicular to the boundary of the half plane. In the latter case we also consider a boundary with ``chemical steps'' \cite{TTD}, which separate the $x$ axis into segments with $+$ and $-$ boundary conditions, and we calculate both the normal and lateral force on the needle. Lateral Casimir forces on colloidal particles in three dimensions exposed to a chemically structured surface have been measured in Ref. \onlinecite{SZHHB}. 

We close by comparing the advantages and disadvantages of the conformal mapping approach of Secs. \ref{finfin} and \ref{five} and of the approach based on the small-particle operator expansion (SPOE). For evaluating the force and torque for arbitrary size to separation ratios, as in Sec. \ref{five}, the former is clearly superior. However, it is limited to two-dimensional critical systems with conformal symmetry, to particle surfaces with uniform boundary conditions, and to the interaction of two particles immersed in the critical medium. The SPOE is only applicable if the particle size is small compared to the interparticle separation and to the correlation length of the medium in which the particles are immersed. However, the SPOE is not limited to two dimensions, is valid in near-critical as well as critical systems, and also applies if there are more than two immersed particles and if the particles have non-uniform boundary conditions. In addition to spherical and nonspherical particles embedded in near critical fluids \cite{ber,e}, the SPOE method has been applied to particles bound to fluctuating surfaces in Ref. \cite{EFT}, where it is called ``effective field theory''.                       
\newpage
\appendix
\section{TRANSLATION AND ROTATION OF ONE OF TWO PARTICLES} \label{shiro}
\label{shiftrot}
A general infinitesimal coordinate transformation
\begin{eqnarray} \label{gentraf}
\hat{\bf r}={\bf r}+{\bf a}({\bf r})
\end{eqnarray}
changes the geometry of a critical system,  including the sizes, shapes, separations, and orientations of embedded particles, from $G$ to $\hat{G}$. The
corresponding change in the universal scaling part
\cite{uniscale} of the free energy is given by
\begin{eqnarray} \label{gentraf'}
F_{\hat{G}}-F_{G} \, = \, - k_B T_{c} \, \int d {\bf r} \sum
\limits_{k,\ell} [\partial a_{k}({\bf r})/\partial r_{\ell}] \, \langle
T_{k\ell}({\bf r}) \rangle_{G}
\end{eqnarray}
to first order in ${\bf a}$, where $T_{k\ell}$ is the stress tensor
\cite{Cardy,normTkl}. 

For two particles in
the $(r_{x},r_{y})$ plane, the force and torque on particle $I$ due to particle $II$ follow directly from the change in free energy 
as particle $I$ is translated by an infinitesimal vector $(dR_{x},
dR_{y})$ or rotated by an infinitesimal angle $d\Phi$ about a
point $(r_{0,x}, r_{0,y})$, while keeping particle $II$ fixed.  Assuming that particles $I$ and $II$
are located above and below the line $r_{y}=\tilde{r}_{y}$,
respectively, we fix particle $II$ by choosing
\begin{eqnarray} \label{fixed}
(a_{x}({\bf r}), \, a_{y}({\bf r})) \, = \, (A_{x}({\bf r}), \,
A_{y}({\bf r})) \times \Theta(r_{y}-\tilde{r}_{y})\,,
\end{eqnarray}
where $\Theta$ is the standard unit step function. To translate and rotate $I$, we choose
\begin{eqnarray} 
&&(A_{x}, \, A_{y}) \, = \, (dR_{x}, \, dR_{y})\,,\label{shift}\\
&&(A_{x}({\bf r}), \, A_{y}({\bf r})) \, = \, (-r_{y}+r_{0,y}\,,
r_{x}-r_{0,x}) \, d \Phi \,,\label{rotate}
\end{eqnarray}
respectively. On substituting
\begin{eqnarray} \label{ggentraf}
\partial a_{k}/\partial r_{\ell} \, = \,[ \partial A_{k}/\partial
r_{\ell}] \times \Theta(r_{y}-\tilde{r}_{y}) \, + \, A_{k} \,
\delta_{\ell,y} \, \delta(r_{y}-\tilde{r}_{y})
\end{eqnarray}
in Eq. (\ref{gentraf'}), the first term on the right hand side does not
contribute, since $\partial A_{k}/\partial r_{\ell}$ vanishes for the
shift and is antisymmetric in $k\ell$ for the rotation while $T_{k\ell}$
is symmetric. Thus,
\begin{eqnarray} \label{gentraf''}
F_{\hat{G}}-F_{G} \, = \, - k_B T_{c} \, \int_{-\infty}^{\infty} d
r_{x} \, J \, , \qquad J \, = \, \sum \limits_{k=x,y} A_{k}(r_{x},
\tilde{r}_{y}) \, \langle T_{ky}(r_{x}, \tilde{r}_{y}) \rangle_{G}
\, .
\end{eqnarray}
Of course, $F_{\hat{G}}-F_{G}$ should not depend on the precise
choice of $\tilde{r}_{y}$, and this property follows from the vanishing of $\partial (\int_{-
\infty}^{\infty} dr_{x} J)/\partial \tilde{r}_{y}$ due to
the continuity equation $\sum_{\ell} \partial \langle T_{k\ell}({\bf r})
\rangle_{G} / \partial r_{\ell} = 0$  at any point ${\bf
r}$ outside the particles. Using the relations 
\begin{eqnarray} \label{Tz}
&& \langle T_{yy}(r_{x}, \tilde{r}_{y}) \rangle_{G} = -\langle T_{xx}(r_{x}, \tilde{r}_{y}) \rangle_{G} = {\rm Re} \, \vartheta(z) , \,  \,  \langle T_{xy}(r_{x},
\tilde{r}_{y}) \rangle_{G} = \langle T_{yx}(r_{x},
\tilde{r}_{y})\rangle_{G} \, = \, {\rm Im} \, \vartheta (z) \, , \nonumber \\
&&\qquad \qquad \qquad \qquad \quad \quad \vartheta(z) \equiv \langle T(z) \rangle_{G}/\pi \, ,
\qquad z=r_{x}+i\tilde{r}_{y}\,
\end{eqnarray}
between the Cartesian components and complex form the of stress tensor (see \cite{Cardy,normTkl}), one finds
\begin{eqnarray} \label{Ixytoz}
&&J \, = \, dR_{x} \, {\rm Im} \, \vartheta(z) \, + \, dR_{y} \,
{\rm Re} \, \vartheta(z)\,,\\
&&J \, = \, d\Phi \, {\rm Re} [(z-z_{0}) \vartheta(z)] \, , \qquad
z_{0}=r_{0x}+ir_{0y}\,,
\end{eqnarray}
for the translation and rotation, respectively.
Together with Eq. (\ref{gentraf''}) and $dr_{x}=dz$, this  implies
\begin{eqnarray}
&&F_{\hat{G}}-F_{G}\, = \, k_{B} T \, \left[ dR_{x} \, {\rm Im} \int
_{{\cal C}_{I}} dz \, \vartheta(z) \, + \, dR_{y} \, {\rm Re} \int
_{{\cal C}_{I}} dz \, \vartheta(z) \right]\,, \label{shiftz}\\
&&F_{\hat{G}}-F_{G}\, = \, k_{B} T \, d\Phi \, {\rm Re} \int _{{\cal
C}_{I}} dz \, (z-z_{0}) \vartheta(z)\,, \label{rotatez}
\end{eqnarray}
where the closed integration contour ${\cal C}_{I}$ goes clockwise around particle $I$, with particle $II$ outside the contour.
In arriving at this result, we first deformed the integration path in Eq. (\ref{gentraf''}) to a counterclockwise loop around needle $I$, as allowed by the analyticity \cite{Cardy} and large $z$ properties of $\vartheta(z)$ and of $(z-z_{0})\vartheta(z)$. We then replaced this integral by minus the integral around the clockwise contour ${\cal C}_{I}$.
 
Equations (\ref{shiftz}) and (\ref{rotatez}) are more general than our derivation and also apply to configurations in which the two particles do not lie above and below a line parallel to the $r_{x}$ axis. The same is true of the corresponding expressions (\ref{force'}) and (\ref{turn'}) for the force and the torque. Two needles can always be separated by a straight line, and after an appropriate global rotation of the system, Eq. (\ref{fixed}) can be applied. On rotating counterclockwise by an arbitrary finite angle $\omega$, $ \langle T(z) \rangle \to e^{-2i \omega}  \langle T(z) \rangle, \, dz \to e^{i\omega} dz, \, z-z_I \to e^{i\omega} (z-z_I)$, and Eqs. (\ref{force'}) and (\ref{turn'}) correctly predict the rotation $f_x +if_y \to e^{i\omega} (f_x +if_y)$ of the force and that the torque is unchanged. Expressions (\ref{shiftz}) and (\ref{rotatez}) hold for two particles of arbitrary shape, even if they are positioned so that no separating straight line exists. This follows from a modified infinitesimal transformation (\ref{fixed}) in which the region onto which the step function $\Theta$ projects is not a half plane. 

For the special case of two widely separated needles we have checked the consistency of Eqs.
(\ref{shiftz}) and (\ref{rotatez}) with the
small-particle operator expansion reviewed in Appendix
\ref{operator}.

\newpage

\section{EXPANSIONS FOR SHORT NEEDLES} \label{SPE}

The Casimir interaction of a needle which is short compared to the distance to other particles and to the boundary can be studied analytically in a power series expansion. In Appendix \ref{smallh} we
consider the small $h$ expansion, where $h$ is the ratio of the inner to outer
radius of the annulus, and provide more details on the derivation of
the distant needle results for force and torque presented in Sec.
\ref{finfin}. In Appendix \ref{operator} we study the interaction of the needles with the small-particle
operator expansion (SPOE).  Since the two methods must lead to identical results,
one can make useful checks.

\subsection{Expanding for small h} \label{smallh}

\subsubsection{Two small needles} \label{htwosmall}

To arrive at the form of $\tau^{(S)}$ for widely separated
needles given in (\ref{tausmall}), we expand the Schwarzian
derivative on the circle ${\cal C}={\cal C}_{c}$, defined below Eq. (\ref{Spq}), in $h$, obtaining
\begin{eqnarray} \label{SCc}
S(w=h^{1/2}C e^{i \varphi}) \, \times \, C^{2} e^{2i\varphi} /6 \,
= \, : \sigma (\varphi) \, = \, \sigma_0 (\varphi) + h^{1/2}
\sigma_1 (\varphi) + h \sigma_2 (\varphi) + {\cal O}(h^{3/2})\,,
\end{eqnarray}
where
\begin{eqnarray} \label{SCc'}
&&\sigma_0 = - C^{2} e^{2i\varphi} e^{-2i \varphi_1} - C^{-2}
e^{-2i \varphi} e^{2i \varphi_3} \,, \nonumber \\
&&\sigma_1 = 2 \{ C^{3} e^{2i\varphi} (e^{-i \varphi_1} - e^{-3i
\varphi_1}) + C^{-3}e^{-2i\varphi} (e^{i\varphi_3} -
e^{3i\varphi_3}) \}\,,
\nonumber \\
&&\sigma_2 = C^{4} e^{2i\varphi}
[-1+e^{-2i\varphi_1}(6-4e^{i\varphi}) +
e^{-4i\varphi_1}(-5+4e^{i\varphi}-2e^{2i\varphi})] \nonumber \\
&& \quad \quad +2e^{-2i\varphi_1} e^{2i\varphi_3} \nonumber \\
&& \quad \quad +C^{-4} e^{-2i\varphi}
[-1+e^{2i\varphi_3}(6-4e^{-i\varphi})+e^{4i\varphi_3}(-5+4e^{-i\varphi}-2e^{-2i\varphi})]
\, .
\end{eqnarray}
The invariance of the right hand sides on exchanging $(C,\varphi,\varphi_1 , \varphi_3) \leftrightarrow (C^{-1},-\varphi,-\varphi_3 , -\varphi_1 )$ presumably
persists in higher order. On the circle ${\cal C}_c$ the prefactor
of the square bracket in the integral (\ref{force''}) reads
\begin{eqnarray} \label{invderiv}
{1\over z'(w)} \, = \, -{1 \over A} h^{2} (1-e^{i\varphi})^{2}
e^{2i\varphi_3} \, [1+h^{1/2} \delta_1 (\varphi) +h \delta_2
(\varphi) + {\cal O}(h^{3/2})]\,,
\end{eqnarray}
where
\begin{eqnarray} \label{invderiv'}
&&\delta_1 = 2 C^{-1} (e^{i\varphi_3} - e^{-i\varphi_3}) \, ,
\nonumber \\
&&\delta_2 = -4C^{-2} + C^{-2} e^{2i\varphi_3} (4-2e^{-i\varphi} +
e^{-2i\varphi}) + C^{2} e^{-2i\varphi_1} (-2e^{i\varphi} +
e^{2i\varphi}) \,,
\end{eqnarray}
and implies
\begin{eqnarray} \label{taufromS}
\tau^{(S)} \equiv \int _{{\cal C}_c} \, d w {1 \over dz/dw} (-) {1
\over 24} S(w)/\pi \, = \, {ih^{5/2} \over 4 \pi A C}
e^{2i\varphi_3} \, I\,,
\end{eqnarray}
where
\begin{eqnarray} \label{taufromS'}
I \, &=& \, \int_{0}^{2\pi} \, d\varphi (e^{-i\varphi} - 2 +
e^{i\varphi}) (1+h^{1/2}\delta_1 + h \delta_2 + ...)(\sigma_0 +
h^{1/2} \sigma_1 + h \sigma_2 + ...) \nonumber \\
&=& \, h \, \int_{0}^{2\pi} \, d\varphi (e^{-i\varphi} - 2 +
e^{i\varphi}) (\sigma_2 + \delta_2 \sigma_0) \, + \, {\cal
O}(h^{3/2}) \nonumber \\
&=& \, 8 \pi h e^{-2i\varphi_1} e^{2i\varphi_3} \, .
\end{eqnarray}
Together with Eq.  (\ref{midmidmu}) this leads to the result for $\tau^{(S)}$ in Eq. (\ref{tausmall}).

Next we derive (\ref{hyptor}) of $\theta^{(S)}$ for two widely separated  needles. Since
in (\ref{thetaw'}) we again integrate $w$ counterclockwise around
the circle ${\cal C}={\cal C}_c\thinspace$, the two required quantities
$\zeta_{n}(w)$ are conveniently obtained by splitting 
the $\tilde{w}$ integration paths in Eq. (\ref{thetaw''}) into three
parts: $[\alpha]$ from $w_1$ or $w_2$ along the outer boundary
circle to the point $-1$; $[\beta]$ from $-1$ along the negative
real axis to $-C h^{1/2}$; $[\gamma]$ from $-C h^{1/2}$ along the
circle ${\cal C}_c$ to the point $w \equiv C h^{1/2} \exp(i
\varphi)$. This yields
\begin{eqnarray} \label{zeta}
\zeta_{n}(w)=\zeta_{n}^{[\alpha]} + \zeta^{[\beta]} +
\zeta^{[\gamma]}(w) \,, \qquad n=1,2
\end{eqnarray}
where
\begin{eqnarray} \label{zeta321}
\zeta_{n}^{[\alpha]} /A&=&{\cal O}(1/h) \, , \quad \zeta^{[\beta]}
/A=-h^{-3/2} (C/2) e^{-2i \varphi_3} + {\cal O}(1/h) \, ,
\nonumber \\
\zeta^{[\gamma]}(w) /A&=& -h^{-3/2} C e^{-2i \varphi_3} \Biggl\{
\Bigl[ 1-h^{1/2}2C^{-1} \Bigl( e^{i \varphi_3} - e^{-i \varphi_3}
\Bigr) \Bigr]\Biggl( {1 \over 1-\Omega}
-{1 \over 2} \Biggr) \nonumber \\
&& \quad + h\Biggl[ 4C^{-2}\Bigl(-1+e^{-2i \varphi_3}
\Bigr)\Biggl( {1 \over 1-\Omega} -{1 \over 2} \Biggr) +
C^{-2}e^{2i \varphi_3}\Biggl({1 \over 1-\Omega} + {1 \over \Omega}
+ {1 \over
2} \Biggr) \nonumber \\
&& \qquad \quad + C^{2} e^{-2i \varphi_1} \Biggl( {1 \over
1-\Omega} - \Omega - {3 \over 2} \Biggr) \Biggr] +{\cal
O}(h^{3/2}) \Biggr\} \, ,
\end{eqnarray}
and $\Omega \equiv e^{i \varphi}$. By construction, only the first term $\zeta_n^{[\alpha]}$ in Eq. (\ref{zeta}) depends on $n$, and only the third term 
$\zeta^{[\gamma]}(w)$ depends on $w$. The $h$-expansion for
$\theta^{(S)}$ is obtained by substituting $1/z'(w)$, $S(w)$ and
$\zeta_{n}$ from Eqs. (\ref{SCc})-(\ref{invderiv'}) and (\ref{zeta}),
(\ref{zeta321}) in Eq. (\ref{thetaw'}). One (readily) finds that
$\zeta_{n}^{[\alpha]}$ and (with more work) that
$\zeta^{[\gamma]}(w)$ only contribute to $\theta^{(S)}$ in orders higher than
$h^{2}$, while $\zeta^{[\beta]}$ makes the
leading contribution $\theta^{(S)} \to \zeta^{[\beta]} \tau^{(S)}$
given in Eq. (\ref{hyptor}), which is of order $h^{2}$. Here Eqs.
(\ref{taufromS}) and (\ref{taufromS'}) have been used in the last
step.

\subsubsection{A small needle in the half plane}
\label{hsmallhalf}

Here we derive, within the $h$-expansion, the contribution $-{\rm Re} \,
\theta^{(T)}$ to the torque acting on a small needle in the half
plane shown in Eq. (\ref{smallhalfthetaT}). For the
integration path ${\cal C}$ in Eq (\ref{thetaw'}), we use the inner boundary
circle $w=h e^{i \varphi}$ and split the integrations for
$\zeta_{n}(w)$, as in Eq. (\ref{zeta}), where for $[\alpha]$ the
integration is as before, while for $[\beta]$ and $[\gamma]$ it
goes from $-1$ to $-h$ and from $-h$ to $w=h e^{i \varphi}$,
respectively. For $[\gamma]$ we integrate over the segment of the
inner circle which does not contain the singular point $w=h$ of
$z'(w)$. Instead of $\varphi$ it is convenient to use the
deviation $\chi=\varphi-\pi$ from $\varphi=\pi$ as the integration
variable on the inner circle, and with the help of Eq. (\ref{Ahalfdzdw'}),
one obtains
\begin{eqnarray} \label{gamma}
&&\zeta^{[\gamma]} (w=he^{i\varphi})/|A| \, \equiv \, -{1 \over h}
\, \int_{0}^{\chi} d\chi' \, {1 \over 4 \cos^{2}(\chi' /2)}
\tilde{\cal P}(\chi') \nonumber \\
&& \;\;\to  \, -{1 \over 2h} \tan(\chi/2) \, + \, h \Bigl[
2(1-\cos\chi)\sin(2\varphi_{1}) + \Bigl( 2 \sin\chi - \tan(\chi/2)
\Bigr)\cos(2\varphi_{1}) \Bigr] \,.
\end{eqnarray}
Here $\tilde{\cal P}$ is the product in Eq. (\ref{Ahalfdzdw'}),
and we have used its behavior for small $h$,
\begin{eqnarray} \label{calP}
&&\tilde{\cal P}(\chi) \equiv \prod \limits_{k=1}^{\infty} {\prod
\limits_{n=1,2} |1+h^{2k-1}e^{i(\chi- \varphi_{n})}|^{2} \over
|1+h^{2k}e^{i\chi}|^{4}} \nonumber \\
&& \; \to \, 1 - 2h^{2} \, \{ [2\sin\chi +
\sin(2\chi)]\sin(2\varphi_{1})+[2\cos\chi +
\cos(2\chi)]\cos(2\varphi_{1}) \} \, .
\end{eqnarray}
To first order in $h$,
\begin{eqnarray} \label{alpha}
\Bigl(\zeta_{1}^{[\alpha]} + \zeta_{2}^{[\alpha]} \Bigr)/(2 |A|)
\, = \, -i \Bigl( 1+e^{-2i\varphi_{1}} \Bigr)
-4he^{-2i\varphi_{1}}\sin \varphi_{1} \, ,
\end{eqnarray}
\begin{eqnarray} \label{alphabeta}
\Biggl({\zeta_{1}^{[\alpha]} + \zeta_{2}^{[\alpha]} \over 2}  \, +
\, \zeta^{[\beta]} \Biggr)/|A| \, = \, - {i \over 2h} \, + \, i h
\Biggl( -1 + {1 \over 2}e^{2i \varphi_{1}} - {5 \over 2}e^{-2i
\varphi_{1}} \Biggr)\,,
\end{eqnarray}
so that
\begin{eqnarray} \label{Realphabeta}
{\rm Re} \, \Biggl({\zeta_{1}^{[\alpha]} + \zeta_{2}^{[\alpha]}
\over 2}  \, + \, \zeta^{[\beta]} \Biggr)/|A| \, = \, -3h \sin(2
\varphi_{1}) \, ,
\end{eqnarray}
and Eq. (\ref{smallhalfthetaT}) then follows from
\begin{eqnarray} \label{thetaprove}
{2\pi \over t(h)} \, {\rm Re} \, \theta^{(T)} \, &\equiv& \, h
\int_{-\pi}^{\pi} d\chi \, 4\cos^{2}(\chi/2) {1 \over \tilde{\cal
P}(\chi)} \,\times\nonumber\\ &&\qquad\times \Biggl\{ {\rm Re} \,\Biggl(
{\zeta_{1}^{[\alpha]}+\zeta_{2}^{[\alpha]} \over 2} +
\zeta^{[\beta]} \Biggr)+ \zeta^{[\gamma]}(he^{i\varphi})
\Biggr\}/|A| \, .
\end{eqnarray}
Inserting Eq. (\ref{gamma}) in Eq. (\ref{thetaprove}), one finds that
$\zeta^{[\gamma]}$ does not contribute to the leading order result
shown on the right hand side of Eq. (\ref{smallhalfthetaT}).

\newpage

\subsection{Operator expansion for a distant needle} \label{operator}

Like a product of two operators in the ``operator-product
expansion'' \cite{WK}, a small particle can be represented by
a sum of operators with appropriate prefactors \cite{ber,e},
see also \cite{ee2006}. Consider a distant needle $J$,
i.e., a needle of short \cite{distsmall} length $D_{J}$ and surface universality class
$H_{J}$, with center at ${\bf r}_{J}$ and directed along the unit
vector ${\bf n}_{J}$. Inserting it into the $d=2$
Ising model at the critical point changes the Boltzmann weight of
the corresponding field theory by a factor
\begin{eqnarray} \label{sme}
e^{-\delta {\cal H}_{J}} \, \propto \, 1+s_{J}\,,
\end{eqnarray}
where $s_{J}$ is the operator series \cite{VED}
\begin{eqnarray} \label{s}
s_{J}&=&\sum_{{\cal O}=\phi,\epsilon} {\cal A}_{\cal O}^{(H_{J})}
\, \Biggl( {D_{J} \over 2} \Biggr)^{x_{\cal O}} \, \Biggl\{
1+\Biggl( {D_{J} \over 2} \Biggr)^{2} \Biggl[{1 \over 16 x_{\cal
O}}\Delta_{{\bf
r}_{J}} + \nonumber \\
&& \; + {3 \over 8 (1+x_{\cal O})} \Biggl( {\cal D}_{J}-{1 \over
2} \Delta_{{\bf r}_{J}} \Biggr)\Biggr] \Biggr\} \, {\cal O}({\bf
r}_{J}) \, - \, {\pi \over 2} \Biggl( {D_{J} \over 2} \Biggr)^2 \,
\tilde{T}(J) \, + \, ...
\end{eqnarray}
Here $\Delta_{\bf r}$ is the Laplacian operator, and the expressions
	\begin{eqnarray} \label{calD}
	{\cal D}_{J}=\sum_{k,\ell=x,y} n_{J,k} \, n_{J,\ell} \, \partial_{r_{J,k}}
	\,
	\partial_{r_{J,\ell}}\, 
	\end{eqnarray}
	and 
	\begin{eqnarray} \label{tildeT}
	\tilde{T}(J)=\sum_{k,\ell=x,y} n_{J,k} \, n_{J, \ell} \, T_{k \ell}({\bf
		r}_{J})
	\end{eqnarray} 
are the second derivative and the component of the stress tensor
\cite{normTkl}, respectively, in the needle direction. In Eq. (\ref{s})
all the operators ${\cal O}$ are subtracted so that their bulk mean values vanish
at the critical point, and  $<s_{J}>_{\rm bulk}=0$. The
operators ${\cal O}=\phi$ and ${\cal O}=\epsilon$ correspond to the order
parameter and energy densities, respectively, and are normalized according to
\begin{eqnarray} \label{norm}
\langle {\cal O}({\bf r}) {\cal O}({\bf r}') \rangle_{\rm
bulk}=|{\bf r}-{\bf r}'|^{-2 x_{\cal O}} \, ,
\end{eqnarray}
with $x_{\phi}=1/8$ and $x_{\epsilon}=1$. The universal quantities
${\cal A}_{\cal O}^{(H_{J})}$ in (\ref{s}) are the amplitudes of the corresponding density profiles $\langle {\cal
O}(r_{x},r_{y}) \rangle_{\rm uhp} = {\cal A}_{\cal O}^{(H_{J})}
r_{y}^{-x_{\cal O}}$ in the upper half plane (uhp) with the
boundary at $r_{y}=0$ belonging to the surface class $H_{J}$. They
are given by \cite{Cardy}
\begin{eqnarray} \label{halfamp}
{\cal A}_{\phi}^{(O)}=0, \, {\cal A}_{\phi}^{(+)}=-{\cal
A}_{\phi}^{(-)}=2^{1/8}, \, {\cal A}_{\epsilon}^{(O)}=-{\cal
A}_{\epsilon}^{(+)}=-{\cal A}_{\epsilon}^{(-)}=1/2 \, .
\end{eqnarray}
The amplitudes ${\cal A}_{\cal O}^{(H)}$ should not be confused with the prefactor ${\cal A}$ of the conformal transformation in Sec. \ref{semiinhalf}. Denoting the angle between the unit vector ${\bf n}_{J}=(n_{J,x},n_{J,y})$ and the $x$ axis by $\Phi_{J}$ and using
complex notation,
\begin{eqnarray} \label{nPhi}
r_{x}+ir_{y}=z, \quad r_{x}-ir_{y}=\bar{z}, \quad  n_x + i n_y =
e^{i \Phi} \, ,
\end{eqnarray}
one obtains  the useful relation
\begin{eqnarray} \label{nnT}
\tilde{T}(J)= \cos(2\Phi_{J}) T_{xx}({\bf r}_{J})+\sin(2\Phi_{J})
T_{xy}({\bf r}_{J}) = -{1 \over 2\pi} \Bigl[ e^{2i\Phi_{J}}
T(z_{J}) + e^{-2i\Phi_{J}} \bar{T} (\bar{z}_{J}) \Bigr] \, .
\end{eqnarray}
Here $T(z)$ and $\bar{T} (\bar{z})$ are components of the complex stress tensor \cite{Cardy}, and Ref. \cite{normTkl} was used in the
last step. Note that the prefactor of the $\tilde{T}(J)$-term in
Eq. (\ref{s}) is independent of the surface universality class $H_{J}$
of the needle, i.e., ``hyper-universal'' \cite{ophype}. The
ellipsis  in Eq. (\ref{s}) represents contributions from higher
descendants of $1, \, \phi, \, \epsilon$, each of which is compatible
with all symmetries of the needle and which, due to their scaling
dimensions, are multiplied by powers of $D_{J}$, greater by at least 2
than the powers shown.

\subsubsection{Two small needles} \label{optwodist}

For two small needles $I$ and $II$ the free energy of interaction
$\delta F$ is determined by \cite{delF}
\begin{eqnarray} \label{FI,II}
e^{-\delta F/(k_{B}T)} \, = \, 1 + \langle s_{I} s_{II}
\rangle_{\rm bulk}\,,
\end{eqnarray}
where, on using (\ref{s})-(\ref{halfamp}),
\begin{eqnarray} \label{complete'}
\langle s_{I} s_{II} \rangle_{\rm bulk} = \pm {\cal E}+{\cal H}+...
\end{eqnarray}
for needle classes $OO$ (upper sign) and $O+$ (lower sign), while
\begin{eqnarray} \label{complete''}
\langle s_{I} s_{II} \rangle_{\rm bulk} = \pm {\cal F}+{\cal E}+{\cal H}+...
\end{eqnarray}
for classes $++$ (upper sign) and $+-$ (lower sign). Here
\begin{eqnarray} \label{calE}
{\cal E}&=&{D_{I}D_{II} \over 16}\Biggl[ 1+\Bigl( {D_{I} \over 8}
\Bigr)^{2} \Bigl( -{1 \over 2} \Delta_{{\rm r}_{I}} + 3{\cal
D}_{I} \Bigr) + \Bigl( {D_{II} \over 8} \Bigr)^{2} \Bigl( -{1
\over 2} \Delta_{{\rm r}_{II}} + 3{\cal D}_{II} \Bigr)\Biggr]{1
\over |{\bf r}_{I}-{\bf r}_{II}|^{2}} \nonumber \\
&=&{D_{I}D_{II} \over 16 |{\bf r}_{I}-{\bf r}_{II}|^2}\Bigl[1+2^{-3}(\beta_{I}+\beta_{II})\Bigr] \, ,
\end{eqnarray}
\begin{eqnarray} \label{calF}
{\cal F}&=&(D_{I}D_{II})^{1/8} \Biggl[ 1+ {D_{I}^{2} \over 12}
\Bigl( \Delta_{{\rm r}_{I}} + {\cal D}_{I} \Bigr)+{D_{II}^{2}
\over 12} \Bigl( \Delta_{{\rm r}_{II}} + {\cal D}_{II} \Bigr)
\Biggr]{1 \over |{\bf r}_{I}-{\bf r}_{II}|^{1/4}} \nonumber \\
&=& \Biggl({D_{I}D_{II} \over |{\bf r}_{I}-{\bf r}_{II}|^2}\Biggr)^{1/8} \Bigl[1+2^{-6}(\beta_{I}+\beta_{II})\Bigr]\, ,
\end{eqnarray}
and the hyper-universal contribution (see Refs. \cite{Cardy,normTkl}) is
\begin{eqnarray} \label{calH}
{\cal H}&=&\Bigl( {\pi \over 2} \Bigr)^{2} \Biggl( {D_{I}D_{II}
\over 4} \Biggr)^{2}\langle \tilde{T}(I) \tilde{T}(II)
\rangle_{\rm bulk} \nonumber \\
&=& 2^{-10} (D_{I}D_{II})^2 \Biggl[ {e^{2i(\Phi_{I}+\Phi_{II})} \over (z_{I}-z_{II})^4} + {\rm c.c.} \Biggr] \, .
\end{eqnarray}
The quantity $\beta_J$ is defined by 
\begin{eqnarray} \label{beta}
\beta_{J}={D_{J}^2 \over |{\bf r}_{I}-{\bf r}_{II}|^2} \Bigl\{ -1+3\Bigl[ \Bigl( {\bf n}_{J}({\bf r}_{I}-{\bf r}_{II}) \Bigr)/|{\bf r}_{I}-{\bf r}_{II}| \Bigr]^2 \Bigr\} \,,
\end{eqnarray}
where the curly bracket depends on the angle between the direction of needle $J$ and the vector between the two needle centers. As expected, all the terms in the free energy remain unchanged if either needle is rotated about its center by 180 degrees.

The force $(f_{x},f_{y})$ on needle $I$ follows from
\begin{eqnarray} \label{ffromdelF}
(f_{x}, \, f_{y})=-\Biggl( {\partial \over \partial r_{I,x}}, \,
{\partial \over \partial r_{I,y}}\Biggr) \, \delta F\,,
\end{eqnarray}
and the torque $\Theta$ from Eq. (\ref{turn}) with $\Phi_{12}
\equiv \Phi_{I}$.

For illustration, consider the symmetric perpendicular (letter T)
needle configuration with needle centers on the $x$ axis and
$r_{I,x}-r_{II,x} >0$, as described in paragraph (A) of section
\ref{map}, and assume that the two needles have equal lengths
$D_{I}=D_{II}\equiv D$. Denoting by $B=|{\bf r}_{I}-{\bf
r}_{II}|/D \equiv (r_{I,x}-r_{II,x})/D \equiv |z_{I,II}|/D$ the
center-to-center distance of the needles in units of $D$, one finds
\begin{eqnarray} \label{syperpEFH}
{\cal E}=2^{-4}B^{-2}+2^{-7}B^{-4} \, , \quad {\cal
F}=B^{-1/4}+2^{-6}B^{-9/4} \, , \quad {\cal H}=-2^{-9}B^{-4} \, .
\end{eqnarray}
The component $f_y$ of the force on needle $I$ vanishes, and 
\begin{eqnarray} \label{perpforce}
D f_{x}/(k_{B}T)=(d/dB)\ln \Bigl[1 +(+1,-1) \times
2^{-4}B^{-2}+(3,-5) \times 2^{-9}B^{-4}\Bigr]
\end{eqnarray}
for needle classes $OO$ (left entry), $O+$ (right entry), and, via \cite{fourphi},
\begin{eqnarray} \label{perpforce'}
D f_{x}/(k_{B}T)=(d/dB)\ln \Bigl[1 \pm \Bigl( B^{-1/4} +
2^{-6}B^{-9/4} \Bigr) +2^{-4} B^{-2} + {\cal O}(B^{-4})\Bigr]
\end{eqnarray}
for classes $++$ (upper sign) and $+-$ (lower sign).

As another example, consider needle configurations mirror symmetric about the imaginary axis, which correspond to class (C) in section
\ref{map}. By symmetry $f_{y}=0$. In terms of the the angle $\alpha=\Phi_{34}-\Phi_{12} \equiv \Phi_{II}-\Phi_{I}$ enclosed by the two needles, Eqs. (\ref{calE})-(\ref{beta}) lead to
\begin{eqnarray} \label{syperpEFH'}
{\cal E}&=&2^{-4}B^{-2}+2^{-6}\{ -1+3[\sin(\alpha/2)]^2 \}B^{-4} \, , \nonumber \\
{\cal F}&=&B^{-1/4}+2^{-5}\{ -1+3[\sin(\alpha/2)]^2 \}B^{-9/4} \, , \quad {\cal H}=+2^{-9}B^{-4} \,
\end{eqnarray}
and
\begin{eqnarray} \label{mirrorforce}
D f_{x}/(k_{B}T)&=&(\partial/\partial B)\ln \Biggl[1 +(1,-1) \times
2^{-4}B^{-2}+ \nonumber \\
&& \qquad \qquad \quad +\{ (-7,9)+(24,-24)[\sin(\alpha/2)]^2 \} \times 2^{-9}B^{-4}\Biggr]
\end{eqnarray}
for $(OO,O+)$ and \cite{fourphi}
\begin{eqnarray} \label{mirrorforce'}
D f_{x}/(k_{B}T)&=&(\partial/\partial B)\ln \Biggl[1 \pm \Biggl( B^{-1/4}+2^{-5}\{ -1+3[\sin(\alpha/2)]^2 \}B^{-9/4} \Biggr) + \nonumber \\
&& \qquad \qquad \quad +2^{-4} B^{-2} + {\cal O}(B^{-4})\Biggr]
\end{eqnarray}
for $++$ (upper sign) and $+-$ (lower sign). The special cases (i) and (ii) of collinear and symmetric parallel needles correspond to $\alpha=\pi$ and $\alpha=0$, respectively. For $0<\alpha<\pi$ needle $II$ exerts a nonvanishing torque $\Theta$ on needle $I$, where $\Theta /(k_B T)=-(\partial/\partial \Phi_{I}) \delta F/(k_{B}T)$ is given by the right hand sides of Eqs. (\ref{mirrorforce}) and (\ref{mirrorforce'}) with $\partial/\partial B$ replaced by $-\partial/\partial \alpha$. For $(OO, O+)$ one finds from Eq. (\ref{mirrorforce}) that $\Theta /(k_B T)=(-1,1)2^{-7} 3 B^{-4} \sin \alpha +{\cal O}(B^{-6})$. The sign of $\Theta$ indicates that the interaction is dominated by the two closer needle halves.

We also consider case (D) in Sec. IV A 1, in which the two needles of equal length $D$ form angles $\Phi_{12} \equiv \Phi_{I} \equiv \Phi$ and $\Phi_{34}=\Phi+\pi$ with the vector $z_{I}-z_{II} > 0$ between their centers on the $x$ axis. For this geometry Eqs. (\ref{FI,II})-(\ref{ffromdelF}) yield    
\begin{eqnarray} \label{antiforcex}
&& \qquad D f_{x}/(k_{B}T)=(\partial/\partial B)\ln (1 + S) \, , \nonumber \\
&&S=\pm \Bigl( 2^{-4} B^{-2}+ \{ -1+3 (\cos\Phi)^2 \}2^{-6} B^{-4} \Bigr) +  \cos(4\Phi) 2^{-9} B^{-4} 
\end{eqnarray}
and
\begin{eqnarray} \label{antiforcey}
&&D f_{y}/(k_{B}T)=\Bigl(\pm 6 \sin(2\Phi) + \sin(4\Phi) \Bigr) 2^{-7} B^{-5} /(1+S)
\end{eqnarray}
for needle classes $OO$ (upper sign) and $O+$ (lower sign). For needle classes $++$ (upper sign) and $+-$ (lower sign) the force components are 
\begin{eqnarray} \label{antiforcex'}
&& \qquad D f_{x}/(k_{B}T)=(\partial/\partial B)\ln (1 + S') \, , \nonumber \\
&&S'= \pm \Bigl( B^{-1/4}+\{ -1+3 (\cos\Phi)^2 \} 2^{-5} B^{-9/4} \Bigr) + 2^{-4} B^{-2} + \nonumber \\
&& \qquad \qquad + \{ -1+3 (\cos\Phi)^2 \}2^{-6} B^{-4}  + \cos(4\Phi) 2^{-9} B^{-4} 
\end{eqnarray}
and
\begin{eqnarray} \label{antiforcey'}
D f_{y}/(k_{B}T)&=&\Bigl[\pm 3\Bigl(\sin(2\Phi)\Bigr)2^{-5} B^{-13/4} + \nonumber \\
&& \; +\Bigl( 6 \sin(2\Phi) + \sin(4\Phi) \Bigr) 2^{-7} B^{-5} \Bigr]/(1+S') \, .
\end{eqnarray}
In Section \ref{five} we found it convenient to rotate this same configuration by an angle $\pi - \Phi$, 
so that needles $I$ and $II$ are antiparallel and parallel to the real axis and
$z_{I}- z_{II}=|{\bf r}_{I}-{\bf r}_{II}|e^{i(\pi-\Phi)}$ , implying 
$[r_{I,x}-r_{II,x}, \, r_{I,y}-r_{II,y}]=|{\bf r}_{I}-{\bf r}_{II}| \times [\sin(\Phi-(\pi/2)), \, \cos(\Phi-(\pi/2))]$. 
For this orientation  
\begin{eqnarray} \label{antiforcexrot}
D f_{x}/(k_{B}T)&=&\Biggl[ \pm \Biggl((2B)^{-3} + (2B)^{-5}[-5+9(\cos\Phi)^{2}] \Biggr) \cos\Phi + \nonumber \\
&& \, \qquad \qquad \qquad \qquad +2^{-7} B^{-5}\cos(5\Phi) \Biggr]/(1+S) \, ,
\end{eqnarray}%
\begin{eqnarray} \label{antiforceyrot}
D f_{y}/(k_{B}T)&=&\Biggl[ \pm \Biggl(-(2B)^{-3} + (2B)^{-5}[2-9(\cos\Phi)^{2}] \Biggr) \sin\Phi - \nonumber \\
&& \, \qquad \qquad \qquad \qquad -2^{-7} B^{-5} \sin(5\Phi) \Biggr]/(1+S) 
\end{eqnarray}
for needle universality classes $OO$ (upper sign) and $O+$ (lower sign), while for $++$ (upper sign) and $+-$ (lower sign) 
\begin{eqnarray} \label{antiforcexrot'}
D f_{x}/(k_{B}T)&=&\Biggl[\pm \Biggl( B^{-5/4} - 2^{-5}3B^{-13/4}[11-17(\cos\Phi)^{2}] \Biggr) (\cos\Phi)/4  + \nonumber \\
&& + \Biggl((2B)^{-3} + (2B)^{-5}[-5+9(\cos\Phi)^{2}] \Biggr) \cos\Phi + \nonumber \\
&& \, \qquad \qquad \qquad \qquad +2^{-7} B^{-5}\cos(5\Phi) \Biggr]/(1+S') \, ,
\end{eqnarray}
\begin{eqnarray} \label{antiforceyrot'}
D f_{y}/(k_{B}T)&=&\Biggl[ \pm \Biggl(- B^{-5/4}+ 2^{-5}3B^{-13/4}[3-17(\cos\Phi)^{2}] \Biggr) (\sin\Phi)/4  + \nonumber \\
&& + \Biggl(-(2B)^{-3} + (2B)^{-5}[2-9(\cos\Phi)^{2}] \Biggr) \sin\Phi - \nonumber \\
&& \, \qquad \qquad \qquad \qquad -2^{-7} B^{-5} \sin(5\Phi) \Biggr]/(1+S') \ . 
\end{eqnarray}
As required by symmetry, $f_{x}$ and $f_{y}$ in Eqs. (\ref{antiforcex})-(\ref{antiforcey'}) are even and odd in $\Phi$, respectively, and in Eqs. (\ref{antiforcexrot})-(\ref{antiforceyrot'}) they are odd and even in $\Phi-(\pi/2)$. For the torque $\Theta$ in case (D) our operator expansion yields
\begin{eqnarray} \label{antitorque}
\Theta /(k_B T)=-\Bigl(\pm 6 \sin(2\Phi) + \sin(4\Phi) \Bigr) 2^{-8} B^{-4} /(1+S)
\end{eqnarray}
for needle classes $OO$ (upper sign) and $O+$ (lower sign) and
\begin{eqnarray} \label{antitorque'}
\Theta /(k_B T)&=&-\Bigl[\pm 3\Bigl(\sin(2\Phi)\Bigr)2^{-6} B^{-9/4} + \nonumber \\
&& \; \quad +\Bigl( 6 \sin(2\Phi) + \sin(4\Phi) \Bigr) 2^{-8} B^{-4} \Bigr]/(1+S') \, .
\end{eqnarray}
for needle classes $++$ (upper sign) and $+-$ (lower sign).

For two small needles with arbitrary lengths $D_{I}, \, D_{II}$
and angles $\Phi_{I}, \, \Phi_{II}$, the SPOE reproduces the
leading force contribution (\ref{flead}) and the leading
hyper-universal contributions (\ref{hypf}) and (\ref{hyptor}) to the
force and torque derived from the $h$-expansion. For the latter
quantities this is apparent from Eqs. (\ref{hypfreeeng}) and (\ref{calH}) since $\delta F^{({\rm hu})} /(k_{B}T)=-{\cal H}$.

\subsubsection{A small and a semi-infinite needle}
\label{opsmallsemi}

The interaction free energy $\delta F$ \cite{delF} of a small needle $I$
and a semi-infinite needle (semi), i.e., the free energy required
to transfer $I$ from the bulk plane to the plane with the
semi-infinite needle, is determined by
\begin{eqnarray} \label{FIsemi}
e^{-\delta F /(k_{B}T)} \, = \, 1+\langle s_{I} \rangle_{\rm semi}\,.
\end{eqnarray}
Here $s_{I}$ is the operator series in Eq. (\ref{s}), and $\langle
\, \rangle_{\rm semi}$ denotes a thermal average in the
$z=r_{x}+ir_{y}$ plane with a semi-infinite needle of class
$H_{\rm semi}$ coinciding with the positive real axis, as in Sec. \ref{finsem}. Since the semi-infinite needle
can be generated from the boundary of the upper half $w$ plane by
the conformal transformation $z=w^{2}$, the averages of the various
operators on the right hand side of Eq. (\ref{FIsemi}) follow
from their counterparts in the half plane. From Eqs. (\ref{nnT}) and
(\ref{stressstress}) and the vanishing of $\langle T(w)
\rangle_{\rm half \, plane}$, we obtain 
\begin{eqnarray} \label{Osemi}
\langle {\cal O}(r_{I,x}, r_{I,y}) \rangle_{\rm semi} &=& {\cal A}_{\cal O}^{(H_{\rm semi})} \Bigl[2|z_{I}| \sin \Bigl( ({\rm arg}z_{I})/2 \Bigr) \Bigr]^{-x_{\cal O}} \nonumber \\
&=&  {\cal A}_{\cal O}^{(H_{\rm semi})} \Bigl[ 2|z_{I}|
\Bigl(|z_{I}|-r_{I,x}\Bigr) \Bigr]^{-x_{\cal O} /2}
\end{eqnarray}
and
\begin{eqnarray} \label{Tsemi}
\langle \tilde{T}(I) \rangle_{\rm semi} = -\cos[2(\Phi_{I}-{\rm
arg}z_{I})]/(64\pi |z_{I}|^{2}) \,,
\end{eqnarray}
where $0<{\rm arg}z_I <2\pi$ and the position vector $z_{I}=r_{Ix}+i r_{Iy}$ is defined below
Eq. (\ref{enclangle'''}). 
The expression
\begin{eqnarray} \label{SPsemforce}
{f_{x}-if_{y} \over k_{B} T} \, = \, {1 \over 1+\langle
s_{I}\rangle_{\rm semi}} \, \Biggl( {\partial \over
\partial \, r_{I,x}} - i {\partial \over \partial \, r_{I,y}}
\Biggr) \langle s_{I} \rangle_{\rm semi}
\end{eqnarray}
for the force, which follows from Eqs. (\ref{FIsemi}) and (\ref{s}), reproduces, in leading
order, the result from the $h$-expansion given below Eq.
(\ref{12semiforce}). The reason is that in
\begin{eqnarray} \label{SPsemiforce}
\Biggl( {\partial \over \partial \, r_{I,x}} - i {\partial \over
\partial \, r_{I,y}} \Biggr) \langle {\cal O}\rangle_{\rm
semi} \, = \, -x_{\cal O} \, {3z_{I}-|z_{I}| \over 2z_{I}(z_{I} -
|z_{I}|)} \, \langle {\cal O} \rangle_{\rm semi} \, ,
\end{eqnarray}
with $\langle {\cal O} \rangle_{\rm semi}$ from Eq. (\ref{Osemi}),
the same fraction appears on the right-hand side as in Eq. (\ref{12semiforce}), and on expressing $\langle {\cal O}
\rangle_{\rm semi}$ via (\ref{exh}) in terms of $h/|z_{12}| \equiv
h/D_{I}$, one may use that
\begin{eqnarray} \label{AAt}
\sum_{{\cal O}=\phi,\epsilon} {\cal A}_{\cal O}^{(H_{I})} \, {\cal
A}_{\cal O}^{(H_{\rm semi})} \, x_{\cal O} \, (4h)^{x_{\cal O}} \,
\to \, t(h \to 0 ) \, \equiv \, \{ h, \, -h, \,
(\sqrt{2}/8)h^{1/8}, \, -(\sqrt{2}/8)h^{1/8} \}
\end{eqnarray}
for $\{ OO, \, O+, \, ++, \, +- \}$. Moreover, in the
cases $++$ and $+-$, the denominator on the right hand side
of (\ref{SPsemforce}) is consistent with the denominators
in Eq. (\ref{t+++-}).

The orientation-dependent contribution to $e^{-\delta F/(k_{B}T)}$
of lowest order in the needle length,
\begin{eqnarray} \label{oriF}
2^{-9} (D_{I}/|z_{I}|)^{2} \, \cos \Bigl[ 2(\Phi_{I}-{\rm
arg}z_{I}) \Bigr] \, ,
\end{eqnarray}
comes from inserting the stress tensor average (\ref{Tsemi}) in
Eq. (\ref{FIsemi}), using  Eq. (\ref{s}), and is independent of the needle classes $H_{I}$
and $H_{\rm semi}$. For universality classes $OO$ and $O+$, the contribution 
(\ref{oriF}) clearly dominates the orientation dependence $\propto
D_{I}^{x_{\epsilon}+2} = D_{I}^{3}$ coming from the ${\cal D}_{I}
{\cal O}$-term in Eq. (\ref{s}), provided that the components $r_{I,x}$ and
$r_{I,y}$ of $z_{I}$ are of the same order. However, on approaching
the limit $r_{I,x} \to + \infty$ with $r_{I,y}$ finite, the
contribution from the ${\cal D}_{I} {\cal O}$-term approaches the
finite orientation dependence of a needle in the half plane (see
Eq. (\ref{Fhalfplane'}) below), while the contribution (\ref{oriF}) vanishes. For
classes $++$ and $+-$ the ${\cal D}_{I} {\cal O}$-term
contributes an orientation dependence preceded by a power law
$D_{I}^{(1/8)+2}$ with an exponent which is only slightly larger than the exponent of the power
$D_{I}^{2}$ in Eq. (\ref{oriF}). Note that Eq. (\ref{oriF}) favors needle
orientations parallel and antiparallel to the vector
$z_{I}$ from the midpoint of the finite needle to the finite end of the
semi-infinite needle.
%
%\newpage
%

\subsubsection{A small needle in the half plane and in the symmetric-parallel configuration}
\label{opsmallhalf}
{}
For a small needle $I$ in the upper half plane (uhp) the free
energy $\delta F$ of interaction {\cite{delF} with the boundary of surface
class $H_{S}$ at $r_{y}=0$ is determined by Eq. (\ref{FIsemi}),
with $\langle \; \rangle_{\rm semi}$ replaced by the average
$\langle \; \rangle_{\rm uhp}$ in the half plane. The expressions for
$\langle {\cal O} \rangle_{\rm uhp}$ given above Eq.
(\ref{halfamp}) and the vanishing of the stress tensor average
imply
\begin{eqnarray} \label{Fhalfplane'}
e^{- \delta F /(k_B T)} &=& 1+\sum_{{\cal O}=\phi,\epsilon} {\cal
A}_{\cal O}^{(H_{I})} {\cal A}_{\cal O}^{(H_{S})} \Biggl({D_{I}
\over 2r_{I,y}}\Biggr)^{x_{\cal O}} \times
\nonumber \\
&& \qquad \quad \times \Biggl[1+\Biggl({D_{I} \over
2r_{I,y}}\Biggr)^{2} {1 \over 16} \Bigl(x_{\cal O}+1-3 x_{\cal O}
\cos (2 \Phi_{I})
 \Bigr)\Biggr] \, .
\end{eqnarray}
Both the force and the torque follow from Eq. (\ref{Fhalfplane'}).

The force $f=-\partial \delta F / \partial r_{I,y}$ with $D_I$ and
$\Phi_{I}$ fixed is given by
\begin{eqnarray} \label{fhalfplane}
D_{I} f/(k_B T)\, = \, (\partial/\partial {\tilde B}) \ln \Biggl\{ 1 \pm {1
\over 4}(2{\tilde B})^{-1} \Biggl[ 1+ (2{\tilde B})^{-2} {1 \over 16} \Bigl( 2-3
\cos(2\Phi_{I}) \Bigr) \Biggr] \Biggr\}
\end{eqnarray}
for classes $OO$ (upper sign) and $O+$ (lower sign), and by
\begin{eqnarray} \label{fhalfplane'}
D_{I} f/(k_B T) &=& (\partial/ \partial {\tilde B}) \ln \Biggl\{ 1 \pm
2^{1/4} (2{\tilde B})^{-1/8} \Biggl[ 1+(2{\tilde B})^{-2} {3 \over 2^{7}}\Bigl(
3-\cos(2\Phi_{I}) \Bigr) \Biggr] + \nonumber \\
&& \qquad \qquad +{1 \over 4}(2{\tilde B})^{-1} \Biggl[ 1+ (2{\tilde B})^{-2} {1
\over 16} \Bigl( 2-3 \cos(2\Phi_{I}) \Bigr) \Biggr] \Biggr\}
\end{eqnarray}
for $++$ (upper sign) and $+-$ (lower sign). Here
${\tilde B}=r_{I,y}/D_{I}$.

The expressions for the torque per $k_B T$, $-\partial (\delta F/k_B T)/
\partial \Phi_{I}$, follow for the various cases of universality classes $H_I \,
H_S$ from the corresponding right hand sides of Eqs.
(\ref{fhalfplane}), (\ref{fhalfplane'}) on replacing $(\partial /
\partial {\tilde B})$ by $(\partial / \partial \Phi_{I})$.

Next we consider a small needle in the symmetric-parallel configuration (B) of Fig. \ref{FigABCD}, assuming $D_I/c\ll 1$ and $D_{II}/c$ arbitrary, where $c=z_I-z_{II}$ is the distance between the needles. The limits $D_{II}/c \to \infty$ and $D_{II}/c \ll 1$ correspond to a small needle in the half plane and configuration (B) with two small needles of different lengths, respectively. The free energy $\delta F$ is determined by Eq. (\ref{FIsemi}) with $\langle \; \rangle_{\rm semi}$ replaced by the average $\langle \; \rangle_{II}$  in the plane containing needle $II$. For a needle $II$ with boundary class $H_{II}$, centered about the origin and extending along the $y$ axis, the profiles of the order parameter and energy density are given by (see, e.g., Appendix A1 in the first paper of Ref. \cite{ee2006}) 	
\begin{eqnarray}
	&&\langle {\cal O}(r_x , 0)  \rangle_{II}^{(H_{II})} = {\cal A}_{\cal O}^{(H_{II})} (D_{II}/2)^{- x_{\cal O}} \Bigl[ \Xi \Bigl( 2 |r_x |/D_{II} \Bigr) \Bigr]^{x_{\cal O}}\,,
	\label{singleneedleprof}\\
	&&\Xi (\xi) \equiv \xi^{-1}  (\xi^2 +1)^{-1/2} \,,\label{funequsymmpar''}
	\end{eqnarray}
	for ${\cal O}=\phi$ and ${\cal O}=\epsilon$, respectively.
Making use of this result and retaining only the the leading monopole contribution in the SPOE, one obtains 
	\begin{eqnarray} \label{funequsymmpar}
	D_{II} f_x /(k_B T)\, = \, 2 (\partial/\partial \xi)\left. \ln \Biggl\{ 1 \pm {1 	\over 4} (D_I /D_{II}) \Xi (\xi) \Biggr\}\right\vert_{\xi=2c/D_{II} }
	\end{eqnarray}
	for classes $OO$ (upper sign), $O+$ (lower sign) and
	\begin{eqnarray} \label{funequsymmpar'}
	D_{II} f_x /(k_B T)\, = \, 2 (\partial/\partial \xi)\left. \ln \Biggl\{ 1 \pm 2^{1/4}  (D_I /D_{II})^{1/8} [\Xi (\xi)]^{1/8} \Biggr\}\right\vert_{\xi=2c/D_{II} }
	\end{eqnarray}
	for $++$ (upper sign), $+-$ (lower sign).

\newpage 

\section{PARALLEL NEEDLES WITH STRONG OVERLAP AND LONG MIRROR-SYMMETRIC NEEDLES} \label{strongover}

First we consider the torque on one of the two antiparallel needles of configuration (D), introduced in Sec. \ref{map} in the limit of ``strong overlap.''   Two needles of equal length oriented parallel to the $x$ axis \cite{rotateD} overlap strongly if the distances $|z_{1}-z_{4}|=|z_{2}-z_{3}|$ between the left ends $z_1$ and $z_4$ and the right ends $z_{2}$ and $z_{3}$ of needles $I$ and $II$, respectively, are much smaller than their lengths $|z_{2}-z_{1}|=|z_{3}-z_{4}|=D$, so that the two needles form boundaries of a long strip of width $W=|r_{1,y}-r_{4,y}|=|r_{2,y}-r_{3,y}|$. On integrating closely around needle $I$, which is located above needle $II$, in  Eq. (\ref{turn'}), the only contributions to the torque $\theta$ come from regions with a width of order $|z_{1}-z_{4}|=|z_{2}-z_{3}|$ near the ends of the needles, i.e., near the ends of the strip . The reason is that (i)  $<T(z)>$ vanishes outside the strip over most of its length, i.e., over most of the upper edge of needle $I$, and (ii) inside the strip $<T(z)>$ is independent of $z$ and equal to its value $\pi \Delta /W^2$ in an infinite strip, with $\Delta$ from \cite{tmeaning}. Thus, by virtue of the odd factor $z-z_{I}=r_{x}-r_{I,x}$ in Eq. (\ref{turn'}), the interval of integration centered about $r_{I,x}=(r_{1,x}+r_{2,x})/2$ and comprising nearly all of the lower edge of needle $I$ gives a vanishing contribution. For large $D/W$ the two end regions are uncorrelated, and each is equivalent to  the end region of a system of two  semi-infinite needles.  Replacing $z-z_{I}$ in the left and right end contributions by $-D/2$ and $D/2$, respectively, we obtain   
\begin{eqnarray} \label{toroverlap}
\pi \theta \to -(D/2) \, {\rm lim}_{d \to + \infty} \Biggl[ \int_{{\cal C}_{I+}(d)} dz <T(z)>_{{\rm si}+} - \int_{{\cal C}_{I-}(d)} dz <T(z)>_{{\rm si}-} \Biggr] \, .
\end{eqnarray}
Here si+ denotes a system of two semi-infinite needles $I_+$ and $II_+$ extending from $z_1$ and $z_4$ to $z_{1}+|\infty|$ and  $z_{4}+|\infty|$, respectively, while si$-$ is the system of two needles $I_-$ and $II_-$ extending from $z_2$ and $z_3$ to $z_{2}-|\infty|$ and  $z_{3}-|\infty|$. The integration path ${\cal C}_{I+}(d)$ goes clockwise around the tip $z_1$ of needle $I_+$, starting at $z=z_{1}+d-i0$ and ending at $z=z_{1}+d+i0$. Similarly, ${\cal C}_{I-}(d)$ goes clockwise around the tip $z_2$ of needle $I_-$, starting at $z=z_{2}-d+i0$ and ending at $z=z_{2}-d-i0$.

On rotating by 180 degrees, the si$-$ system is mapped onto the si+ system,  with needle $I_-$ mapped onto needle $II_+$, i.e. $z_2$ onto $z_4$, and needle $II_-$ mapped onto $I_+$, i.e., $z_3$ onto $z_1$. 
Since exchanging the universality classes in a two needle system does not change the stress tensor average \cite{localT}, it is the same for the si+ and rotated si- systems. Moreover, the rotation changes $dz \to -dz$ while no prefactor arises in front of $T$, and Eq. (\ref{toroverlap}) yields 
\begin{eqnarray} \label{toroverlap'}
\pi \theta \to -(D/2) \, {\rm lim}_{d \to + \infty} \int_{{\cal C}_{I+}(d)+ {\cal C}_{II+}(d)} dz <T(z)>_{{\rm si}+} \, .
\end{eqnarray}
Here the path ${\cal C}_{II+}(d)$ encircles the tip $z_4$ of needle $II_+$ clockwise, starting at $z_{4}+d-i0$ and ending at $z_{4}+d+i0$. The integration path in Eq. (\ref{toroverlap'}) becomes connected, leading to a vanishing result, if one adds both a vertical segment from $z_{4}+d+i0$ to $z_{1}+d+r_{4,x}-r_{1,x}-i0$ and a horizontal segment from $z_{1}+d+r_{4,x}-r_{1,x}-i0$ to $z_{1}+d-i0$ to the integration path. Since for both segments $<T(z)>_{{\rm si}+}$ equals its value inside the infinite strip, and since the vertical segment leads to a purely imaginary result, the torque $\Theta=-{\rm Re} \thinspace\theta$ on needle $I$ is given by $-1/\pi$ times the contribution of the horizontal segment, with the result 
\begin{eqnarray} \label{toroverlapequal'}
\Theta=-(D/2)(r_{1,x}-r_{4,x})\Delta /W^2\,,
\end{eqnarray}
where $\Delta$ depends on the universality classes of the two needles, as specified in Ref. \cite{tmeaning}.  

We now calculate the component $f_x$  of the force on needle $I$ due to needle $II$. For this, it is convenient to place the origin at the center of reflection of the needle configuration by setting $z_{3}=-z_{1}$, $z_{4}=-z_{2}$ and to integrate along a path ${\cal C}_I$ in Eq. (\ref{force'}) midway between the needles along the real axis from $z=+ \infty$ to $z=- \infty$, closing the path with a semicircle of infinite radius which does not contribute to the integral. Since $<T(z=r_{x})>=<T(z=-r_{x})>$, and since Im$<T(z=r_{x})>$ vanishes except near the ends of the needle, the desired integral over Im$<T(z=r_{x})>$ equals twice the corresponding integral with $I$ and $II{}$ replaced by their semi-infinite counterparts $I_+$ and $II_+\thinspace$. In this way we obtain
\begin{eqnarray} \label{parforce}
{f_{x} \over k_{B} T}={\pi \over W} \Biggl({1 \over 48} {1+3b-3b^{2}-b^{3} \over 1+3b+3b^{2}+b^{3}} - {1-b \over 1+b} {\tilde t} \Biggr) \,,  
\end{eqnarray}
where ${\tilde t}$ is given in Eq. (\ref{t12}). Here $b$ is positive and related by
\begin{eqnarray} \label{paraxlambda}
(r_{1,x}-r_{4,x})/W= {1 \over 2 \pi} ( 2 \ln b +b-1/b )
\end{eqnarray}
to the ratio $(r_{1,x}-r_{4,x})/W$ of the parallel and perpendicular components of the vector between the two left needle ends. As expected,  $f_{x}/(k_{B} T)$ is an odd function of $r_{1,x} -r_{4,x}$ and tends to $\Delta /W$, $-\Delta /W$, and 0 in the cases $b \to +\infty, \, 0$, and 1 in which the ratio on the left hand side of (\ref{paraxlambda}) tends to $+\infty, \, -\infty$, and 0, respectively. 

To derive Eqs. (\ref{parforce}) and  (\ref{paraxlambda}), we first generate the geometry of parallel semi-infinite needles $I_+$ and $II_+$ from the upper half $w$ plane by means of the conformal transformation \cite{Kober153} 
\begin{eqnarray} \label{paratrafo}
z(w)={W \over \pi} \Biggl[ {w^{2} \over 2b}+w\Biggl(1-{1 \over b} \Biggr) + {1 \over 4} \Biggl({1 \over b}+b\Biggr) -1 - \ln {w \over \sqrt{b}}+ {i\pi \over 2}  \Biggr] \, .
\end{eqnarray}
Together with (\ref{paraxlambda}) this transformation conveniently places the tips of $I_+$ and $II_+$ symmetrically about the origin, at $z=z(1) \equiv z_1 = r_{1,x} +iW/2$ and $z=z(-b) \equiv z_4 =-z_1$, respectively. The integration path mentioned just above Eq. (\ref{parforce}), which is midway between the semi-infinite needles $I_+$ and $II_+\thinspace$,  corresponds, according to Eqs. (\ref{force'}) and (\ref{force''}), to the imaginary axis of the upper half $w$ plane. Similarly, the integral over Im$<T(z=r_{x})>$ corresponds to the integral of a real rational function of $|w|$ from 0 to $+ \infty$ and leads to a force $f_x$ on needle $I_+$ due to $II_+$ which is exactly half of $f_x$ in Eq. (\ref{parforce}) \cite{paralimit}.

Finally we consider the mirror-symmetric needle configuration (class  (C) of Sec. \ref{map}) and argue that in this case the torque $\Theta$ also increases linearly with the needle length $D$ for $D \to \infty$. First we place needles $I$ and $II$ so that $z_2 =1, \, z_1 =D+1$ and $z_4 = e^{i \alpha} z_2, \, z_3 =e^{i \alpha} z_1$. We also  introduce an auxiliary  ``wedge'' configuration of two corresponding semi-infinite needles which extend from $z=0$ to $|\infty|$ and to $e^{i \alpha} |\infty|$, dividing the $z$-plane into two wedges of opening angles $\alpha$ and $2\pi-\alpha$. To evaluate $\theta$ in (\ref{turn'}), we choose ${\cal C}_I$ so that it encircles needle $I$ closely and subtract and add $<T(z)>_{\rm wedge}$ to $<T(z)>$. This leads to $\theta=\delta \theta+{\tilde \theta}$, where
\begin{eqnarray} \label{thetamirror}
\delta \theta = \int_{{\cal C}_I} dz \Bigl[ <T(z)> - <T(z)>_{\rm wedge} \Bigr] \Biggl( r_{x}-\Bigl( {D \over 2}+1 \Bigr) \Biggr) \, , \nonumber \\
{\tilde \theta} = \int_{1}^{D+1} dr_{x} <T(r_{x}+i0)-T(r_{x}-i0)>_{\rm wedge} \Biggl( r_{x}-\Bigl( {D \over 2}+1 \Bigr) \Biggr) \, .
\end{eqnarray}
Since $<T(z)>_{\rm wedge}\,\propto z^{-2}$ \cite{Cardy}, its average in the integral for ${\tilde \theta}$ is proportional to $r_{x}^{-2}$, and calculating the integral reveals the leading behavior ${\tilde \theta} \propto D$ for $D>>1$. Since the square bracket becomes arbitrarily small, for $z=r_{x}+i0$ and $z=r_{x}-i0$, in the ``central'' region $1<<r_{x}<<D$, the quantity $\delta \theta$  represents the contribution to the torque from the ends of the needles, and $\delta \theta$ can be written as a sum of two expressions.  One of these, $\delta_{<} \theta$, corresponds to semi-infinite needles extending from 1 to $|\infty|$ and from $e^{i\alpha}$ to $e^{i\alpha} |\infty|$. The other contribution, $\delta_{>} \theta$, corresponds to needles extending from 0 to $D$ and from 0 to $e^{i \alpha} D$. In the case of $\delta_{<} \theta$, the difference $ <T(z)> - <T(z)>_{\rm wedge}$ for $r_{x}>>1$ is proportional to $r_{x}^{-2-(2\pi/\alpha)}$, if $z=r_{x}+i0$, and to $r_{x}^{-2-(2\pi/(2\pi-\alpha))}$, if $z=r_{x}-i0$. This follows from Eqs. (\ref{tn1})-(\ref{integrand2}) for the mirror-symmetric case considered here with $b=1$. Thus, only $r_{x}$ values of order 1 contribute, and $\delta_{<} \theta$ is proportional to $D$ for $D \to \infty$. As for $\delta_{>} \theta$, its needle geometry can be mapped either by the dilatation $z/D \to z$ to needles of length 1 or by the inversion $D/z \to z$ to the needle geometry of $\delta_{<}$. Either way, one realizes that $\delta_{>} \theta$ is of order 1. The plausible assumption that the $D$ dependence from ${\tilde \theta}$ and $\delta_{<} \theta$ does not cancel leads to the predictions  $\theta\propto D$ and $\Theta\propto D$ for $D \to \infty$, in agreement with the numerical results for case (C) in Sec. \ref{five}. 

For a needle $I$ with ends at $e^{i \Phi}$ and $(D+1) e^{i \Phi}$ in the upper half plane, similar arguments also imply $\Theta\propto D$ for $D \to \infty$. This is consistent with the numerical results for the torque in Fig. \ref{FigE433} for $\Phi$ close to $\pi /3.$   
\newpage
\section{NEEDLES WITH MIXED BOUNDARY CONDITIONS} \label{diffrims}
\subsection{Half plane with inhomogeneous boundary conditions} \label{stressinhbound}

We begin with a discussion  of $\langle T(w) \rangle_{u_1 , u_2, ... , u_N}$ in the upper half $w$ plane with boundary conditions on the real axis that alternate between + and $-$ at the $N$ points $u_1,u_2,...,u_N$. If, for example, the boundary condition for $- \infty < u < u_1$ is $+$, then it is $-$ for $u_1 < u < u_2$, + for $u_2 < u < u_3$, etc. The stress tensor for such mixed boundary conditions is of interest in its own right and is also the starting point for studying the Casimir interaction of needles with mixed boundary conditions.

For $N=0$, $\langle T(w) \rangle$ vanishes, and for $N=1, \, 2$ \cite{BX}
\begin{eqnarray} \label{1+2T}
\langle T(w) \rangle_{u_1}={\tilde{t} \over (w-u_1)^2}, \quad \langle T(w) \rangle_{u_1, u_2}=\tilde{t} \Biggl( {1 \over w-u_1} - {1 \over w-u_2} \Biggr)^2 \, , \quad \tilde{t}=\tilde{t}_{+-} \equiv 1/2\,. 
\end{eqnarray}
For $N=3$ 
\begin{eqnarray} \label{3T}
&&\langle T(w) \rangle_{u_1 , u_2 , u_3} \, = \, {1 \over (u_{12})^{-1} - (u_{13})^{-1} + (u_{23})^{-1}} \times \nonumber \\
&& \qquad \qquad \times\Bigl\{ [12] - [13] + [23] + (12,3) - (13,2) + (23,1)\Bigr\} \, , \nonumber \\
&& \qquad [ab] \, \equiv \, {\langle T(w) \rangle_{u_a, u_b} \over u_{ab} } \, , \qquad (ab,c) \, \equiv \, {1 \over u_{ab}} \, \langle T(w) \rangle_{u_c} \, ,
\end{eqnarray}
and for $N=4$
\begin{eqnarray} \label{4T}
&&\langle T(w) \rangle_{u_1 , u_2 , u_3 , u_4} \, = \, {1 \over (u_{12}u_{34})^{-1} - (u_{13}u_{24})^{-1} + (u_{23}u_{14})^{-1}} \times \nonumber \\
&& \qquad \qquad \times \Bigl( [12]/u_{34}-[13]/u_{24}+[14]/u_{23}+[23]/u_{14} -[24]/u_{13}+[34]/u_{12} \Bigr) \, , 
\end{eqnarray}
respectively, where $u_{ab}=u_a - u_b$. For $N$ an arbitrary even integer $\geq 4$
\begin{eqnarray} \label{arbitraryT}
\langle T(w) \rangle_{u_1, u_2 , ..., u_N}&=&\Biggl( {\rm Pf}^{(N)} {1 \over u_{ij}} \Biggr)^{-1} \times {\partial \over \partial \lambda} {\rm Pf}^{(N)} \Biggl( {1 \over u_{ij}} +\lambda [ij] \Biggr) \Bigg|_{\lambda=0} \nonumber \\
&=&\Biggl( {\rm Pf}^{(N)} {1 \over u_{ij}} \Biggr)^{-1}\times \sum \limits_{1 \leq a < b \leq N} (-1)^{a+b+1} \, [ab] \; {\rm Pf}_{ab}^{(N \to N-2)} {1 \over u_{ij}}  \, .
\end{eqnarray}
Here ${\rm Pf}^{(N)}A_{ij}$ is the Pfaffian \cite{Pfaff} of the $N \times N$ antisymmetric matrix with elements $A_{ij}= - A_{ji}$, the sum in (\ref{arbitraryT}) contains ${1\over 2}N(N-1)$ terms, and ${\rm Pf}_{ab}^{(N \to N-2)} A_{ij}$ is the Pfaffian of the $(N-2) \times (N-2)$ matrix obtained from the $N \times N$ matrix by removing the $a$th and $b$th rows and columns. In the limit $u_N \to \infty$ Eq. (\ref{arbitraryT}) yields the stress tensor for an arbitrary odd number $N-1$ of switches. Equation (\ref{4T}) follows from Eq. (\ref{arbitraryT}) for $N=4$ and Eq. (\ref{3T}) from Eq. (\ref{4T}) in the limit $u_4\to\infty$. Since the operator $T$ is even in the order parameter field $\phi$, $\langle T(w) \rangle_{u_1, u_2 , ..., u_N}$ is unchanged on exchanging $+$ and $-$ in the boundary conditions. 

Equation  (\ref{arbitraryT}) follows from the result 
\begin{eqnarray} \label{phiphiinh}
&&\langle \phi(w_1 , \bar{w}_1)  \, \phi(w_2 , \bar{w}_2) \rangle_{u_1 , u_2 , ..., u_N} = \Biggl( {\rm Pf}^{(N)} {1 \over u_{ij}} \Biggr)^{-1} \times \nonumber \\
&& \qquad \qquad \times \langle \phi(w_1 , \bar{w}_1)  \, \phi(w_2 , \bar{w}_2) \rangle \, {\rm Pf}^{(N)} \Biggl[ {1 \over u_{ij}} {\langle \phi(w_1 , \bar{w}_1)  \, \phi(w_2 , \bar{w}_2) \rangle_{u_i , u_j} \over  \langle \phi(w_1 , \bar{w}_1)  \, \phi(w_2 , \bar{w}_2) \rangle} \Biggr] \, .
\end{eqnarray}
of Burkhardt and Guim \cite{BG} for the two-point correlations of the order parameter in the presence of mixed boundary conditions.
In this expression the angular brackets without subscripts denote thermal averages for a homogeneous $+$ or $-$boundary condition. In the limit that $w_1$ is much closer to $w_2$ than to the boundary, Eq. (\ref{phiphiinh}) must be consistent with the OPE 
\begin{eqnarray} \label{phiphiOPE}
\phi(w_1 , \bar{w}_1)  \, \phi(w_2 , \bar{w}_2) \, &\to& \, |w_{12}|^{-1/4} \Biggl\{1 - {1 \over 2} |w_{12}| \epsilon(w , \bar{w}) + {1 \over 4} \Bigl[ w_{12}^2 T(w) + \bar{w}_{12}^2 \bar{T}(\bar{w}) \Bigr] + {\cal O}(|w_{12}|^3) \Biggr\} \, , \nonumber \\
w_{12} \equiv w_1 - w_2 \, &,& \quad w \equiv (w_1 + w_2)/2 \, , 
\end{eqnarray}
see, e.g., Eq. (2.39) and Sec. IIIC in Ref. \cite{e}. Sustituting the expansion  (\ref{phiphiOPE}) in all the averages in Eq.  (\ref{phiphiinh}) and comparing the coefficients of $ |w_{12}|^{-1/4} w_{12}^2$ on the right and left hand sides leads to Eq. (\ref{arbitraryT}).  

\subsection{Interaction of semi-infinite needles with mixed boundary conditions } \label{intsemisemi}

 Under the mapping  (\ref{twoneedlemapping}) of the upper half $w$ plane onto the $z$ plane with two embedded semi-infinite needles, the intervals $-\infty < u < -b, \, -b < u < 0$ and $0 < u < 1, \, 1 < u < +\infty$, which we denote by (i), (ii) and (iii), (iv),  map onto the edges of the  semi-infinite needles $II$ and $I$, respectively.
%We know from the previous Subsection how the stress tensor average %$<T(w)>$ in the force equations (\ref{stressstress})-(\ref{force''}) 
%depends on the distribution of surface universality classes + and $-$ %along the $u$-axis and consider the following distributions:}
In the notation of the preceding Subsection, we consider the following distributions of surface universality classes $+$ and $-$ along the $u$-axis:   

(1) $N=0$   

(2) $N=2, \, u_1 = -b, \, u_2 = 1$

(3) $N=3, \, u_1 = -b, \, u_2 = 0, \, u_3 = 1$

(4) $N=1, \, u_1 = 0$

(5) $N=1, \, u_1 = 1$\\
\noindent In cases (1) and (4), which were considered in Sec. \ref{semisemi}, the boundary conditions on the two edges of each needle are the same. In cases (2), (3), and (5), on the other hand, one or both of the needles has a different boundary condition on each of its two edges. The stress tensor averages given above allow us to calculate the force between these needles.  

We illustrate the approach in the particularly simple case of {\it collinear} semi-infinite needles generated by the mapping (\ref{twoneedlemapping}) with $b=1$ and $\alpha = \pi$.  Needles $I$ and $II$ occupy the portions  $- \infty < x < -|z(1)|= - 4{\cal B}$ and $0<x<+\infty$, respectively, of the $x$-axis, and the four intervals of the boundary of the $w$ plane map onto the upper and lower edges of needles $I$ and $II$ according to
\begin{eqnarray} \label{maptorims}
({\rm i}) \to II_{\rm lower}, \, ({\rm ii}) \to II_{\rm upper}, \, ({\rm iii}) \to I_{\rm upper}, \, ({\rm iv}) \to I_{\rm lower} \, . 
\end{eqnarray}

Starting with $+$ at $u=- \infty$, Eq. (\ref{maptorims}) implies
\begin{eqnarray} \label{distribclasses}
\left( \begin{array}{cc}
I_{\rm upper} & II_{\rm upper} \\
I_{\rm lower} & II_{\rm lower}
\end{array} \right) =
\left( \begin{array}{cc}
+ & + \\
+ & +
\end{array} \right) \, ,
\left( \begin{array}{cc}
- & - \\
+ & +
\end{array} \right) \, ,
\left( \begin{array}{cc}
+ & - \\
- & +
\end{array} \right) \, ,
\left( \begin{array}{cc}
- & + \\
- & +
\end{array} \right) \, ,
\left( \begin{array}{cc}
+ & + \\
- & +
\end{array} \right) \, 
\end{eqnarray}
in cases (1)-(5), respectively.
 The force acting on needle $I$ follows from Eqs. (\ref{force}) and (\ref{force''}),  the collinear needle mapping $z(w)$, and the averages $\langle T(w) \rangle_{...}$ in Eqs. (\ref{1+2T}) and (\ref{3T}). The component $f_x$  is given by \cite{intform}    
\begin{eqnarray} \label{collfor}
16 |z(1)| f_x /(k_B T) = 1, \; 1+16 \tilde{t}=9, \; -7, \; 1-32 \tilde{t}=-15, \, 1-8 \tilde{t}=-3; \quad \tilde{t} \equiv 1/2   
\end{eqnarray}
in cases (1)-(5), respectively, and the component $f_y$ vanishes in all the cases except (5), where $|z(1)|f_y /(k_B T)=-2 \tilde{t} /\pi \equiv -1/\pi$.  

It is remarkable that in case (2) of Eqs. (\ref{distribclasses}) and (\ref{collfor}), the attraction is 9 times stronger than in case (1)$\thinspace$! To help understand this result, note that for the same  {\em nonvanishing} distance $|z(1)|$ between the needle tips, the free energy is greater in case  (2) than in case (1), since in case (2) the spins change direction near the needle tips. However, when the tips touch, the free energy is the same in cases (1) and (2), since the  upper and lower halfs of the $z$ plane are decoupled. Thus, the free energy varies more rapidly with the tip separation in case (2).

\subsection{Semi-infinite needle perpendicular to an infinite needle}  \label{intsemiinh}

Next we consider a semi-infinite needle $I$ in the upper half $z$ plane oriented {\em perpendicular} to an infinite needle $II$ on the $x$ axis, as described by Eq. (\ref{infandsemiinfmapping}) with $\alpha=\pi/2$. The tip of needle $I$ is at $z=z(1)=4{\cal A}i\thinspace$, and the pre-image of the origin $z=0$ is at $w=u=-1$. Allowing for both a homogeneous boundary ($+$ for all $x$) and a boundary with a ``chemical step'' at the origin (i.e., a mixed boundary with $+$ for $x<0$ and $-$ for $x>0$), and allowing for different boundary conditions on the right and left edges of needle $I$, we consider the six cases
\begin{eqnarray} \label{distribclasses'}
\left( \begin{array}{cc}
I_{\rm left} & I_{\rm right} \\
II_{\rm left} & II_{\rm right}
\end{array} \right) =
\left( \begin{array}{cc}
+ & + \\
+ & +
\end{array} \right) \, ,
\left( \begin{array}{cc}
+ & - \\
+ & -
\end{array} \right) \, ,
\left( \begin{array}{cc}
- & + \\
+ & -
\end{array} \right) \, ,
\left( \begin{array}{cc}
- & - \\
+ & +
\end{array} \right) \, ,
\left( \begin{array}{cc}
- & + \\
+ & +
\end{array} \right) \, ,
\left( \begin{array}{cc}
+ & + \\
+ & -
\end{array} \right) \, .
\end{eqnarray}
The stress tensor averages $\langle T(w) \rangle$ in the first five cases are the five defined in the first paragraph of Subsec. \ref{intsemisemi} with $b=1$ and in the  sixth case $N=2, u_1 =-b=-1, u_2 =0$, corresponding to
\begin{eqnarray} \label{maptorims'}
({\rm i}) \to II_{\rm left}, \, ({\rm ii}) \to II_{\rm right}, \, ({\rm iii}) \to I_{\rm right}, \, ({\rm iv}) \to I_{\rm left} \, . 
\end{eqnarray}
Together with Eqs. (\ref{is1}), (\ref{force}), and (\ref{force''}) this leads to \cite{intform} 
\begin{eqnarray} \label{perphalffor}
32 |z(1)| f_y /(k_B T) &=& -3, \; -3(1+16 \tilde{t})=-27, \; 37, \; -3+128 \tilde{t}=61, \nonumber \\
&& 3(-1+16 \tilde{t})=21, \; -3+32 \tilde{t}=13; \quad \tilde{t} \equiv 1/2\,. 
\end{eqnarray}
The parallel force component $f_x$ vanishes in cases (1)-(5), and in case (6), $|z(1)|f_x /(k_B T)=-2 \tilde{t} \equiv -1$. The factor 9 increase in attraction on going from (1) to (2) has an explanation similar to the one below Eq. (\ref{collfor}).

In principle, one can calculate the force for arbitrary configurations of two semi-infinite or infinite needles with an arbitrary configuration of ``chemical steps'' with this approach.

As a final example, we consider the Casimir force exerted on the semi-infinite needle $I$ by the boundary $II$ of the upper half $z$ plane in the presence of chemical steps at two arbitary points
\begin{eqnarray} \label{twosteps}
x_1 \equiv X_1 |z(1)|  \, < \, x_2 \equiv X_2 |z(1)|\,,
\end{eqnarray}
which separate the $x$ axis into regions with $+ ,\, - ,\, +$ boundary conditions. Needle $I$ has boundary condition $+$ on both of its edges and extends along the $y$ axis from $y=|z(1)|$ to $y=+ \infty$. The arrangement is reminiscent of an atomic force microscope probing an inhomogeneous boundary. The force follows from the mapping (\ref{infandsemiinfmapping}) with $\alpha=\pi/2$ and the stress tensor in (\ref{1+2T}) with $N=2$ and $u_1 =-|u_1|, \, u_2 = - |u_2|$, where
\begin{eqnarray} \label{twosteps'}
|u_j|=1+2X_j^2 - 2X_j \sqrt{1+X_j^2} \, ; \quad j=1,2 \, ,
\end{eqnarray}
and the calculation yields.
\begin{eqnarray} \label{forcetwosteps}
&&|z(1)| {f_x \over k_B T} = 4 \tilde{t} \Biggl\{- {2 \over |u_1|-|u_2|} \Biggl[ {|u_1|^{3/2} \over 1+|u_1|}-{|u_2|^{3/2} \over 1+|u_2|} \Biggr] + \sum\limits_{j=1}^2 {|u_j|^{1/2} (|u_j| +3) \over 2 (1+|u_j|)^2} \Biggr\}\,, \nonumber \\
&&|z(1)| {f_y \over k_B T} = - {3 \over 32} +4 \tilde{t} \Biggl( {1 \over 1+|u_1|}-{1 \over 1+|u_2|} \Biggr)^2\,,
\end{eqnarray}
where $|u_j|$ is defined in Eq. (\ref{twosteps'}). 

To get a feeling for the result, we discuss two special cases.

(A) Boundary with a single step: In the limit $x_2 \to + \infty$, i.e. $|u_2| \to 0$, only the single step on the boundary at $x_1$ remains.  It separates regions with $+$ and $-$ boundary conditions to its left and right, respectively. The corresponding force on needle $I$ is
\begin{eqnarray} \label{forceonestep}
&&|z(1)| {f_x \over k_B T} = - 2 \tilde{t} {|u_1|^{1/2} (3|u_1| +1) \over  (1+|u_1|)^2}\,,  \nonumber \\
&&|z(1)| {f_y \over k_B T} = - {3 \over 32} +4 \tilde{t} {|u_1|^2 \over (1+|u_1|)^2}\,,
\end{eqnarray}
with $|u_1|$ given by Eq. (\ref{twosteps'}). While the parallel force component $f_x$ is negative for all $x_1$, the perpendicular component $f_y$ changes sign from positive to negative on increasing $x_1$ beyond a critical value of the order of $|z(1)|$. This is expected, since needle $I$ with its $+$ edges is attracted to the the $+$ region and repelled by the $-$ region of the boundary. For $x_1 \to [- \infty, \, 0, \, + \infty]$, $|z(1)|f_x /(k_B T) \to - \tilde{t}[3/|X_1|, \, 2, \, 1/X_1]$, and $|z(1)|f_y /(k_B T) \to -(3/32) + \tilde{t} [4-(2/X_1^2), \, 1,  \, 1/(4 X_1^4)]$. For $x_1 =0$ one recovers case (6) defined below Eq. (\ref{distribclasses'}), and Eq. (\ref{forceonestep}) reproduces the corresponding force components given in the paragraph containing Eq. (\ref{perphalffor}). For $x_1 \to - \infty$ and $x_1 \to +\infty$, Eq. (\ref{forceonestep}) approaches the force in cases (4) and (1) of Eq. (\ref{perphalffor})

(B) Boundary with two steps at equal distances from the needle: Since the configuration, with steps at $\pm x_1$  separating the $x$ axis into regions $+,-,+$ is mirror symmetric about the $y$-axis, the parallel component $f_x\thinspace$ of the force vanishes. The perpendicular component follows from Eqs. (\ref{twosteps'}) and (\ref{forcetwosteps}), which  yield $|u_2|=1/|u_1|$ and
\begin{eqnarray} \label{twosymmsteps}
|z(1)| {f_y \over k_B T} = - {3 \over 32} + 4 \tilde{t} {X_1^2 \over 1+X_1^2} \, .
\end{eqnarray}
For $x_1 =0$ the boundary with two steps reduces to a homogeneous $+$ boundary, and we are back to case (1) of Eqs. (\ref{distribclasses'}) and (\ref{perphalffor}). For a large distance between the steps, $|x_1| \gg |z(1)|$, Eq. (\ref{twosymmsteps}) yields
\begin{eqnarray} \label{twodiststeps}
|z(1)| {f_y \over k_B T} \to - {3 \over 32} + 4 \tilde{t} - {4\tilde{t} \over X_1^2} \, .
\end{eqnarray}
Here the first two terms on the right hand side represent the force exerted on the needle by a homogeneous $-$ boundary, and the third term is contributed by the $+$ boundaries beyond the two distant steps. As expected, the latter contribution is twice the corresponding contribution $-2\tilde{t} /X_1^2$ of a single distant step, given below Eq. (\ref{forceonestep}). 

For switches of the boundary universality class between $+$ and $O$ instead of $+$ and $-$, $\langle T(w) \rangle_{u_1}$ and 
$\langle T(w) \rangle_{u_1, u_2}$ are again given by Eq. (\ref{1+2T}), \cite{BX}, but with $\tilde{t} = \tilde{t}_{+O} =1/16$ instead of  $\tilde{t} = \tilde{t}_{+-} =1/2$ .
Thus, all of the results of this Appendix which are based on the stress tensor for $N=1$ or $N=2$ hold, with the appropriate value of  $\tilde{t}$, for $+O$ as well as $+-$ switches in the boundary conditions. 
%Replacing in the cases discussed above the universality class $-$ by class $O$ %the forces on needle $I$ follow from replacing in the above forces the %coefficient $\tilde{t} \equiv \tilde{t}_{+-} =1/2$ by $\tilde{t} \equiv %\tilde{t}_{+O}=1/16$.   
%
\newpage
\newpage
\begin{figure}[HereABCD]
\begin{center}$
\begin{array}{cc}
\includegraphics[width=5.0in]{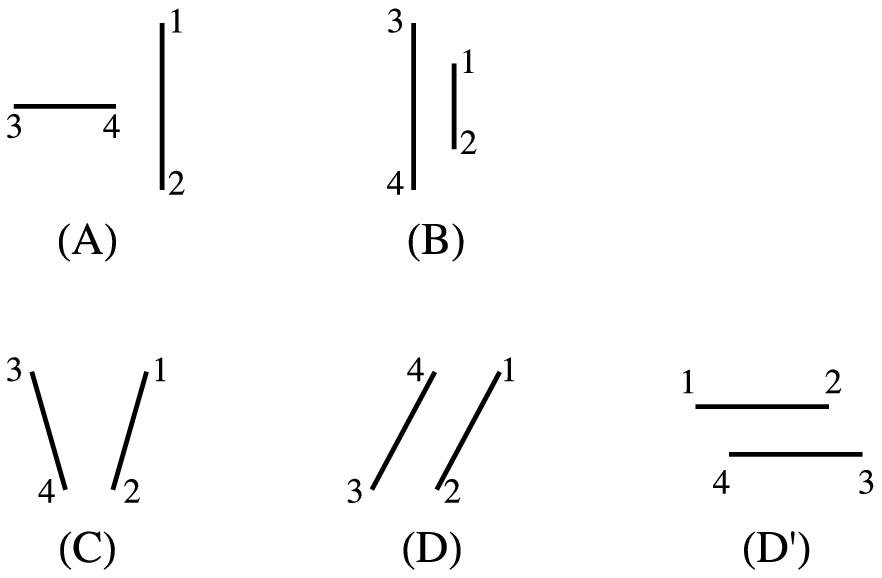}
\end{array}$
\caption{Simple configurations of two needles defined in Sec. \ref{map} for which the six mapping parameters $C$, $h$, $\varphi_1$, ..., $\varphi_4$ of the conformal mapping are restricted to subspaces of lower dimension. Configuration (D') is the same as (D), apart from a rotation to orient the needles along the $x$ axis. Our results for the force and torque in configurations (A), (B), (C), and (D') are presented in Figs. \ref{FigAfx}-\ref{FigDtheta}.}\label{FigABCD}
\end{center}
\end{figure}
\newpage
\begin{figure}[HereA]
\begin{center}$
\begin{array}{cc}
\includegraphics[width=3.0in]{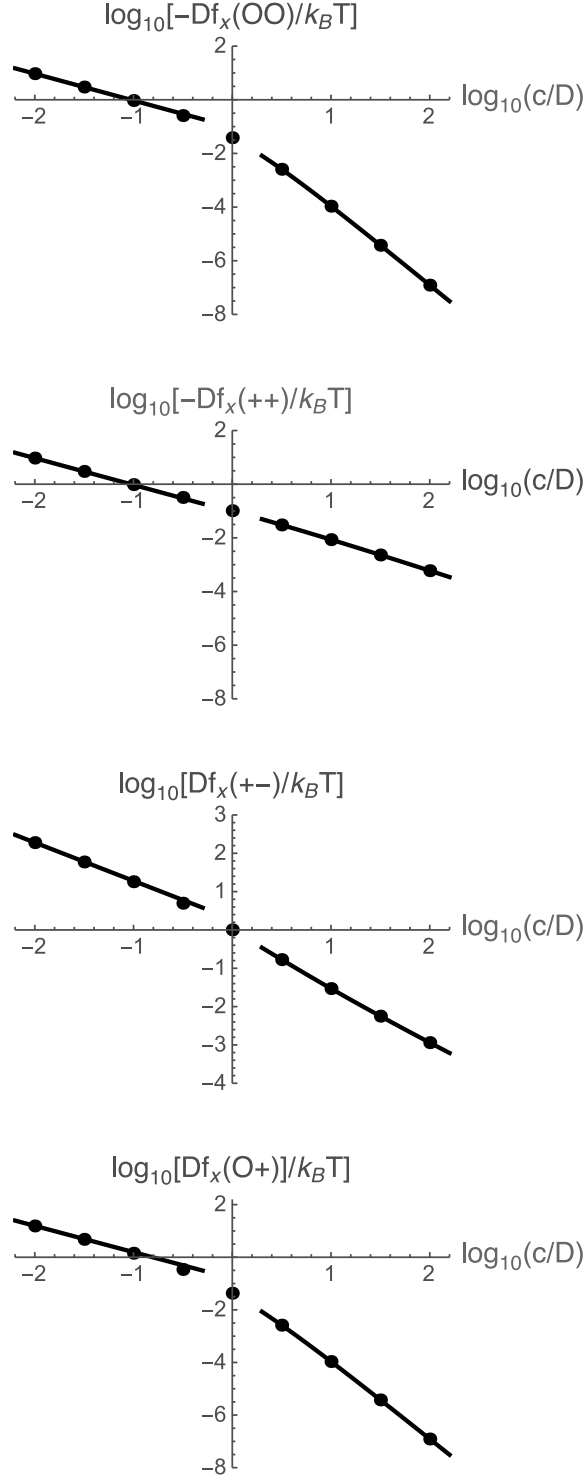}
\end{array}$
\caption{Component $f_x$ of the force exerted on needle $I$ by needle $II$ for needles of equal length $D$ in the symmetric-perpendicular configuration (A) shown in Fig. \ref{FigABCD}. Here $c=z_I-z_4$ is the distance from the right tip of needle $II$ to the midpoint of needle $I$.  The points indicate exact numerical results, and the two curves show the asymptotic form for large and small $c/D$. The force component $f_y$ and the torque vanish due to symmetry. For more details see Sec. \ref{five}. }
\label{FigAfx}
\end{center}
\end{figure}

\newpage
\begin{figure}[HereB]
\begin{center}$
\begin{array}{cc}
\includegraphics[width=2.9in]{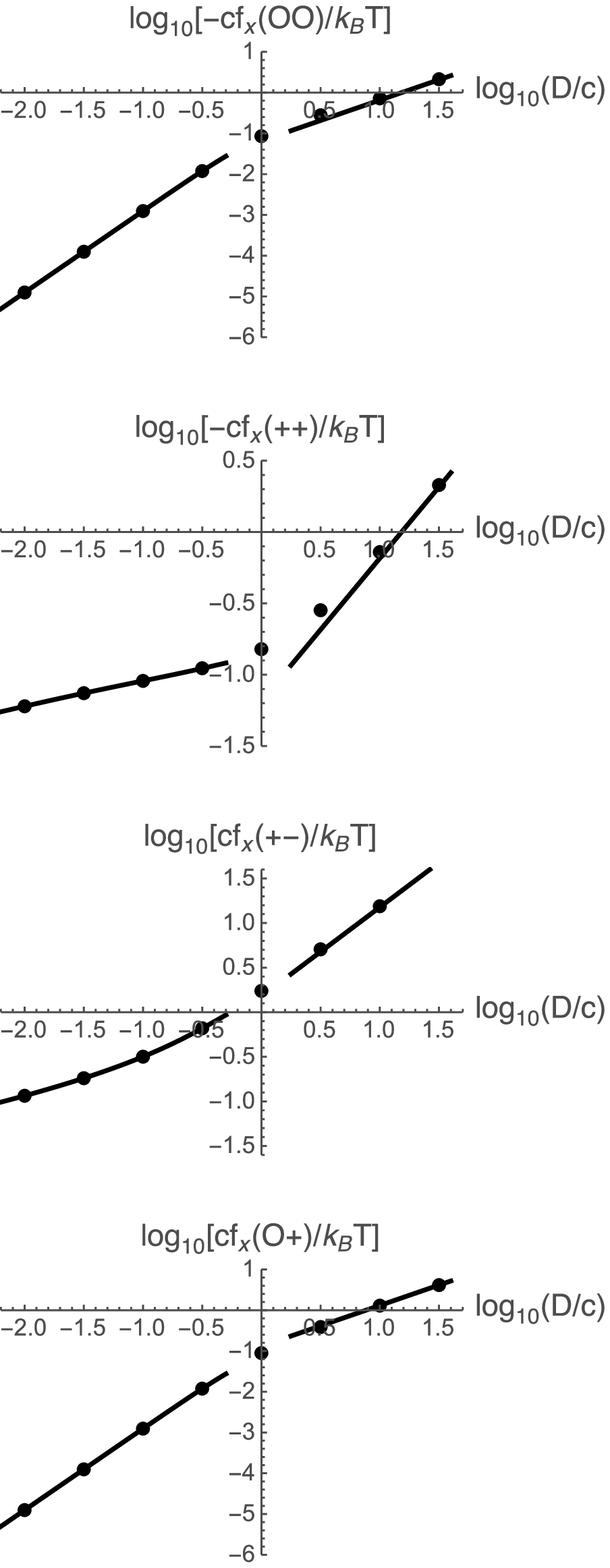}\quad
\includegraphics[width=2.9in]{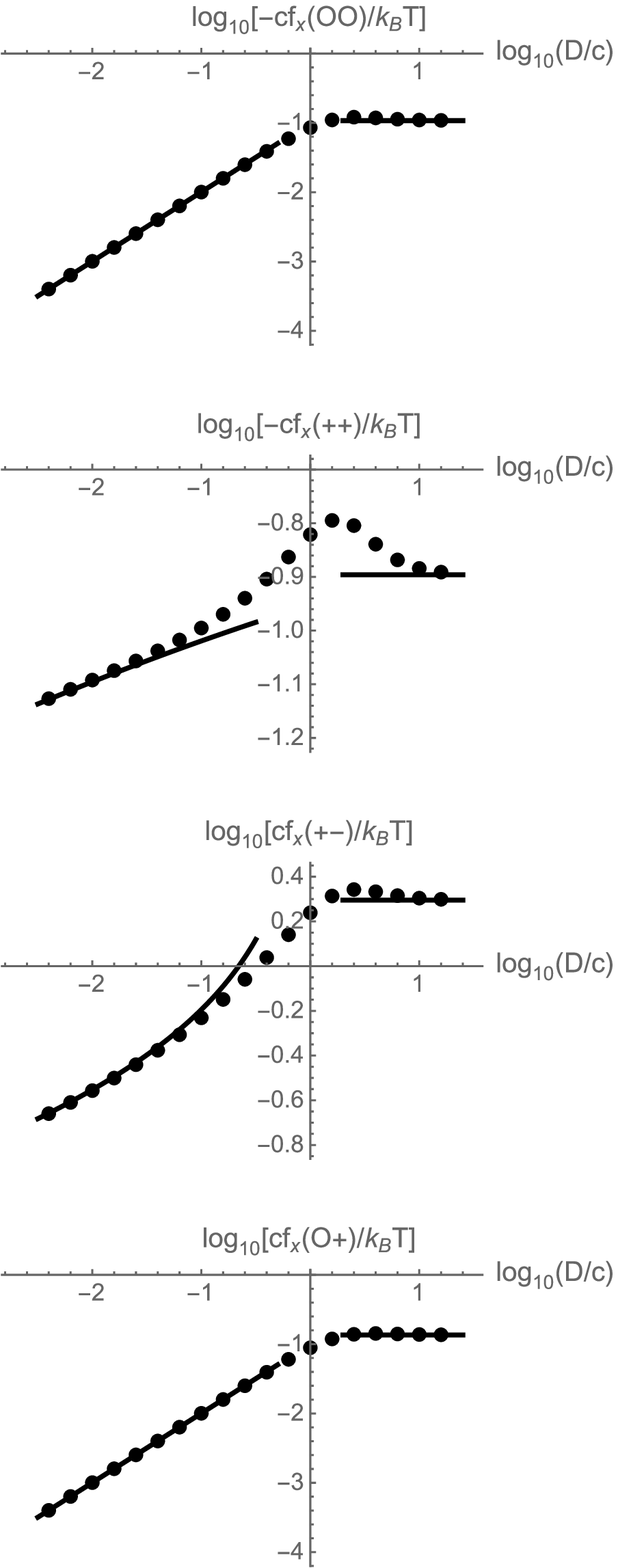}
\end{array}$
\caption{Component $f_x$ of the force exerted on needle $I$ by needle $II$ for needles with separation $c=z_I-z_{II}$ in the symmetric-parallel configuration (B) in Fig. \ref{FigABCD}. The results in the left and right columns are for needles of the same length $D$ and for needles with different lengths, $c$ and $D$, respectively. The points indicate exact numerical results, and the curves show the asymptotic form for large and small $D/c$. The force component $f_y$ and the torque vanish due to symmetry. For more details see Sec. \ref{five}.}
\label{FigBfx}
\end{center}
\end{figure}

\newpage
\begin{figure}[HereCfxtheta]
\begin{center}$
\begin{array}{cc}
\includegraphics[width=3.0in]{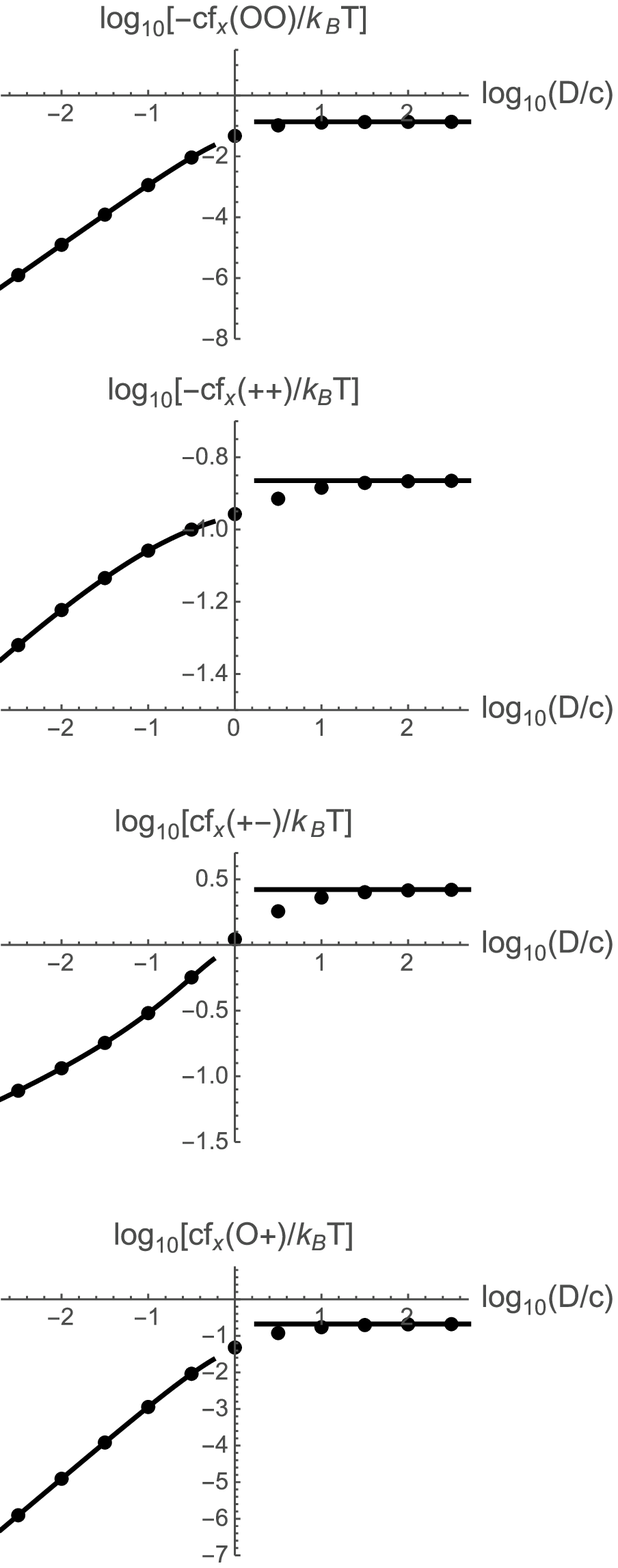}\quad
\includegraphics[width=3.0in]{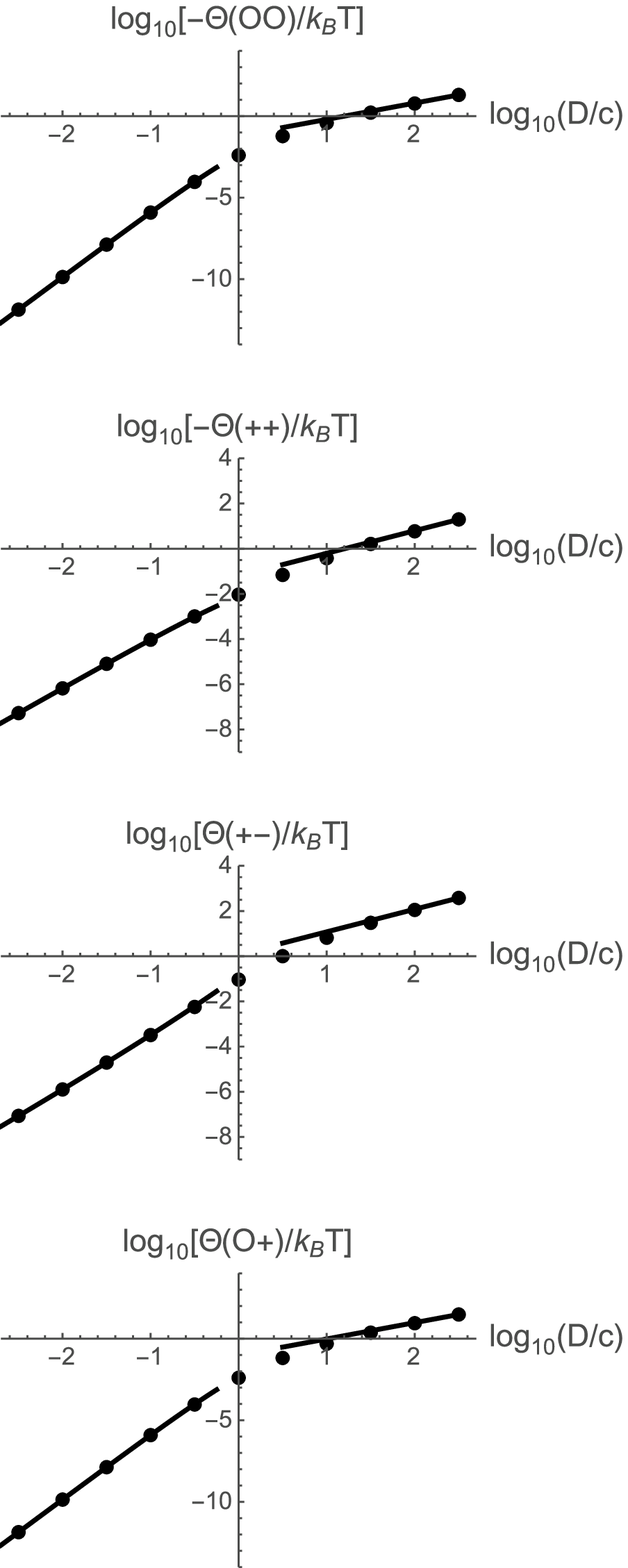}

\end{array}$
\caption{Force $f_x$ and torque $\Theta$ exerted on needle $I$ by needle $II$ for needles of equal length $D$ in the mirror-symmetric configuration (C) shown in Fig. \ref{FigABCD}. Here $c=z_2-z_4$ is the distance between the closest points of the needles, and the angle between them is $\pi/5$. The points indicate exact numerical results, and the two curves show the asymptotic form for large and small $D/c$. The force component $f_y$ vanishes due to symmetry. For more details see Sec. \ref{five}.}
\label{FigCfxtheta}
\end{center}
\end{figure}
\newpage
\begin{figure}[HereDfxfy]
\begin{center}$
\begin{array}{cc}
\includegraphics[width=3.0in]{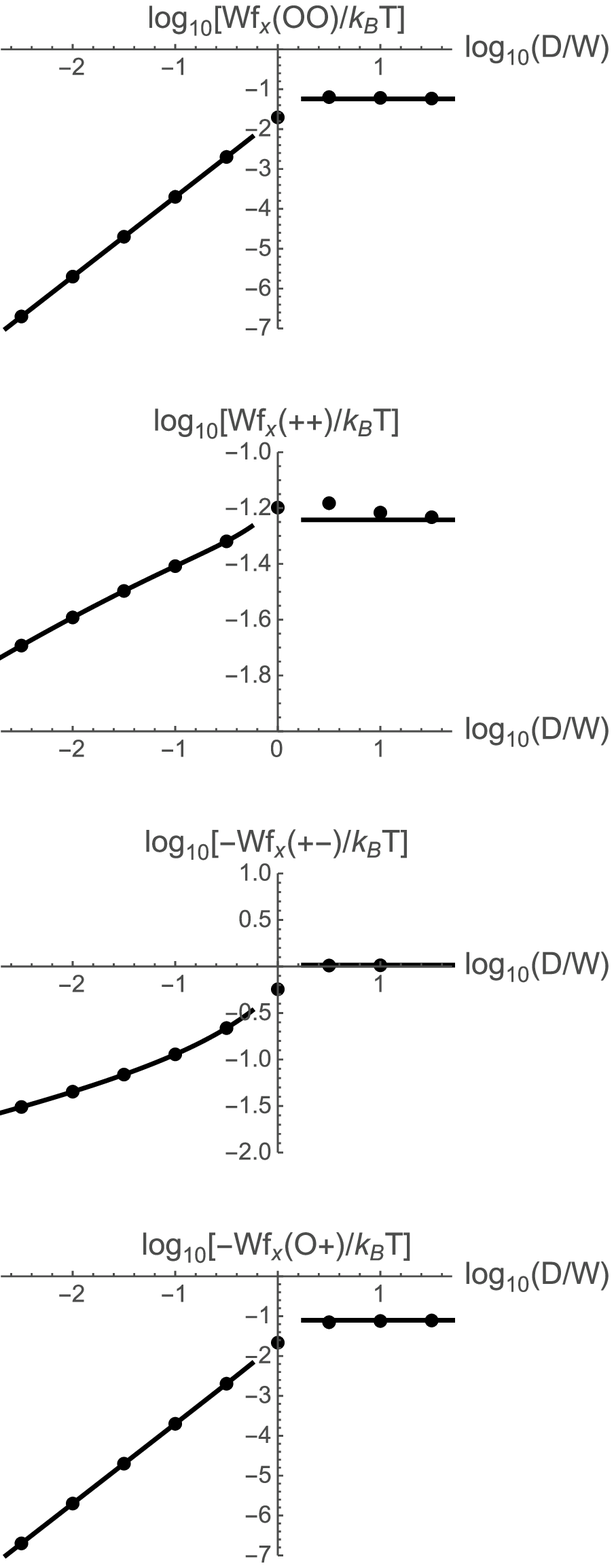}\quad
\includegraphics[width=3.0in]{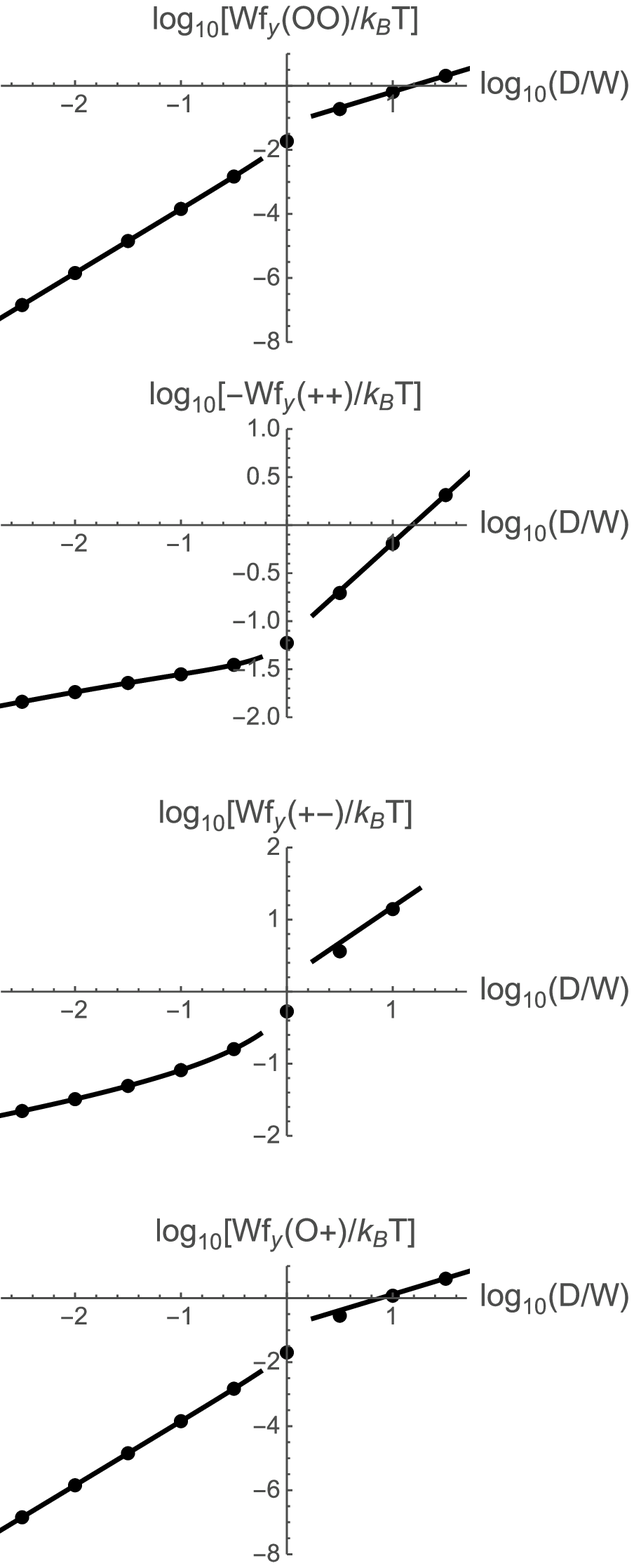}
\end{array}$
\caption{Components $f_x,\,f_y$ of the force exerted on needle $I$ by needle $II$ for needles of equal length $D$ in the antiparallel configuration (D') shown in Fig. \ref{FigABCD}. Here $W=r_{1,y}-r_{4,y}$ is the vertical separation of the needles, and results are shown for the fixed ratio 
$\left(r_{4,x}-r_{1,x}\right)/W=1.4$. The points indicate exact numerical results, and the two curves show the asymptotic form for large and small $D/W$. For more details see Sec. \ref{five}.}
\label{FigDfxfy}
\end{center}
\end{figure}
\newpage
\begin{figure}[HereDtheta]
\begin{center}$
\begin{array}{cc}
\includegraphics[width=3.3in]{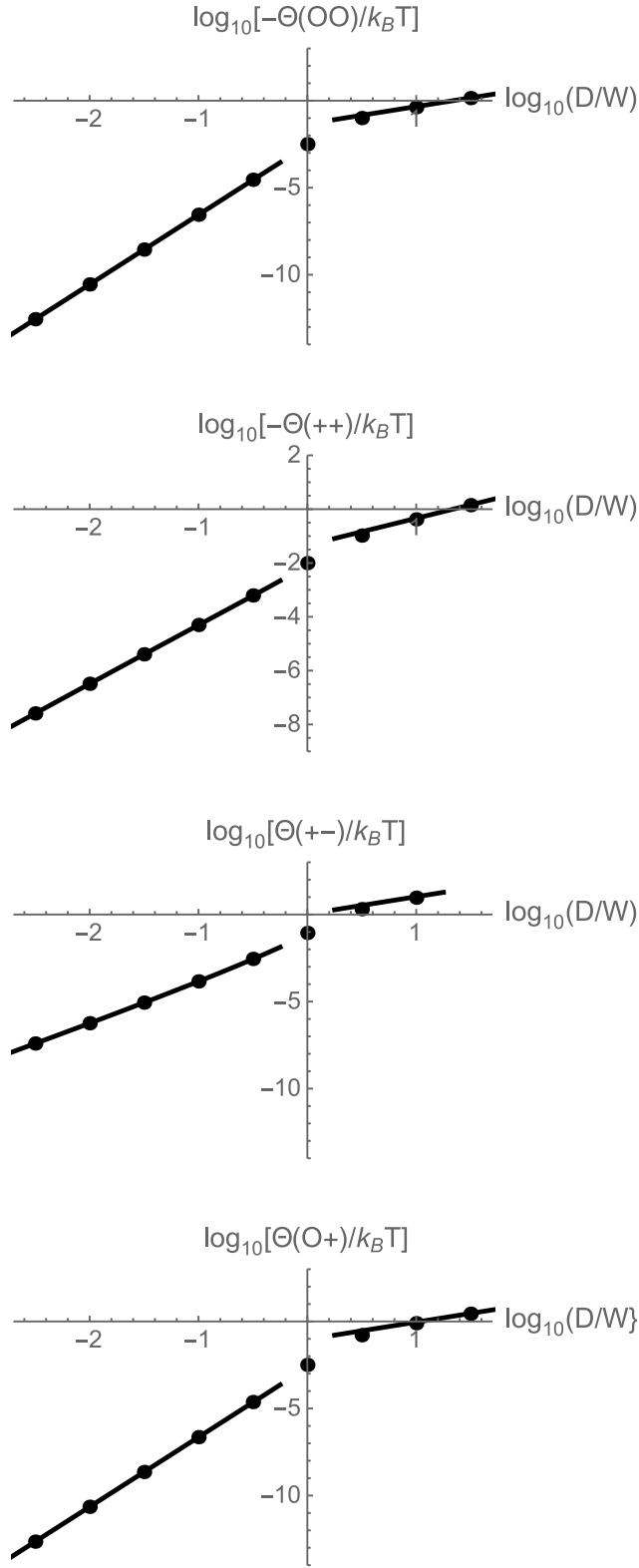}
\end{array}$
\caption{Torque $\Theta$ on needle $I$ for the same antiparallel configuration considered in Fig. \ref{FigDfxfy}.}
\label{FigDtheta}
\end{center}
\end{figure}
\newpage
\begin{figure}[HereE10]
\begin{center}$
\begin{array}{cc}
\includegraphics[width=3.0in]{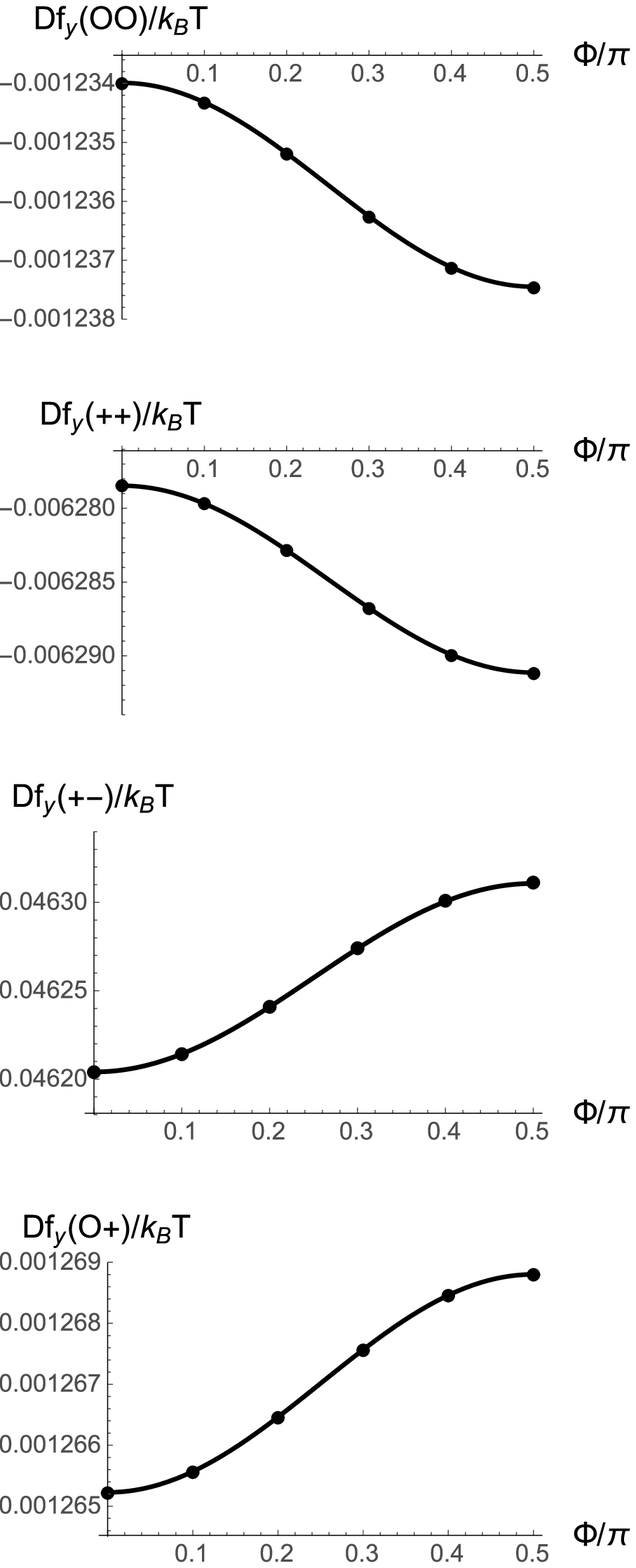}\quad
\includegraphics[width=3.0in]{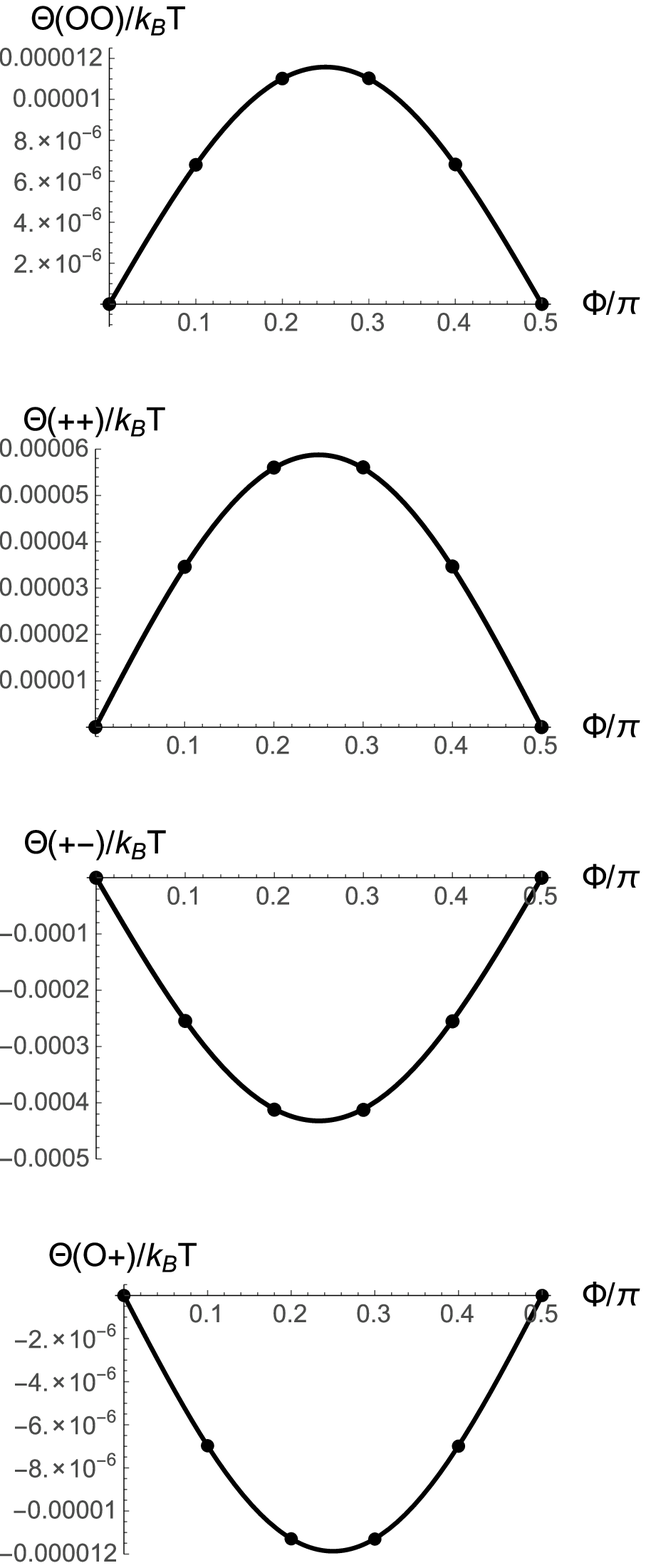}
\end{array}$
\caption{Dependence of the force $f_y$ and torque $\Theta$ on a needle of length $D$ in the upper half plane on the angle $\Phi$ between the needle and the boundary. The reduced distance $\tilde{B}=r_{I,y}/D$ of the needle midpoint from the boundary has the value $\tilde{B}=10$ . The points indicate exact numerical results, and the curves show the asymptotic predictions (\ref{fhalfplane}) and (\ref{fhalfplane'}) of the operator expansion for a distant needle, i.e. for large $\tilde{B}$. The force component $f_x$ vanishes. For more details see Sec. \ref{five}.}
\label{FigE10}
\end{center}
\end{figure}
\newpage
\begin{figure}[HereE1]
\begin{center}$
\begin{array}{cc}
\includegraphics[width=3.0in]{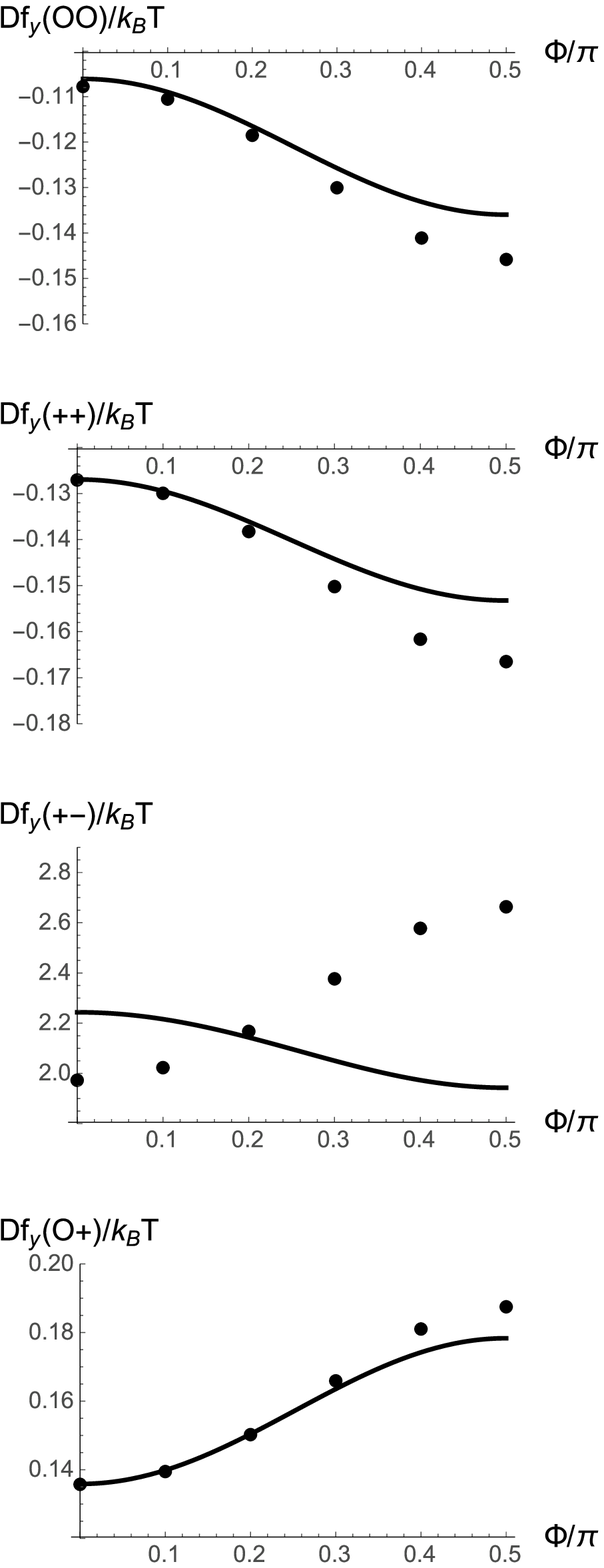}\quad
\includegraphics[width=3.0in]{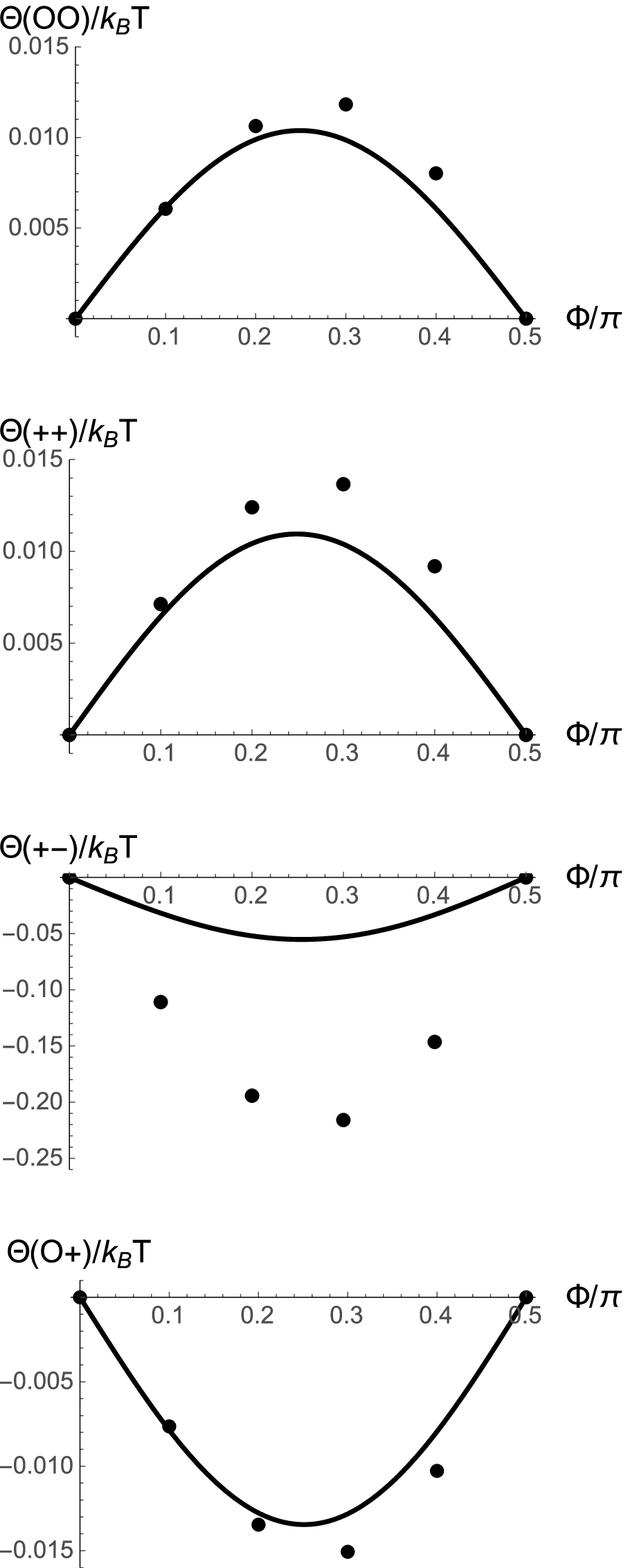}
\end{array}$
\caption{Same as Fig. \ref{FigE10} except that the reduced distance has the value $\tilde{B}=1$ instead of 10. Since the distant needle condition $\tilde{B}\gg1$ is not satisfied, the exact numerical results (points) deviate significantly from the the predictions (curves) of the operator expansion for a distant needle. For more details see Sec. \ref{five}.}
\label{FigE1}
\end{center}
\end{figure}
\newpage
\begin{figure}[HereE433]
\begin{center}$
\begin{array}{cc}
\includegraphics[width=2.8in]{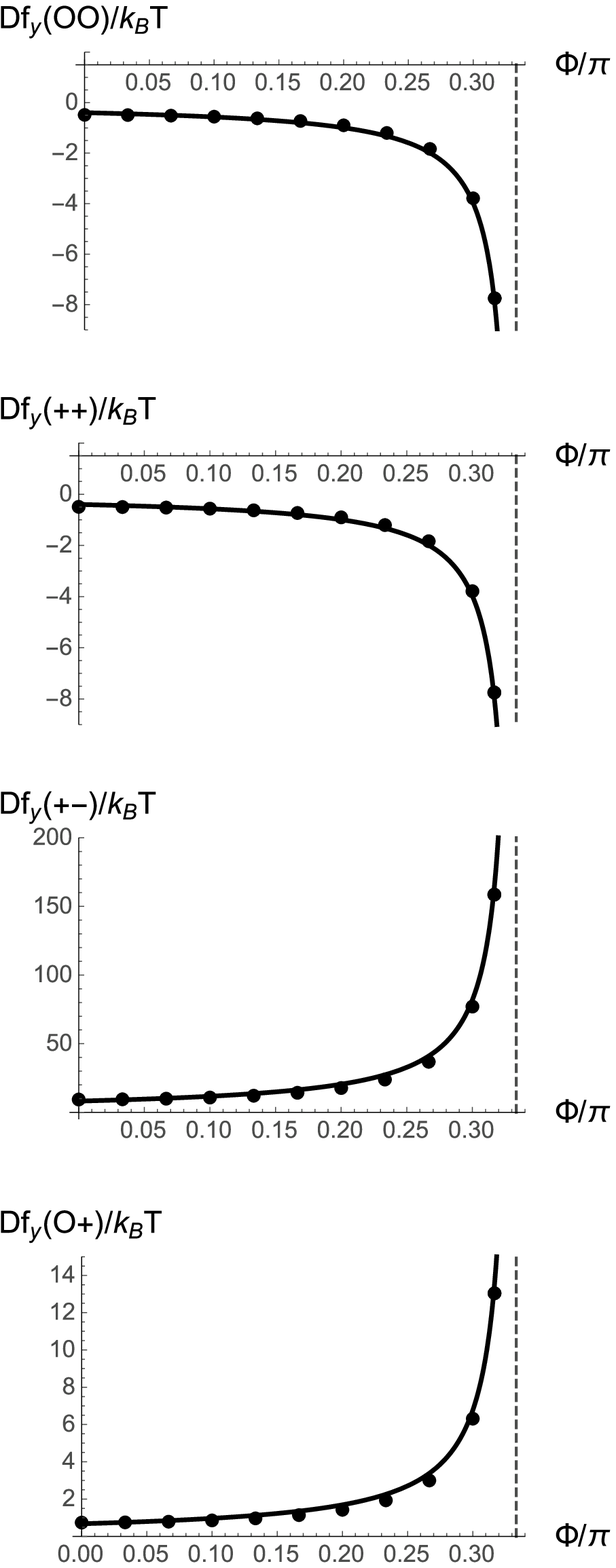}\quad
\includegraphics[width=2.8in]{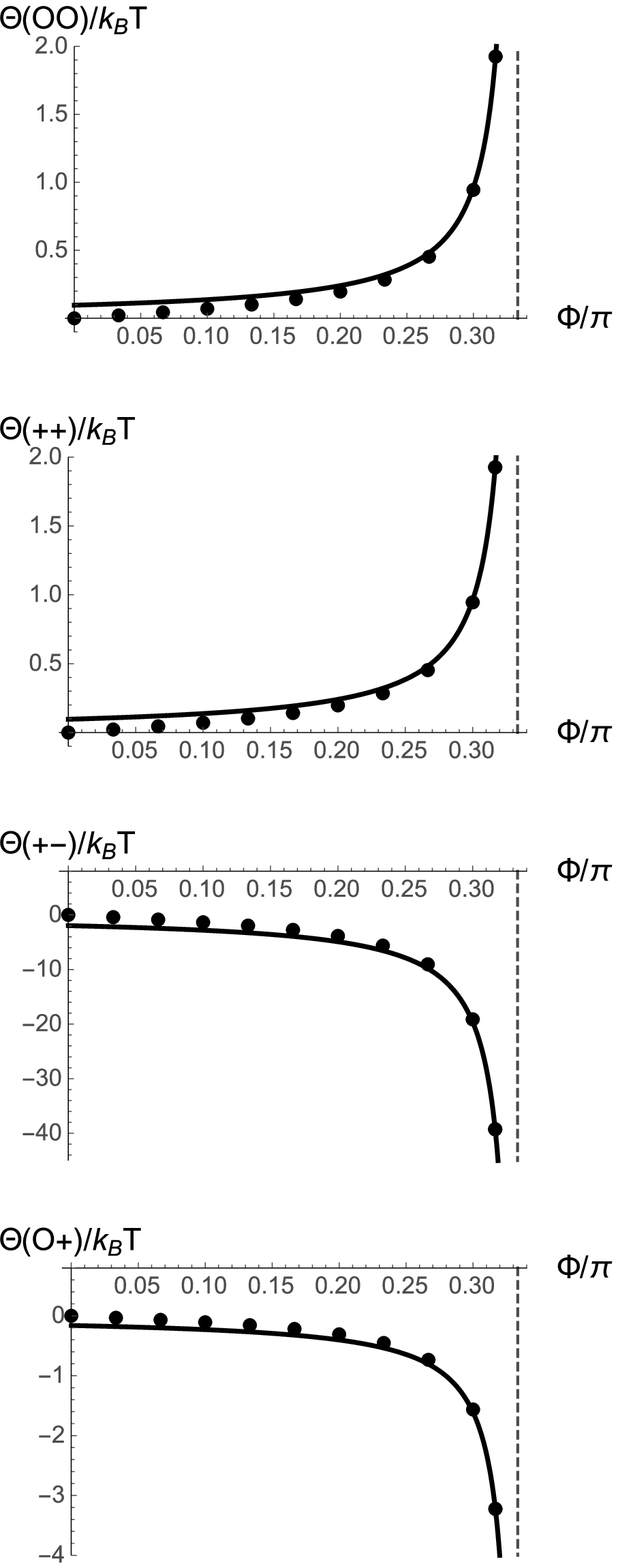}
\end{array}$
\caption{Same as Fig. \ref{FigE10} except that $\tilde{B}=\sqrt{3}/4=0.433$. Both $f_y$  and $\Theta$ diverge as $\Phi$ approaches $\pi/3$, the angle at which the needle tip touches the boundary. For $\Phi$ close to $\pi/3$, the exact numerical results (points) for $f_y$ agree with the asymptotic expression (curves) given in the last paragraph of Sec. \ref{five}. Like $f_y$, $\Theta$ also appears to diverge as $\left({\pi\over 3}-\Phi\right)^{-1}$. The exact numerical results for $\Theta$ are compared with fits of the form $\Theta_{\rm fit}(\Phi)=A({\pi\over 3}-\Phi)^{-1}$, with $A$ chosen to reproduce the rightmost point in each graph.}
\label{FigE433}
\end{center}
\end{figure}

\begin{thebibliography}{99}
\bibitem{distsmall}We use the terms ``distant'', ``small'', and ``short'' interchangeably to describe a needle that is short compared with its distance from the rest of the system but large on a microscopic scale.
\bibitem{delF}The free energy of interaction $\delta F$ is the free energy required to move the two needles from infinite separation to the configuration of interest.
\bibitem{intform}A useful formula for evaluating the integrals in Sec. \ref{infseminf} and Appendix \ref{diffrims} is\\ $\int_{0}^{\infty} dU U^{\mu -1} [(U+b)(U-a+i0)]^{-1} =-\pi \csc (\pi \mu) [b^{\mu -1} +i^{2\mu} a^{\mu-1}]/(a+b)\thinspace,$ where $a$ and $b$ are positive.
\bibitem{but} For finite needles that nearly touch, $h$ is very close to $1$. This regime can be studied analytically with the help of ``Jacobi's imaginary transformation,'' which relates $\vartheta_1$ with parameter $h$ to $\vartheta_1$ with parameter $h' \equiv {\rm exp}\Bigl( -\pi^2 / \ln (1/h) \Bigr)$, since $h'\to 0$ for $h\to 1$. For details see \S 22 of Akhiezer's book \cite{Akh}.
\bibitem{rotateD}Rotating configuration (D) of Sec. \ref{map} counterclockwise by an angle of $\pi-\Phi_{12}$, where $0<\Phi_{12}<\pi$, yields a configuration (D') with needle vectors  $z_{12}=-|z_{12}|$ and $z_{34}=|z_{34}|$ antiparallel and parallel to the real axis, respectively, and with needle $I$ located above needle $II$, as shown in Fig. \ref{FigABCD}-(D) and \ref{FigABCD}-(D'). In Sec. \ref{five} and Appendix \ref{strongover} this new orientation is more convenient. The angle $\Phi_{12}$ is related to the new $x$ and $y$ components by $\cot \Phi_{12} = - (r_{1,x} -r_{4,x})/W$, $W \equiv r_{1,y} -r_{4,y} = |r_{1,y} -r_{4,y}|$.
\bibitem{tmeaning}At bulk criticality the free energy of an annulus bounded by two
concentric circles with radii $h \varrho$ and $\varrho$ contains a
scale-free contribution $f_{\rm annulus}(h)$ 
which is independent of $\varrho$, depends only on $h$, and
vanishes for $h \to 0$. Since $h(d/dh)f_{\rm annulus}/(k_{B}T)=-t(h)$, this contribution is directly related to $t(h)$ in Eqs. (\ref{Tann})-(\ref{Spq}). While $t(h\ll 1)$ is given in Eqs. (\ref{tOOO+}) and (\ref{t+++-}) and is relevant for the small particle expansion, $t(h \to 1) = -2 \pi (1-h)^{-2} \Delta$ is determined by the interaction between the parallel boundaries of an infinitely long strip, where the Derjaguin approximation is exact \cite{ber}. Here $\Delta=\pi[-1,-1,23,2]/48$ is the amplitude \cite{Cardy} in the Casimir contribution $\Delta/W$ to the free energy per $k_B T$ and per unit length of the strip with width $W$ and boundary universality classes $OO,++,+-,O+$ and is related \cite{BX} via $\Delta=\pi ({\tilde t}-1/48)$ to the amplitude ${\tilde t}$ in Eq. (\ref{t12}). This is consistent with $h \to 1$ for needles much longer than their minimum separation.
\bibitem{localT}That $\langle T(w) \rangle$ in (\ref{Tann}) does not change on swapping the surface universality classes of the two boundary circles follows from its transformation under the inversion $w \to h/w$. This result and Eq. (\ref{stressstress}) imply that $<T(z)>$ for a given configuration of two needles is also unaffected by an exchange of their surface universality classes.
\bibitem{fxcoll}Only in exceptional cases does the force between two finite needles have a simple analytic form, free of special functions, for arbitrary distance to size ratios. One such exception is the collinear configuration (i) of Sec. \ref{map}, where the force on needle $I$ is given by $f_x /(k_B T) = \pm \{8c[(c/D)+1][(c/D)+2] \}^{-1}$ for $O+$ (upper sign) and $OO$ (lower sign) boundaries. Here $c=z_2 - z_4 \equiv r_{2,x}-r_{4,x}$ is the distance between the closest points of the needles. This result follows from stress tensor averages $\langle T(z) \rangle$ given in Appendix B of Ref. \cite{VED} on using our Eqs. (\ref{Tz}) and (\ref{shiftz}).
\bibitem{uniscale}Since translating or rotating
particle $I$ relative to particle $II$ as described in
(\ref{fixed})-(\ref{rotate}) does not alter the nonuniversal contributions to the bulk and surface free energies and to the free energy of needle ends, $F_{\hat{G}}-F_{G}$ can be
identified with the change in the total free energy.
\bibitem{paralimit}The expression for $f_x$ for parallel semi-infinite needles derived in Appendix \ref{strongover} also follows from the results of Sec. \ref{semisemi} for the force between needles which form a nonvanishing angle $\alpha>0$. Before taking the limit $\alpha \searrow 0$, one replaces the positive real constant ${\cal B}$ in Eqs. (\ref{tn1})-(\ref{fxy}) by $|{\cal B}|e^{-i\alpha /2}$, so that needles $I$ and $II$ form angles $\alpha /2$ and $-\alpha /2$ with the $x$ axis and from the outset $f_x$ is an odd function of the difference of the $x$ coordinates of the needle tips.
\end{thebibliography}
\end{document}